\documentclass[twocolumn,aps,pra,superscriptaddress,nofootinbib]{revtex4-2}
\usepackage{bm,physics,amsmath,embedfile}

\usepackage[normalem]{ulem}

\usepackage[T1]{fontenc}
\usepackage{tikz}

\usepackage[utf8]{inputenc}
\usepackage{xcolor}
\usepackage{lipsum}
\usepackage{mathtools} 
\usepackage{amssymb}   
\usepackage{appendix}
\usepackage{braket}
\usepackage{graphicx}
\usepackage{hyperref}
\usepackage{multirow}
\usepackage{hhline}
\usepackage{comment}

\usepackage{breqn}

\makeindex

\begin{document}
\title{Emergence of a Bose polaron in a small ring threaded by the Aharonov-Bohm flux}

	\author{Fabian Brauneis}
	\email{fbrauneis@theorie.ikp.physik.tu-darmstadt.de}
	\affiliation{Technische Universit\"{a}t Darmstadt$,$ Department of Physics$,$ 64289 Darmstadt$,$ Germany}
	
	\author{Areg Ghazaryan}
	\affiliation{Institute of Science and Technology Austria (ISTA)$,$ Am Campus 1$,$ 3400 Klosterneuburg$,$ Austria} 
	
	\author{Hans-Werner Hammer}
	\affiliation{Technische Universit\"{a}t Darmstadt$,$ Department of Physics$,$ 64289 Darmstadt$,$ Germany}
	\affiliation{ExtreMe Matter Institute EMMI and Helmholtz Forschungsakademie
  Hessen f\"ur FAIR (HFHF)$,$ GSI Helmholtzzentrum f\"ur Schwerionenforschung GmbH$,$ 64291 Darmstadt$,$ Germany}
	
	\author{Artem G. Volosniev}
	\email{artem.volosniev@ist.ac.at}
	\affiliation{Institute of Science and Technology Austria (ISTA)$,$ Am Campus 1$,$ 3400 Klosterneuburg$,$ Austria}

	\begin{abstract}
	\vspace{1em}
\noindent \textbf{Abstract}
\vspace{1em}

	The model of a ring threaded by the Aharonov-Bohm flux underlies our understanding of a coupling between 
	gauge potentials and matter. 
    The typical formulation of the model is based upon a single particle picture, and should be extended when interactions with other particles become relevant.
	Here, we illustrate such an extension for a particle in an Aharonov-Bohm ring subject to interactions with a weakly interacting Bose gas. We show that the ground state of the system can be described using the Bose polaron concept -- a particle dressed by interactions with a bosonic environment. We connect the energy spectrum to the effective mass of the polaron, and demonstrate how to change currents in the system by tuning boson-particle interactions. Our results suggest the Aharonov-Bohm ring as a platform for studying coherence and few- to many-body crossover of quasi-particles that arise from an impurity immersed in a medium.  
	\end{abstract}

\maketitle


\vspace{1em}
\noindent \textbf{Introduction}
\vspace{1em}

\noindent In the idealized model of a ring threaded by the Aharonov-Bohm (AB) flux,
 a particle moves in a region with zero fields, and the presence of an electromagnetic potential manifests itself only in a minimal substitution $-i\partial/\partial x\to -i\partial/\partial x + \Phi$, where the position-independent parameter $\Phi$ determines the strength of the flux.
This model provides insight into many physical phenomena. For example, it illustrates the significance of potentials in quantum mechanics~\cite{Aharonov1959}, geometric phases~\cite{Berry1984}, the Josephson effect and persistent currents~\cite{Bloch1968,BUTTIKER1983}. Foundations of the AB physics are based upon a single-particle picture~\cite{Aronov1987,VIEFERS2004}, which already has the power to explain some experiments qualitatively such as spectroscopy in semiconductor rings~\cite{Lorke2000}. However, one-body studies do not take into account interactions with other particles, in particular, with the environment. Therefore, they should be extended for realistic systems. In this paper, we discuss such an extension assuming a one-dimensional bosonic environment.

\begin{figure}
\centering
    \includegraphics[width=1\linewidth]{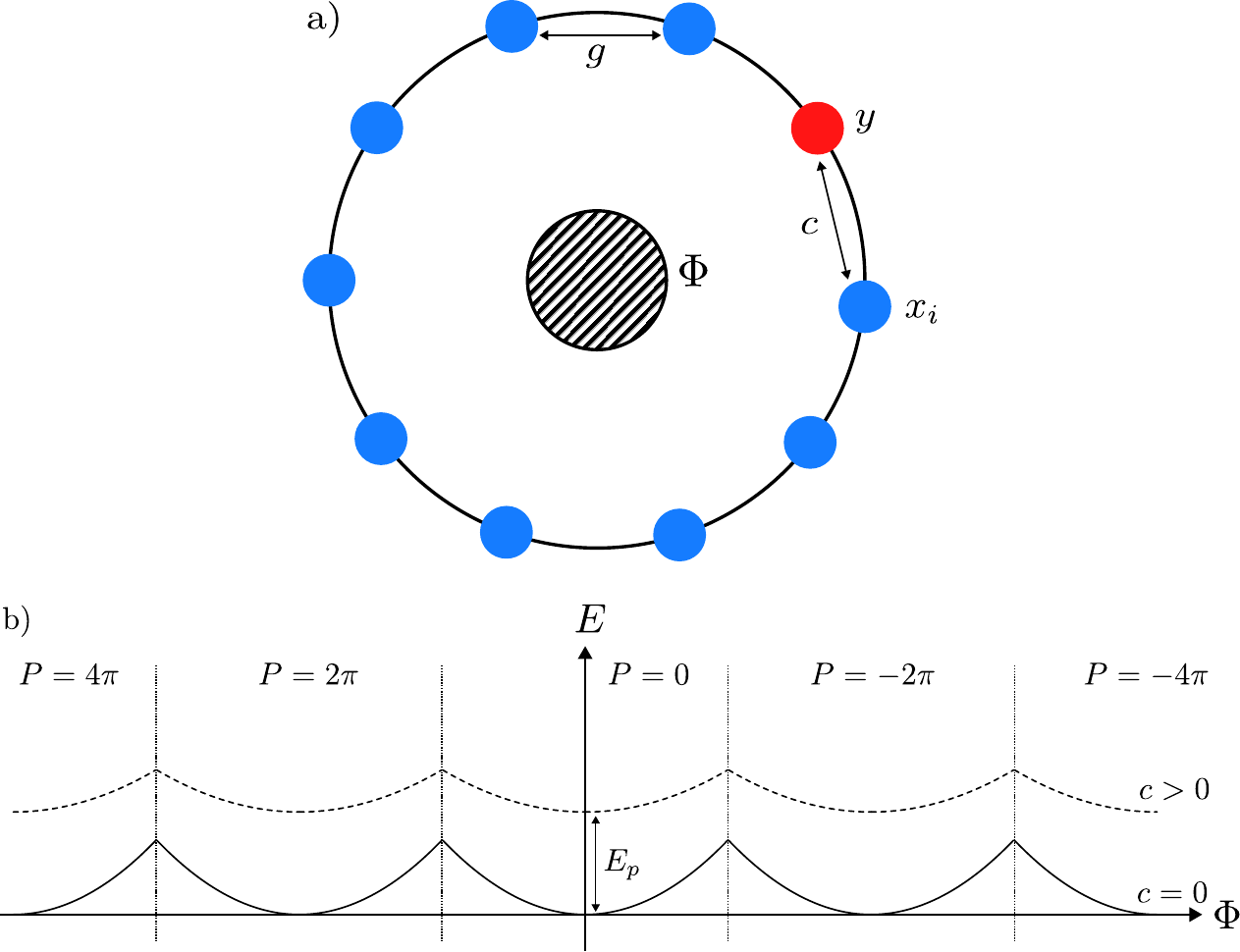}
    \caption{{\bf Sketch of the system and the ground state energy}. \\
    a): Sketch of the system. The bosons are shown as blue balls at the positions $x_i$; the impurity is a red ball at $y$. The AB flux is $\Phi$. The strength of the boson-boson (boson-impurity) interaction is given by $g$ ($c$). b):  Sketch of the ground-state energy as a function of the flux without ($c=0$) and with ($c>0$) the interaction with the bosonic environment (leading to an energy shift $E_p$ at $\Phi=0$). The total momentum of the system is $P/L$. The shift of the energy spectrum, $E_p$, and the change of the curvature can be parameterized by effective one-body parameters, see the text for details.}
    \label{fig:EnergySketch}
\end{figure}

Before we proceed, let us briefly review known few-body physics in the AB ring. If all particles are identical, then
the flux couples only to the total angular momentum of the system. It can change the global minimum of the energy leaving the internal [i.e., in relative coordinates] dynamics intact, see, e.g.,~\cite{Groeling1993,VIEFERS2004,MANNINEN2012,Naldesi2022}. In short, there is no interplay between particle-particle interactions and the AB flux for identical particles. This conclusion holds true also for distinguishable [by spin or quasi-spin] particles with identical masses and AB fluxes. In this case the strength of the AB flux can however change the symmetry of the ground state, see, e.g., Pecci et al.~\cite{Pecci2022}.

For particles with non-identical charges and/or masses, such as electrons and holes~\cite{Romer2000,Kyriakou2010},
the internal structure of a one-dimensional system is coupled to the AB flux. At the two-body level, this coupling can modify the threshold for binding (which may preclude formation of excitons for weakly attractive potentials in one dimension~\cite{Moulopoulos2004}) or lead to formation of dark excitonic states~\cite{Ghazaryan2011}. Systems with more than two particles are less explored, to the best of our knowledge. 

In this paper, we study one of the simplest two-component many-body models --  a particle (impurity) coupled to the AB flux that interacts with a Bose gas.
The system is motivated by recent cold-atom experiments on Bose polarons~\cite{spethmann2012,Catani2012Experiment,Jin2016,Arlt2016,Ardilla2019,Yan2020,Skou2021Polaron}, and by theoretical and experimental progress in realizing ring-shape potentials and artificial gauge fields with neutral cold atoms. For reviews of these advances see~\cite{Dalibard2011,Amico2021,Amico_2022}. Ring-shaped condensates with effective gauge potentials have so far not been engineered together with impurities. As we show below such a combination may lead to rich physics. Note that recent advances in engineering ring shaped potentials~\cite{RingShapedPolaritons1, RingShapedPolaritons2, RingShapedPolaritons3} and tunable gauge fields~\cite{PolaritonsGauge} suggest experiments with polaritons as other set-ups to test our results. 

The focus of the paper is on `dressing' the impurity -- a typical question addressed in many-body physics -- which determines properties of the system such as transport and `magnetization'. As such,
our results complement previous works that investigated small systems using few-body methods and approaches.

One of the main findings of our work is that the system can be described using ideas developed for the one-dimensional Bose-polaron problem~\cite{Casteels2012Impurity,Petkovic2016,Schecter2016,Parisi2017BosePolaron,Grusdt2017BosePolaron,Volosniev2017BosePolaron,Pastukhov2017Impurity,Kain2018Static,Mistakidis2019QuenchBosePolarons,Jager2020Deformation}. This connection leads to a number of useful conclusions.
First, previous studies of the Bose polaron contribute insight into properties of our system, and provide an intuitive interpretation of our results. This insight can be also useful for understanding numerical lattice simulations where electron-phonon interactions are taken into account, see, e.g., Monisha et al.~\cite{Monisha2016}. Second, persistent currents can be an experimental measure of validity of the Bose-polaron concept in one dimension. In particular, they can be used to investigate phase coherence of the polaron across the AB ring -- a necessary condition for the existence of persistent currents. 
Third, the AB ring provides a conceptual model for
defining the effective mass in a finite-size system, allowing one to better understand a few- to many-body crossover of  one-dimensional systems. 
In particular, our work paves the way for studying this crossover beyond  the standard testbed -- the ground-state energy~\cite{wenz2013}.


\vspace{1em}
\noindent \textbf{Results and Discussion}
\vspace{1em}

\noindent \textbf{System.} We study a one-dimensional system of $N$ bosons and a single impurity atom, see Fig.~\ref{fig:EnergySketch}. The system is in a ring of length $L$, which corresponds to periodic boundary conditions. The position of the impurity ($i$th boson) is given by $L y$ ($L x_i$); the mass of the impurity (a boson) is $m$ ($M$). We assume that {\it only} the impurity is coupled to the AB flux $\Phi/L$. 
For neutral particles, $\Phi$ is not generated by
a magnetic flux threading the ring. Instead, other techniques are used~\cite{Dalibard2011,Goldman2014}, e.g., stirring with a weak external potential with speed $v$, in which case $\Phi=mvL/\hbar$. Note that the more general case, which might be more suitable for experimental realization, where the artificial flux is coupled to both particle species can be easily incorporated in our model, see Suppl. Note~1 for the flux coupled to bosons. 

The Hamiltonian in first quantization reads
\begin{equation}
    \mathcal{H}=h+H+V_{ib}+V_{bb}
    \label{eq:Hamiltonian},
\end{equation}
where $h = \frac{\hbar^2}{2mL^2}\left(-i \partial/\partial y+ \Phi\right)^2$ describes the impurity; for the bosons, we have $H=-\frac{\hbar^2}{2M L^2}\sum_i\partial^2/\partial x_i^2$. 
The impurity-boson, $V_{ib}$, and boson-boson, $V_{bb}$, interactions are parameterized by delta-function potentials 
\begin{equation}
V_{ib}=\frac{c}{L}\sum\limits_{i=1}^N\delta(x_i-y), \qquad V_{bb}=\frac{g}{L}\sum\limits_{i,j}\delta(x_i-x_j),
\end{equation}
where $c$ and $g$ define the strength of interactions. For simplicity, we shall use the system of units in which $\hbar=M=1$. In the main part of the paper, 
a boson and the impurity have identical masses, $m=M$. [A mass imbalance does not change the main conclusions of our study, see Suppl. Note~2]. For a fixed value of $N$, dimensionless parameters that determine all physical properties are $c/g$ and $\gamma=g L/N$. For our numerical simulations, we shall use $\gamma=0.2$, which corresponds to a weakly-interacting Bose gas amenable to the mean-field treatment discussed below.  We focus on $c>0$ to avoid bound states~\cite{Gunn1988,Kolomeisky2004,Brauneis2022} that are beyond the polaron physics. Note that the case with $\gamma=0.2$ and $N=19$ for $\Phi=0$ was considered in~\cite{Yang2022} providing us with a reference point to benchmark our numerical calculations. 

In what follows, we shall use the Hamiltonian from Eq.~(\ref{eq:Hamiltonian}) in our analysis. However, it is worthwhile noting that the parameter $\Phi$ can in principle be excluded from this Hamiltonian via a gauge transformation $\Psi\to e^{i\Phi y}\Psi$, where $\Psi$ is the wave function. The effect of the flux is then incorporated in a `twisted' boundary condition that demands that the wave function acquires a phase $e^{i 2\pi\Phi}$ after a full turn~\cite{Byers1961,Imry1986}. Such a condition implies that the energy spectrum must be a periodic function with period $\Phi/(2\pi)$ as shown in Fig.~\ref{fig:EnergySketch}. 
Note that for general multi-component systems (e.g, strongly interacting Bose-Fermi mixtures) a smaller period of the ground-state energy is also possible, see, e.g.,~\cite{Pecci2022,Naldesi2022}. As we demonstrate below, this does not happen for an impurity in a weakly-interacting Bose gas whose low-energy spectrum resembles that of a single particle. 

The Hamiltonian $\mathcal{H}$ with $\Phi=0$ corresponds to one of the most studied one-dimensional models~\cite{cazalilla2011,Guan2013Review,Sowinski2019Review,Mistakidis2022}.  Therefore, we can use the already known methods to tackle our problem with $\Phi\neq 0$. We choose to work in the frame co-moving with the impurity (see below), where 
the mean-field approach (MFA) and flow equations (IM-SRG) provide powerful theoretical tools for our investigation (see Methods). These methods allow us to investigate the effect of the AB flux on the properties of the one-dimensional polaron problem beyond previous studies~\cite{CHEN1998537, ZHOU1996167}, which investigated relevant molecular-crystal models. In particular, we can  define and study flux-independent properties of the Bose polaron (e.g., the effective mass) in a finite system.

\vspace{1em}
\noindent {\bf Co-moving frame.} The total momentum of the system is conserved since all interactions are translation-invariant. Therefore, we eliminate the impurity coordinate by writing the wave function as (cf.~\cite{Gross1962})
\begin{equation}
    \Psi(y, \{x_i\})=\Tilde{\Psi}( \{z_i\})e^{iPy},
\end{equation}
where $z_i=\theta(y-x_i)+x_i-y$ [$\theta(x)$ is the Heaviside step function], and $P/L$ is the total momentum.  The transformation $\{y,x_i\}\to \{z_i\}$ can be seen as a coordinate-space analogue of the Lee-Low-Pines transformation~\cite{Lee1953}.  The parameter $P$ is quantized to fulfill the periodic boundary conditions: $P=2\pi n$, where $n$ is an integer. Note that the transformation to the co-moving frame has been already used to study the few- to many-body transition in the ground state of the Bose-polaron problem with $\Phi=0$~\cite{Volosniev2017BosePolaron}. Here, we study this transition for non-vanishing $P$ where a continuous parameter $\Phi$ provides a bridge between the discrete values of $P$.

The Schr{\"o}dinger equation in the co-moving frame  reads as follows
\begin{equation}
    \begin{split}
        &\bigg[-\frac{1}{2}\left(\sum_i\frac{\partial}{\partial z_i}\right)^2+\frac{1}{2}\sum_i\frac{\partial^2}{\partial z_i^2}+ \frac{(P+\Phi)^2}{2}\\
        &+L^2V_{ib} + L^2 V_{bb} + i(P+\Phi)\sum_i\frac{\partial}{\partial z_i}\bigg]\Tilde{\Psi} =  E\Tilde{\Psi},
        \end{split}
        \label{eq:SchroedingerEquation}
\end{equation}
where $E$ is the dimensionless energy of the system [to obtain the dimensionful energy one needs to multiply $E$ by $\hbar^2/(m L^2)$].
Note that $\Phi$ and $P$ enter this equation together as a sum $\mathcal{P}=P+\Phi$. $\mathcal{P}$ is a continuous variable that can be seen as an effective total momentum that determines total currents in the system. This observation will be crucial for interpreting our results in terms of an effective one-body picture, see below. Note that $\tilde \Psi_{\mathcal{P}}^*$ solves the Schr{\"o}dinger  Eq.~(\ref{eq:SchroedingerEquation}) with $-\mathcal{P}$, which is a manifestation of time-reversal symmetry.

\begin{figure}
\centering
    \includegraphics[width=1\linewidth]{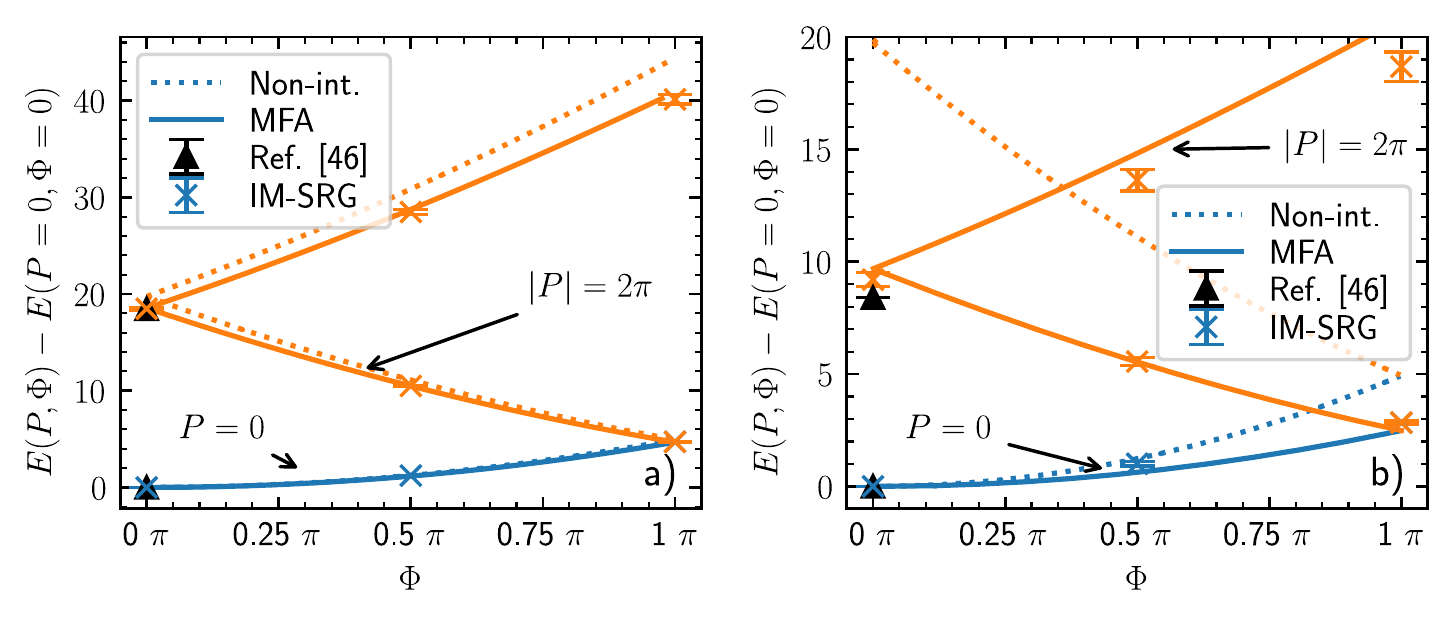}
    \caption{{\bf The  energy spectrum as a function of the AB flux, $\Phi$.}\\
    Different colors correspond to different values of the momentum: Blue $P=0$ and orange $|P|=2\pi$. The parameters of the system are $N=19$, $\gamma=0.2$, and $c/g=1$ [a)], $c/g=5$ [b)]. The data are obtained using the mean-field ansatz (solid curves, MFA) and the in-medium similarity renormalization group (crosses with errorbars, IM-SRG), black triangles show results from Yang et al.~\cite{Yang2022} for quantized momenta. The dotted curves show the energy of the non-interacting system.}
    \label{fig:Energy_c_02}
\end{figure}


\vspace{1em}
\noindent {\bf Energy spectrum.} The energy spectrum for $\Phi=P=0$ and finite values of $N$ was calculated in Volosniev et al.~\cite{Volosniev2017BosePolaron}. Therefore, in what follows we only calculate $E(P,\Phi)-E(P=0,\Phi=0)$, where $E(P,\Phi)$ is the energy of the Hamiltonian for a given value of the total momentum, $P$, and the AB flux, $\Phi$.
$E(P=0,\Phi=0)$ approaches the ground-state energy of the Bose-polaron problem in the thermodynamic limit ($N,L\to\infty$ with a fixed value of $N/L$), see also Suppl. Note~4.  Due to the periodicity of the energy (cf.~Fig.~\ref{fig:EnergySketch}), it is enough to focus on fluxes $-\pi\leq \Phi\leq \pi$. Furthermore, the energy spectrum is symmetric with respect to $\Phi\to-\Phi$ due to time-reversal symmetry.   Therefore, in what follows we shall calculate the lowest-energy states for fixed values of $P$, so called Yrast states (cf.~\cite{VIEFERS2004}), and currents only for $0\leq \Phi\leq \pi$. This also fixes the values of the flux needed to observe our findings experimentally.

{\it Energy for $\Phi\neq 0$}. We illustrate the energies calculated with MFA and IM-SRG in Fig.~\ref{fig:Energy_c_02}. For
$P=0$ and $|P|=2\pi$ both methods agree reasonably well on the energy, demonstrating that the MFA is a useful analytical tool to describe the system. We observed a worse agreement for $|P| \geq 2\pi$. The failure of the mean-field approach is expected for high values of $P$ as there are various ways to distribute momentum between the bosons and the impurity, see also Suppl. Note~3.

Let us give a few general remarks about Fig.~\ref{fig:Energy_c_02}. For $\Phi=\pm\pi$ there is a level crossing between two Yrast states with $P=0$ and $|P|=2\pi$. It is a consequence of the rotational symmetry of the problem. If a defect is introduced into a system, then it will lead to an avoided crossing, see below where we discuss the role of defects. In Fig.~\ref{fig:Energy_c_02}, we also present the ground-state energy of a non-interacting impurity ($c=0$), $E=(P+\Phi)^2/2$. We see that the solid curves are always below this value.  The effect is more pronounced for stronger impurity-boson interactions -- compare the left and right panels of Fig.~\ref{fig:Energy_c_02}. These features can be easily understood using the concept of a polaron and its effective mass.

{\it Effective mass.} For the thermodynamic limit with $\Phi=0$, one finds that the low-energy spectrum of the system is quadratic in the total momentum (see, e.g.,~\cite{Mistakidis2019Effective,Jager2020Deformation} for one-dimensional Bose polarons):
\begin{equation}
    \lim_{P\to0} \left[E(P,\Phi=0)-E(P=0,\Phi=0)\right]= \frac{P^2}{2 m_{\mathrm{eff}}^{\mathrm{TD}}},
    \label{eq:eff_mass_TD}
\end{equation}
where we introduce an effective mass, $m_{\mathrm{eff}}^{\mathrm{TD}}$; other definitions of the effective mass are discussed in Suppl. Note~5.
Eq.~(\ref{eq:eff_mass_TD}) is a cornerstone of the polaron concept and effective one-body descriptions of mobile impurities. Note that for small systems, the limit in Eq.~(\ref{eq:eff_mass_TD}) should be re-defined since $P$ is discrete. 

The parameter $\Phi$ is continuous. The mean-field solution as well as time-reversal symmetry suggest that $E(\Phi, P=0)$ is proportional to $\Phi^2$.  By analogy to the Bose-polaron problem, we can define the effective mass of an impurity in a small AB ring
\begin{equation}
\lim_{P, \Phi\to0} [E(P=0,\Phi)-E(P=0,\Phi=0)]=\frac{\Phi^2 }{2m_{\mathrm{eff}}}.
\label{eq:eff_mass_finite}
\end{equation}
This expression connects our problem to the body of knowledge developed by solving polaron problems. The connection allows one to make predictions about the behavior of an impurity in the AB ring. For example, the effective mass is an increasing function of $c$. Therefore, one reduces the current associated with the impurity by increasing $c$, see the discussion below.  

\begin{figure}
\centering
    \includegraphics[width=1\linewidth]{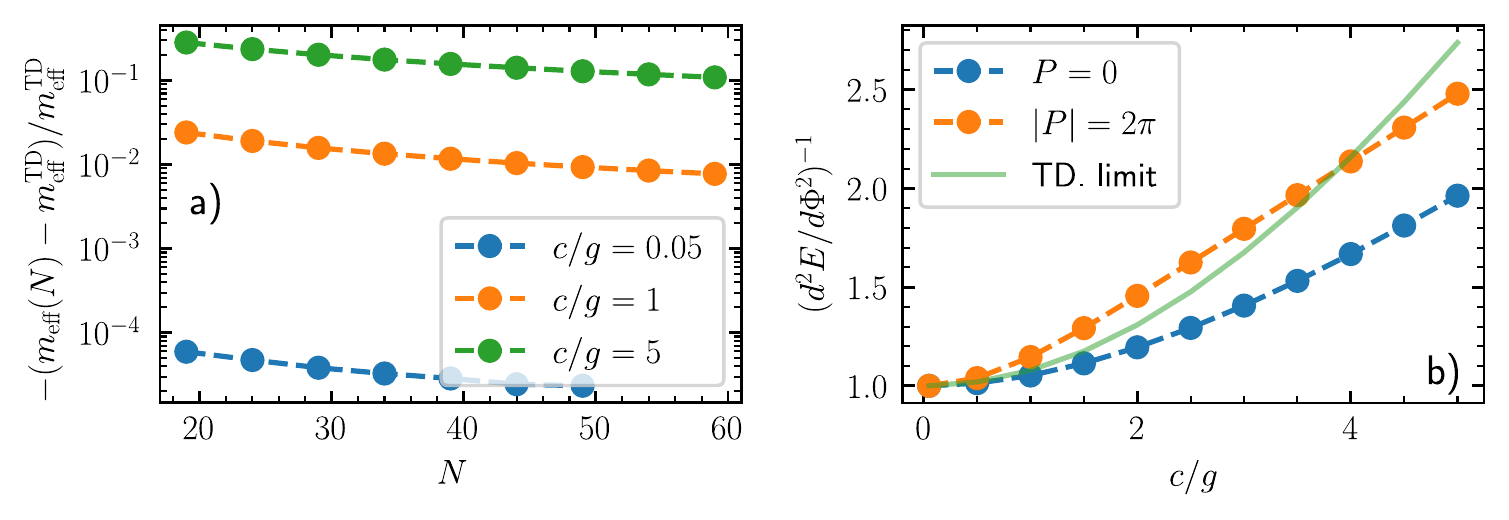}
    \caption{{\bf Effective mass for a finite-size system.}\\
    a): Convergence of the effective mass to its thermodynamic limit for $\gamma=0.2$ and different values of $c/g$. b): Inverse of $\frac{d^2E}{d\Phi^2}$ for $\Phi\to0$ with $P=0$ and $|P|=2\pi$. The parameters of the system are $N=19$, $\gamma=0.2$.  The green curve shows the known result for $P\to 0$ in the thermodynamic limit (TD. limit)~\cite{Jager2020Deformation,Mistakidis2019Effective}. The data in both panels are obtained using the mean-field approach.}
    \label{fig:Non-trivialRedistribution}
\end{figure}

In addition, Eq.~(\ref{eq:eff_mass_finite}) demonstrates that the AB ring can be a physical testbed for studying the few- to many-body crossover
in one-dimensional Bose-polaron problems. We illustrate this crossover for the effective mass in 
Fig.~\ref{fig:Non-trivialRedistribution}. For weak interactions ($c/g=0.05$), we see that the effective mass 
converges to the thermodynamic limit quickly. This is not the case for strong interactions ($c/g=5$), meaning that many bosons are needed to screen the impurity for large values of $c/g$. Although, our analysis suggests that the effective mass converges somewhat slower than the energy towards the thermodynamic limit (see also Suppl. Note~4), the basic mechanism is the same: A high compressibility of a weakly-interacting Bose gas requires a large number of bosons to screen a strongly interacting impurity. Note that the number of bosons needed for screening heavily depends on the parameter $\gamma$. In particular, in the limit $\gamma\to\infty$, the system fermionizes and the impurity is screened by a handful of particles~\cite{wenz2013,Astrakharchik2013Impurity,Levinsen_2015}. This observation highlights the fact that the few- to many-body crossover should be studied separately for fermions and weakly-interacting bosons.

Finally, we note that the effective mass computed with Eq.~(\ref{eq:eff_mass_finite}) describes only the Yrast curve with $P=0$ well. To illustrate this, we calculate the second derivative of the energy in the limit $\Phi\to0$ using the MFA. For a non-interacting impurity, this derivative is given by $1/m$ for all values of $P$. For an interacting impurity, this is not the case. 
The second derivative for $P=0$ is by definition given by $1/m_{\mathrm{eff}}$. The right panel of
Fig.~\ref{fig:Non-trivialRedistribution} shows that
the effective mass increases for stronger impurity-boson repulsion in agreement with our expectations.  
The figure also shows that for the $|P|=2\pi$ state additional effects come into play and change the second derivative.
The physical picture behind these effects will become more clear below, when we consider currents. The difference between `effective masses' defined for $P=0$ and $|P|=2\pi$ illustrates a shortcoming of the use of the quasiparticle picture for a small AB ring. However, even then, the polaron picture explains the qualitative features of the spectrum well.


\begin{figure}
\centering
    \includegraphics[width=1\linewidth]{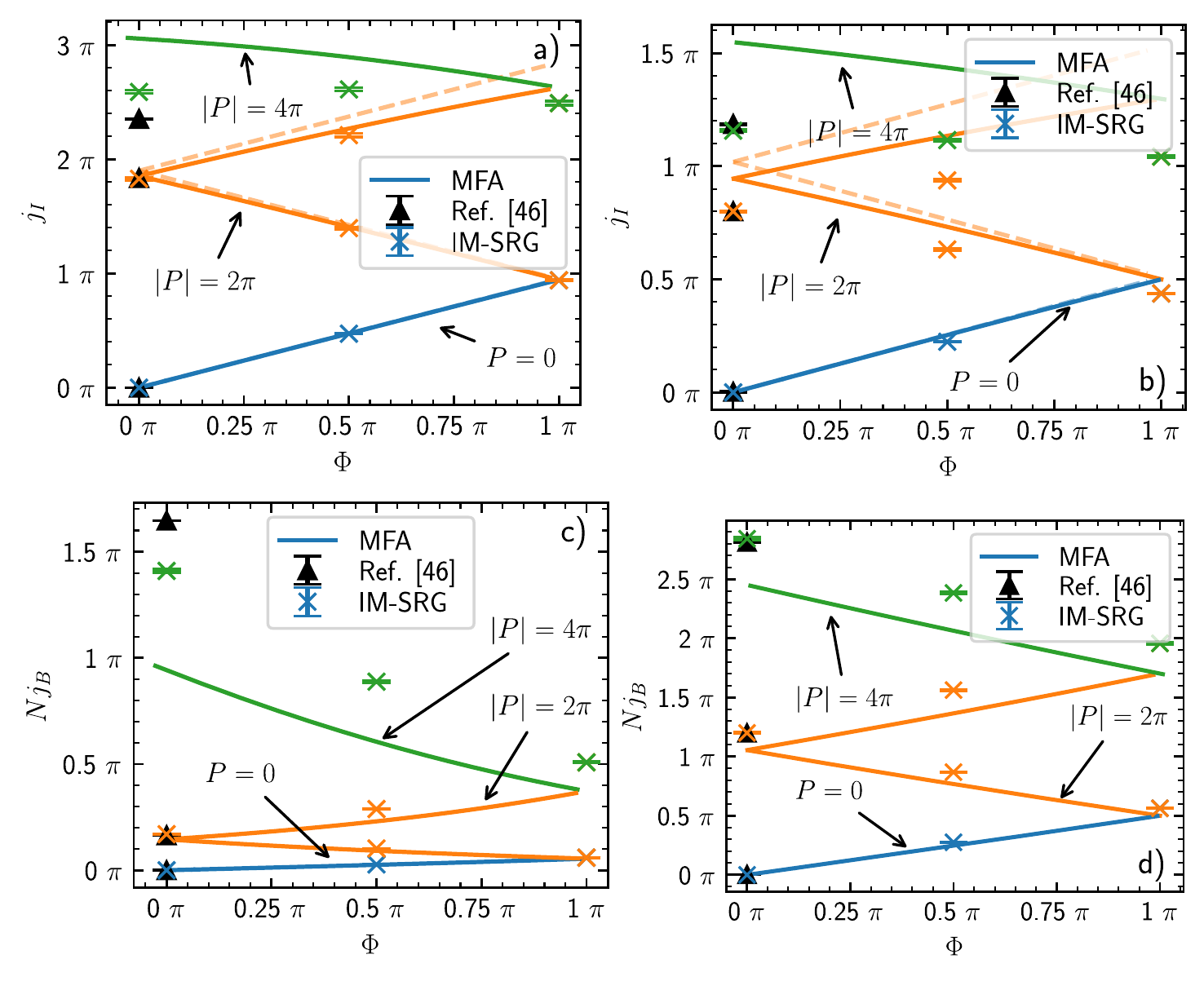}
    \caption{{\bf  Currents in the system.}\\
    a) and b) present the impurity current. c) and d) demonstrate the bosonic current. Blue, orange, and green colors are for $P=0$, $|P|=2\pi$, and $|P|=4\pi$, correspondingly. The parameters of the system are $N=19$ and $\gamma=0.2$. a) and c) show results for $c/g=1$, whereas b) and d) are for $c/g=5$. Data were obtained with the mean-field ansatz (solid lines, MFA) and in-medium similarity renormalization group (crosses, IM-SRG). Black triangles are the results of Yang et al.~\cite{Yang2022} for $\Phi=0$. Dashed lines in a) and c) show the impurity current in the polaron approximation $j_I=\mathcal{P}/m_{\mathrm{eff}}$. Here, $m_{\mathrm{eff}}$ is calculated within the MFA using Eq.~(\ref{eq:eff_mass_finite}).}
    \label{fig:Currents}
\end{figure}

\vspace{1em}
\noindent{\bf Currents.}
The AB flux in our system induces currents that can be defined via the continuity equations for the impurity and the bosons in the laboratory frame:
\begin{equation}
    \frac{\partial \rho_I}{\partial t}=-\frac{\partial j_I}{\partial y}; \qquad \frac{\partial \rho_B}{\partial t}=-\frac{\partial j_B}{\partial x},
\end{equation}
where $t L^2$ is time,  $\rho_{I}$ ($\rho_{B}$) is the probability density of the impurity (bosonic) cloud.
The (local) probability currents are defined as
\begin{align}
    &j_{I}=-\frac{i}{2}\int\mathrm{d}x_1...\mathrm{d}x_N \left(\Psi^*\frac{\partial \Psi}{\partial y}-\Psi\frac{\partial \Psi^*}{\partial y}\right) + \Phi \rho_I,
    \label{eq:DefinitionImpurityCurrent}\\
    &j_{B}=-\frac{i}{2}\int\mathrm{d}y\mathrm{d}x_2...\mathrm{d}x_N \left(\Psi^*\frac{\partial \Psi}{\partial x_1}-\Psi\frac{\partial \Psi^*}{\partial x_1}\right).
\end{align}

The rotational symmetry implies that $j_I, j_B,\rho_I$ and $\rho_B$ are position-independent, allowing us to work with the integral quantities, e.g., $\rho_I=\int \rho_I \mathrm{d}y/(2\pi)$, which is more convenient. For example, using these quantities, it is easy to show that $j_I+N j_B=\mathcal{P}$. Therefore, the total current -- the current that corresponds to the total density $\rho_I+N\rho_B$ -- is given by $\mathcal{P}=P+\Phi$.
Note that even though the AB flux is coupled only to the impurity, boson-impurity interactions also generate a current of bosons. We illustrate these currents for $c/g=1$ and $c/g=5$ in Fig.~\ref{fig:Currents} (other parameters are $N=19$, $\gamma=0.2$). The increase in the boson-impurity interaction leads to an increase in the bosonic current. This observation is most easily explained using the Bose-polaron picture.

Using the Hellmann–Feynman theorem, it is straightforward  to show that 
\begin{equation}
    j_I=\frac{\partial E}{\partial \Phi},
    \label{eq:Hell_Feynm}
\end{equation}
which coincides with the standard definition of the current in a one-body problem, see, e.g.,~\cite{VIEFERS2004}. [Note that this expression provides an indirect way for measuring currents by studying the energy landscape of the problem with RF spectroscopy (cf. Scazza et al.~\cite{Scazza2017}).] For the polaron picture with $P=0$, this leads to
$j_I=\Phi/m_{\mathrm{eff}}$ connecting the current (and transport properties) of the impurity to its effective mass. The bosonic current in the same approximation is given by $N j_{B}=(1-1/m_{\mathrm{eff}})\Phi$. The bosonic current generated by the AB flux follows the impurity, and leads to renormalization of its mass. 
We conclude that in the polaron picture the currents depend linearly on $\Phi$, with the slope fully determined by the effective mass. 

The region of validity of this picture is determined by the boson-impurity interaction, see the dashed lines in panels a) and c) of Fig.~(\ref{fig:Currents}).  For $c/g=1$ we observe that the polaron approximation, in which the energy is related to the AB flux via Eq.~(\ref{eq:eff_mass_finite}), is accurate for $|\mathcal{P}|\lesssim2\pi$, but for stronger interactions, $c/g=5$, it is appropriate merely for $|\mathcal{P}|\lesssim\pi$. For even stronger interactions, Eq.~(\ref{eq:eff_mass_finite}) is accurate only in the limit $\mathcal{P}\to0$.
Our interpretation here is that the coherent propagation of the impurity is not possible for strong boson-impurity interaction and large fluxes. Indeed, for $c\to\infty$ an impurity can exchange its position with a boson in a coherent manner only at timescales given by $1/c$. Thus strong (fast) impurity currents excite bosons.
This leads to a non-linear increase of the currents with $\Phi$,
see also~\cite{Yang2022} and Suppl. Note~3. To quantify these effects, it is convenient to rely on the second derivative of the energy (effective mass), which is larger for $|P|=2\pi$, see Fig.~\ref{fig:Non-trivialRedistribution} (note that the bosonic current is related to the impurity current via $Nj_B=P+\Phi-j_I$, i.e., one can reach the same conclusion by considering $Nj_B$ instead of $j_I$).
The IM-SRG results show a somewhat stronger generation of bosonic currents than the MFA, but the qualitative picture stays the same.   

In addition limits of validity of the polaron picture can be investigated by considering states with higher values of $|P|$. For example, Fig.~\ref{fig:Currents} shows that for $|P|=4\pi$, the current of the impurity (almost) does not depend on $\Phi$. This current is critical in a sense that by changing the flux of the impurity one generates only the current of bosons. The value of the critical current, $j^{cr}_I$, is decreased by increasing $c$, in agreement with the mean-field studies~\cite{Hakim1997,Smith2019}. 

The critical current can be seen as an analogue of the critical velocity of a classic impurity that moves in a superfluid (cf. Landau critical velocity). Using this analogy, we can understand why the mean-field approximation in the co-rotating frame does not provide the correct value of the critical current. The MFA does not describe accurately the excitations of the Bose gas when it is decoupled from the impurity. In particular,
the MFA leads to an incorrect phononic dispersion relation and implies that the critical velocity can be larger than the speed of sound for small values of $c/g$~\cite{Smith2019}, which is unphysical. Furthermore, it does not capture type-II excitations of the Lieb-Liniger gas~\cite{Lieb1963_2}, which define the lowest energy state for a given value of the momentum of the Bose gas, see~\cite{Syrwid2021} for tutorial. Note that our IM-SRG method also does not capture these states well -- the flow equations diverge when the type-II excitations become relevant. 
For some additional details, see Suppl. Note 3.

\begin{figure}
    \centering
    \includegraphics[width=1\linewidth]{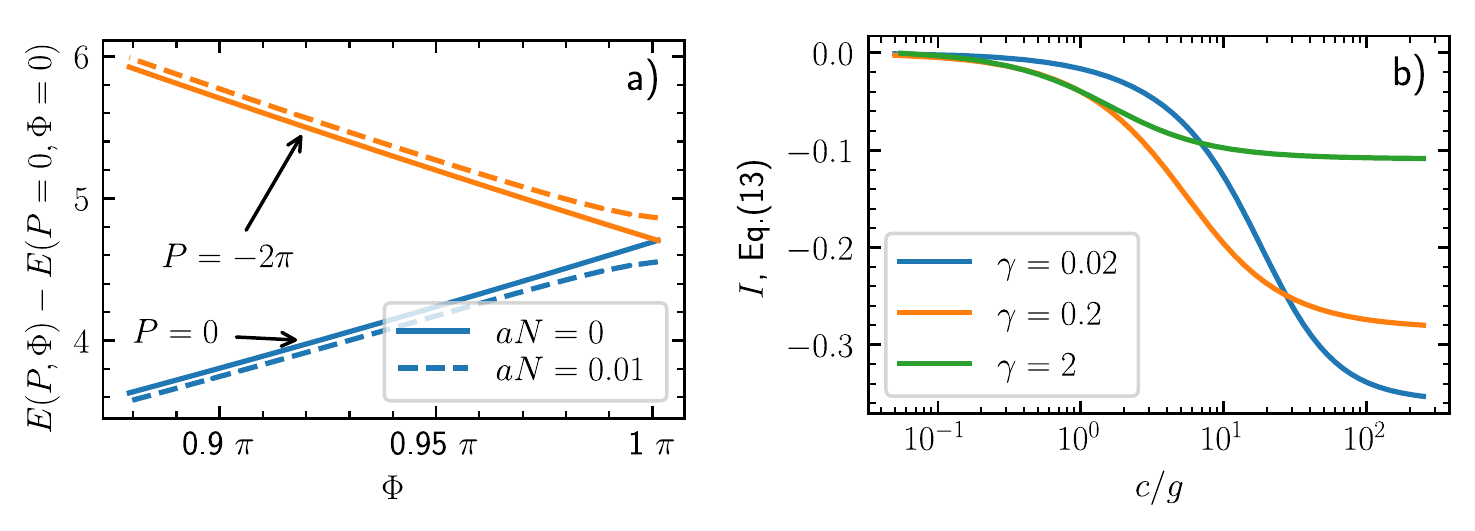}
    \caption{{\bf Effect of a defect on the energy crossing.}\\
a): The energy for $P=0$ and $P=-2\pi$ as a function of the flux in with and without a defect.  Solid lines are for the unperturbed system; dashed lines correspond to $aN=0.01$. Other parameters of the system are $N=19$, $\gamma=0.2$ and $c/g=1$. b): Eq.~\eqref{eq:IntegralSplitting} in the thermodynamic limit as a function of $c/g$ for different boson-boson interactions $\gamma$. The data were obtained with the mean-field ansatz.}
\label{fig:Perturbation}
\end{figure}

\vspace{1em}
\noindent{\bf Role of defects.}
The rotational symmetry of the problem makes the Yrast energy spectrum of Fig.~\ref{fig:Energy_c_02} double degenerate at $\Phi=\pm \pi$. In realistic systems, the symmetry is typically broken due to the presence of defects, leading to avoided crossings in the energy spectrum (cf.~Fig.~\ref{fig:Perturbation}). At the maxima of the avoided crossings one-body currents defined via $\partial E/\partial \Phi$ vanish affecting transport properties of the system~\cite{Aronov1987}. We also note that the simplest experimental realization of the AB flux in cold-atom set-ups can be achieved with a rotating weak link~\cite{Wright2013}, which utilizes the equivalence between the Coriolis force in a non-inertial frame and the Lorentz force on a charged particle in a uniform magnetic field. The rotating link introduces a `defect' potential into the problem whose effect can be studied using the methods discussed in this section.

Our two-component set-up offers unique possibilities to modify currents that are not present in a single-body AB physics. 
To illustrate this, we add to the Hamiltonian a small perturbation:
\begin{equation}
    \mathcal{H}_W=\mathcal{H}+W\,,
\end{equation}
where $\mathcal{H}$ is the original Hamiltonian from Eq.~\eqref{eq:Hamiltonian} and 
\begin{equation}
    W=\frac{a}{L}\sum_i\delta(x_i).
\end{equation}
This additional term describes a short-range potential coupled {\it exclusively} to the Bose gas. The current of the impurity is sensitive to $W$ only via the boson-impurity interaction, and therefore the avoided crossing should contain information about the boson-impurity correlation function. 

 As long as $a$ is small ($a\to0$), we can assume that the defect has only a minor influence on our system unless the system is close to the degeneracy point, $\Phi=\pm \pi$. Close to these points, we calculate the dimensionless energy using degenerate state perturbation theory:
\begin{equation}
    \mathcal{H}_W\simeq \begin{pmatrix}
        E_0 + L^2\braket{\Psi_0|W|\Psi_0} & L^2\braket{\Psi_0|W|\Psi_1} \\
        L^2\braket{\Psi_1|W|\Psi_0} & E_1 + L^2\braket{\Psi_1|W|\Psi_1} 
    \end{pmatrix}, \nonumber
\end{equation}
where $E_0\simeq \Phi^2/(2m_{\mathrm{eff}})$ ($E_1\simeq \Phi^2/(2m_{\mathrm{eff}})$) is the energy of the Yrast state with $P=0$ ($|P|=2\pi$); $\Psi_0$ ($\Psi_1$) is the corresponding eigenstate. 
Within the MFA, the matrix elements read as 
\begin{equation}
    \braket{\Psi_i|W|\Psi_j}=\frac{a}{L}N\alpha^{N-1}\int e^{i(P_j-P_i)y}f^*_{i}(y)f_j(y)dy, \nonumber
\end{equation}
where $\alpha=\int f^*_i(z)f_j(z)dz$, and subscripts determine the Yrast state, e.g., $i=0$ corresponds to $P=0$.

To provide insight into the avoided crossing,
we focus on $\Phi=\pm \pi$. In this case $f_1=f_0$, and $\braket{\Psi_i|W|\Psi_j}$ depends only on the density in the co-moving frame (or equivalently on the impurity-boson correlation function in the laboratory frame). This density, hence, the splitting of the energy levels, is sensitive to the values of $c$. For example, if $c=0$, then the defect should destroy the rotational invariance of the Bose gas only. Indeed, in this case, $|f|^2$ is constant and $\braket{\Psi_i|W|\Psi_j}=0$. 

Panel a) of Fig.~\ref{fig:Perturbation} illustrates the avoided crossing for a small value of $a$. Note that the energy of the system, $E(P,\Phi)$, increases for $a>0$. This effect does not appear in the figure as we only show the energy difference. The interesting part is that in the presence of $W$ the energies of the first and second Yrast state no longer cross. The splitting of the energies, $\Delta E=2aLN I$, is determined by the integral 
\begin{equation}
    I=\int e^{i2\pi y}|f_0(y)|^2\mathrm{d}y,
    \label{eq:IntegralSplitting}
\end{equation}
which can be estimated using the density in the thermodynamic limit at $\Phi=0$~\cite{Volosniev2017}:
\begin{equation}
|f_0(y)|^2\simeq\frac{L\mu}{g (N-1)}\mathrm{tanh}^2\left(\sqrt{\mu\kappa}L\delta\left[\frac{(z-L/2)}{\delta L}+\frac{1}{2}\right]\right), \nonumber
\label{eq:wave_func_thermod}
\end{equation}
where
\begin{align}
\label{eq:thermod_d}
\delta &\simeq 1+\frac{2 d}{\sqrt{\gamma\kappa}N}, \qquad d=\frac{1}{2}\mathrm{asinh}\left(\frac{2\rho}{c}\sqrt{\frac{\gamma}{\kappa}}\right),\\
\mu &\simeq \gamma\rho^2\frac{N-1}{N}\left(1-2\frac{\mathrm{tanh}(d)-1}{\sqrt{\gamma \kappa}N}\right). 
\label{eq:energy_thermod}
\end{align}
Panel~b) of Fig.~\ref{fig:Perturbation} shows that by increasing the value of $c$, one increases the energy splitting. Since the interactions are of zero range, the value of $I$ converges to a constant value for $c\to\infty$, which depends on the strength of boson-boson interactions, $\gamma$. Note that for small values of $\gamma$ the impurity can modify the density of the bosons significantly, leading to larger values of $I$.

Finally, we note an interference effect that appears if we place a second small perturbation into the system:
\begin{equation}
    W=\frac{a}{L}\sum_i\left(\delta(x_i)+\delta(x_i+d)\right).
\end{equation}
In analogy to above, we can define a coupling integral as
\begin{equation}
    I=\left(1+e^{i 2\pi d}\right)\int e^{i 2 \pi y}|f_{0}(y)|^2\mathrm{d}y.
\end{equation}
If $d=1/2$, this matrix element vanishes and the energy levels cross again (within the lowest order of perturbation theory). This happens because the perturbations are placed opposite to one another, which effectively restores the rotational symmetry in this case. This effect can be seen also for more than two perturbations, as long as they are placed in a symmetric order on the ring (for example three defects in a form of an equilateral triangle).

\vspace{1em}
\noindent{\bf Conclusions}
\vspace{1em}

\noindent To summarize, we studied an impurity coupled to the AB flux in a Bose gas. We argued that i) the system can be described using the ideas developed for the Bose polaron, ii) the AB ring can be a testbed for studies of the few- to many-body crossover in cold-atom polaron problems. In particular, observation of persistent currents in the AB ring with an impurity can shed light onto coherence properties of the Bose polaron. Note that the 1D world has inherent phase fluctuations, which can be captured using the IM-SRG approach (see also Suppl. Note~3). These fluctuations should be necessarily taken into account when studying persistent currents of polarons.

Our investigation of currents shows that the AB ring can provide a platform for studying a few-body analogue of the critical velocity in a Bose-polaron problem.
Furthermore, if we assume that the critical current, $j_I^{cr}$, does not depend on $\Phi$, then (according to Eq.~(\ref{eq:Hell_Feynm})) the energy of the system is $E=E^{cr}+j_I^{cr}\Phi$. This expression connects the bosonic current and the energy of the system motivating a study of few-body precursors of collective excitations in a Bose gas.

Finally, we note that $\partial^2 E/(\partial \Phi^2)$ relates to the inverse of the effective mass, which defines transport properties of a polaron.  This relation bears some similarity to the Thouless conductance in a disordered medium~\cite{Thouless1974}. It might be interesting to explore this connection further, in particular, for a weakly interacting light impurity that within the Born-Oppenheimer approximation experiences a disorder potential created by heavy bosons. 

\vspace{1em}
\noindent {\bf Methods}
\vspace{1em}

{\it Mean-field approach.} We use two methods to investigate the system. The first one is the mean-field approach (MFA) in relative coordinates. It assumes that all bosons occupy one state in the frame co-moving with the impurity, so that the total wave function of the system can be approximated as  
\begin{equation}
    \Tilde{\Psi}(z_1, z_2, ..., z_N)=\prod\limits_{i=1}^N f(z_i),
\end{equation}
where $f(z)$ is a normalized function determined by minimizing the Hamiltonian. The variational procedure leads to the  Gross-Pitaevskii equation, which can be solved semi-analytically, see Cominotti et al.~\cite{Cominotti2014} and Suppl. Note~3. 

{\it Flow equations.} The MFA is a well-established approach by now whose accuracy for stationary impurity problems has been shown by comparing to numerical quantum Monte Carlo~\cite{Volosniev2017BosePolaron,PANOCHKO2019} and state-of-the-art RG methods~\cite{Cominotti2014,Volosniev2017BosePolaron,Jager2020Deformation,Brauneis2021}. The MFA in time-dependent problems was discussed in~\cite{Jager2021,Koutentakis2022}.
In spite of those previous tests of MFA, we still find it necessary to validate it for the problem at hand. To this end, we shall use flow equations in the form of the so-called in-medium similarity renormalization group method (IM-SRG). This is an {\it ab initio} method that has been employed in condensed matter and nuclear physics~\cite{Kehrein2006,Tsukiyama2011,HERGERT2016} (for applications for one-dimensional problems with impurities, see~\cite{Volosniev2017BosePolaron,Brauneis2021,Brauneis2022}). For convience of the reader, we provide a brief introduction into IM-SRG and further compare its results to the MFA in Suppl. Note~3.

\vspace{1em}
\noindent {\bf Data Availability}
\vspace{1em}

The data that support the findings of this study are available from the corresponding author upon reasonable request.

\vspace{1em}
\noindent {\bf Code Availability}
\vspace{1em}

The code used for this study is available from the corresponding author upon reasonable request.

\vspace{2em}

\noindent \textbf{Acknowledgements}: 
We would like to thank Jonas Jager for sharing his data with us in early stages of this project. We thank Joachim Brand and Ray Yang for sharing with us data from Yang et al.~\cite{Yang2022}.
This work has received funding from the DFG Project No. 413495248 [VO 2437/1-1] (F. B., H.-W. H., A. G. V.). 
We acknowledge support by the Deutsche Forschungsgemeinschaft (DFG – German Research Foundation) and the Open Access Publishing Fund of Technical University of Darmstadt.

\vspace{1em}
\noindent {\bf Author Contributions}
\vspace{1em}

A. G. V. and F. B.  designed the project; F. B. performed  calculations under the supervision of A. G. V. and H.-W. H;  F. B., A. G., H.-W. H. and A. G. V. discussed the interpretation of the results; A. G. V. and F. B. wrote the initial draft of the manuscript; H.-W. H. and A. G. contributed to its final version.

\vspace{1em}
\noindent {\bf Competing interests}
\vspace{1em}

The authors declare no competing interests.

\newpage
\section*{References}
\bibliography{refs}

\begin{thebibliography}{21}%
\makeatletter
\providecommand \@ifxundefined [1]{%
 \@ifx{#1\undefined}
}%
\providecommand \@ifnum [1]{%
 \ifnum #1\expandafter \@firstoftwo
 \else \expandafter \@secondoftwo
 \fi
}%
\providecommand \@ifx [1]{%
 \ifx #1\expandafter \@firstoftwo
 \else \expandafter \@secondoftwo
 \fi
}%
\providecommand \natexlab [1]{#1}%
\providecommand \enquote  [1]{``#1''}%
\providecommand \bibnamefont  [1]{#1}%
\providecommand \bibfnamefont [1]{#1}%
\providecommand \citenamefont [1]{#1}%
\providecommand \href@noop [0]{\@secondoftwo}%
\providecommand \href [0]{\begingroup \@sanitize@url \@href}%
\providecommand \@href[1]{\@@startlink{#1}\@@href}%
\providecommand \@@href[1]{\endgroup#1\@@endlink}%
\providecommand \@sanitize@url [0]{\catcode `\\12\catcode `\$12\catcode
  `\&12\catcode `\#12\catcode `\^12\catcode `\_12\catcode `\%12\relax}%
\providecommand \@@startlink[1]{}%
\providecommand \@@endlink[0]{}%
\providecommand \url  [0]{\begingroup\@sanitize@url \@url }%
\providecommand \@url [1]{\endgroup\@href {#1}{\urlprefix }}%
\providecommand \urlprefix  [0]{URL }%
\providecommand \Eprint [0]{\href }%
\providecommand \doibase [0]{https://doi.org/}%
\providecommand \selectlanguage [0]{\@gobble}%
\providecommand \bibinfo  [0]{\@secondoftwo}%
\providecommand \bibfield  [0]{\@secondoftwo}%
\providecommand \translation [1]{[#1]}%
\providecommand \BibitemOpen [0]{}%
\providecommand \bibitemStop [0]{}%
\providecommand \bibitemNoStop [0]{.\EOS\space}%
\providecommand \EOS [0]{\spacefactor3000\relax}%
\providecommand \BibitemShut  [1]{\csname bibitem#1\endcsname}%
\let\auto@bib@innerbib\@empty
\bibitem [{\citenamefont {Viefers}\ \emph {et~al.}(2004)\citenamefont
  {Viefers}, \citenamefont {Koskinen}, \citenamefont {{Singha Deo}},\ and\
  \citenamefont {Manninen}}]{VIEFERS2004}%
  \BibitemOpen
  \bibfield  {author} {\bibinfo {author} {\bibfnamefont {S.}~\bibnamefont
  {Viefers}}, \bibinfo {author} {\bibfnamefont {P.}~\bibnamefont {Koskinen}},
  \bibinfo {author} {\bibfnamefont {P.}~\bibnamefont {{Singha Deo}}},\ and\
  \bibinfo {author} {\bibfnamefont {M.}~\bibnamefont {Manninen}},\ }\bibfield
  {title} {\bibinfo {title} {Quantum rings for beginners: energy spectra and
  persistent currents},\ }\href
  {https://doi.org/https://doi.org/10.1016/j.physe.2003.08.076} {\bibfield
  {journal} {\bibinfo  {journal} {Physica E: Low-dimensional Systems and
  Nanostructures}\ }\textbf {\bibinfo {volume} {21}},\ \bibinfo {pages} {1}
  (\bibinfo {year} {2004})}\BibitemShut {NoStop}%
\bibitem [{\citenamefont {Yang}\ \emph {et~al.}(2022)\citenamefont {Yang},
  \citenamefont {Čufar}, \citenamefont {Pahl},\ and\ \citenamefont
  {Brand}}]{Yang2022}%
  \BibitemOpen
  \bibfield  {author} {\bibinfo {author} {\bibfnamefont {M.}~\bibnamefont
  {Yang}}, \bibinfo {author} {\bibfnamefont {M.}~\bibnamefont {Čufar}},
  \bibinfo {author} {\bibfnamefont {E.}~\bibnamefont {Pahl}},\ and\ \bibinfo
  {author} {\bibfnamefont {J.}~\bibnamefont {Brand}},\ }\bibfield  {title}
  {\bibinfo {title} {Polaron-depleton transition in the yrast excitations of a
  one-dimensional {Bose} gas with a mobile impurity},\ }\bibfield  {journal}
  {\bibinfo  {journal} {Condensed Matter}\ }\textbf {\bibinfo {volume} {7}},\
  \href {https://doi.org/10.3390/condmat7010015} {10.3390/condmat7010015}
  (\bibinfo {year} {2022})\BibitemShut {NoStop}%
\bibitem [{\citenamefont {Sacha}\ and\ \citenamefont
  {Timmermans}(2006)}]{Sacha2006}%
  \BibitemOpen
  \bibfield  {author} {\bibinfo {author} {\bibfnamefont {K.}~\bibnamefont
  {Sacha}}\ and\ \bibinfo {author} {\bibfnamefont {E.}~\bibnamefont
  {Timmermans}},\ }\bibfield  {title} {\bibinfo {title} {Self-localized
  impurities embedded in a one-dimensional {Bose}-{Einstein} condensate and
  their quantum fluctuations},\ }\href
  {https://doi.org/10.1103/PhysRevA.73.063604} {\bibfield  {journal} {\bibinfo
  {journal} {Phys. Rev. A}\ }\textbf {\bibinfo {volume} {73}},\ \bibinfo
  {pages} {063604} (\bibinfo {year} {2006})}\BibitemShut {NoStop}%
\bibitem [{\citenamefont {Cominotti}\ \emph {et~al.}(2014)\citenamefont
  {Cominotti}, \citenamefont {Rossini}, \citenamefont {Rizzi}, \citenamefont
  {Hekking},\ and\ \citenamefont {Minguzzi}}]{Cominotti2014}%
  \BibitemOpen
  \bibfield  {author} {\bibinfo {author} {\bibfnamefont {M.}~\bibnamefont
  {Cominotti}}, \bibinfo {author} {\bibfnamefont {D.}~\bibnamefont {Rossini}},
  \bibinfo {author} {\bibfnamefont {M.}~\bibnamefont {Rizzi}}, \bibinfo
  {author} {\bibfnamefont {F.}~\bibnamefont {Hekking}},\ and\ \bibinfo {author}
  {\bibfnamefont {A.}~\bibnamefont {Minguzzi}},\ }\bibfield  {title} {\bibinfo
  {title} {Optimal persistent currents for interacting bosons on a ring with a
  gauge field},\ }\href {https://doi.org/10.1103/PhysRevLett.113.025301}
  {\bibfield  {journal} {\bibinfo  {journal} {Phys. Rev. Lett.}\ }\textbf
  {\bibinfo {volume} {113}},\ \bibinfo {pages} {025301} (\bibinfo {year}
  {2014})}\BibitemShut {NoStop}%
\bibitem [{\citenamefont {Abramowitz}\ and\ \citenamefont
  {Stegun}(1972)}]{abramowitz1972handbook}%
  \BibitemOpen
  \bibinfo {editor} {\bibfnamefont {M.}~\bibnamefont {Abramowitz}}\ and\
  \bibinfo {editor} {\bibfnamefont {I.~A.}\ \bibnamefont {Stegun}},\ eds.,\
  \href {http://scans.hebis.de/HEBCGI/show.pl?12433957_toc.pdf} {\emph
  {\bibinfo {title} {Handbook of mathematical functions : with formulas,
  graphs, and mathematical tables}}},\ \bibinfo {edition} {9th}\ ed.\ (\bibinfo
  {address} {New York},\ \bibinfo {year} {1972})\BibitemShut {NoStop}%
\bibitem [{\citenamefont {Volosniev}\ and\ \citenamefont
  {Hammer}(2017{\natexlab{a}})}]{Volosniev2017BosePolaron}%
  \BibitemOpen
  \bibfield  {author} {\bibinfo {author} {\bibfnamefont {A.~G.}\ \bibnamefont
  {Volosniev}}\ and\ \bibinfo {author} {\bibfnamefont {H.-W.}\ \bibnamefont
  {Hammer}},\ }\bibfield  {title} {\bibinfo {title} {{Analytical approach to
  the {Bose}-polaron problem in one dimension}},\ }\href
  {https://doi.org/10.1103/PhysRevA.96.031601} {\bibfield  {journal} {\bibinfo
  {journal} {Phys. Rev. A}\ }\textbf {\bibinfo {volume} {96}},\ \bibinfo
  {pages} {031601} (\bibinfo {year} {2017}{\natexlab{a}})}\BibitemShut
  {NoStop}%
\bibitem [{\citenamefont {Kehrein}(2006)}]{Kehrein2006}%
  \BibitemOpen
  \bibfield  {author} {\bibinfo {author} {\bibfnamefont {S.}~\bibnamefont
  {Kehrein}},\ }\href@noop {} {\emph {\bibinfo {title} {The Flow Equation
  Approach to Many-Particle Systems}}}\ (\bibinfo  {publisher}
  {Springer(Berlin)},\ \bibinfo {year} {2006})\BibitemShut {NoStop}%
\bibitem [{\citenamefont {Volosniev}\ and\ \citenamefont
  {Hammer}(2017{\natexlab{b}})}]{Volosniev2017}%
  \BibitemOpen
  \bibfield  {author} {\bibinfo {author} {\bibfnamefont {A.~G.}\ \bibnamefont
  {Volosniev}}\ and\ \bibinfo {author} {\bibfnamefont {H.-W.}\ \bibnamefont
  {Hammer}},\ }\bibfield  {title} {\bibinfo {title} {Flow equations for cold
  {Bose} gases},\ }\href {https://doi.org/10.1088/1367-2630/aa9011} {\bibfield
  {journal} {\bibinfo  {journal} {New Journal of Physics}\ }\textbf {\bibinfo
  {volume} {19}},\ \bibinfo {pages} {113051} (\bibinfo {year}
  {2017}{\natexlab{b}})}\BibitemShut {NoStop}%
\bibitem [{\citenamefont {Brauneis}\ \emph {et~al.}(2021)\citenamefont
  {Brauneis}, \citenamefont {Hammer}, \citenamefont {Lemeshko},\ and\
  \citenamefont {Volosniev}}]{Brauneis2021}%
  \BibitemOpen
  \bibfield  {author} {\bibinfo {author} {\bibfnamefont {F.}~\bibnamefont
  {Brauneis}}, \bibinfo {author} {\bibfnamefont {H.-W.}\ \bibnamefont
  {Hammer}}, \bibinfo {author} {\bibfnamefont {M.}~\bibnamefont {Lemeshko}},\
  and\ \bibinfo {author} {\bibfnamefont {A.~G.}\ \bibnamefont {Volosniev}},\
  }\bibfield  {title} {\bibinfo {title} {{Impurities in a one-dimensional
  {Bose} gas: the flow equation approach}},\ }\href
  {https://doi.org/10.21468/SciPostPhys.11.1.008} {\bibfield  {journal}
  {\bibinfo  {journal} {SciPost Phys.}\ }\textbf {\bibinfo {volume} {11}},\
  \bibinfo {pages} {008} (\bibinfo {year} {2021})}\BibitemShut {NoStop}%
\bibitem [{\citenamefont {Brauneis}\ \emph {et~al.}(2022)\citenamefont
  {Brauneis}, \citenamefont {Backert}, \citenamefont {Mistakidis},
  \citenamefont {Lemeshko}, \citenamefont {Hammer},\ and\ \citenamefont
  {Volosniev}}]{Brauneis2022}%
  \BibitemOpen
  \bibfield  {author} {\bibinfo {author} {\bibfnamefont {F.}~\bibnamefont
  {Brauneis}}, \bibinfo {author} {\bibfnamefont {T.~G.}\ \bibnamefont
  {Backert}}, \bibinfo {author} {\bibfnamefont {S.~I.}\ \bibnamefont
  {Mistakidis}}, \bibinfo {author} {\bibfnamefont {M.}~\bibnamefont
  {Lemeshko}}, \bibinfo {author} {\bibfnamefont {H.-W.}\ \bibnamefont
  {Hammer}},\ and\ \bibinfo {author} {\bibfnamefont {A.~G.}\ \bibnamefont
  {Volosniev}},\ }\bibfield  {title} {\bibinfo {title} {Artificial atoms from
  cold bosons in one dimension},\ }\href
  {https://doi.org/10.1088/1367-2630/ac78d8} {\bibfield  {journal} {\bibinfo
  {journal} {New Journal of Physics}\ }\textbf {\bibinfo {volume} {24}},\
  \bibinfo {pages} {063036} (\bibinfo {year} {2022})}\BibitemShut {NoStop}%
\bibitem [{\citenamefont {Hergert}\ \emph {et~al.}(2013)\citenamefont
  {Hergert}, \citenamefont {Binder}, \citenamefont {Calci}, \citenamefont
  {Langhammer},\ and\ \citenamefont {Roth}}]{Hergert2013}%
  \BibitemOpen
  \bibfield  {author} {\bibinfo {author} {\bibfnamefont {H.}~\bibnamefont
  {Hergert}}, \bibinfo {author} {\bibfnamefont {S.}~\bibnamefont {Binder}},
  \bibinfo {author} {\bibfnamefont {A.}~\bibnamefont {Calci}}, \bibinfo
  {author} {\bibfnamefont {J.}~\bibnamefont {Langhammer}},\ and\ \bibinfo
  {author} {\bibfnamefont {R.}~\bibnamefont {Roth}},\ }\bibfield  {title}
  {\bibinfo {title} {Ab initio calculations of even oxygen isotopes with chiral
  two-plus-three-nucleon interactions},\ }\href
  {https://doi.org/10.1103/PhysRevLett.110.242501} {\bibfield  {journal}
  {\bibinfo  {journal} {Phys. Rev. Lett.}\ }\textbf {\bibinfo {volume} {110}},\
  \bibinfo {pages} {242501} (\bibinfo {year} {2013})}\BibitemShut {NoStop}%
\bibitem [{\citenamefont {Hergert}\ \emph {et~al.}(2014)\citenamefont
  {Hergert}, \citenamefont {Bogner}, \citenamefont {Morris}, \citenamefont
  {Binder}, \citenamefont {Calci}, \citenamefont {Langhammer},\ and\
  \citenamefont {Roth}}]{Hergert2014}%
  \BibitemOpen
  \bibfield  {author} {\bibinfo {author} {\bibfnamefont {H.}~\bibnamefont
  {Hergert}}, \bibinfo {author} {\bibfnamefont {S.~K.}\ \bibnamefont {Bogner}},
  \bibinfo {author} {\bibfnamefont {T.~D.}\ \bibnamefont {Morris}}, \bibinfo
  {author} {\bibfnamefont {S.}~\bibnamefont {Binder}}, \bibinfo {author}
  {\bibfnamefont {A.}~\bibnamefont {Calci}}, \bibinfo {author} {\bibfnamefont
  {J.}~\bibnamefont {Langhammer}},\ and\ \bibinfo {author} {\bibfnamefont
  {R.}~\bibnamefont {Roth}},\ }\bibfield  {title} {\bibinfo {title} {Ab initio
  multireference in-medium similarity renormalization group calculations of
  even calcium and nickel isotopes},\ }\href
  {https://doi.org/10.1103/PhysRevC.90.041302} {\bibfield  {journal} {\bibinfo
  {journal} {Phys. Rev. C}\ }\textbf {\bibinfo {volume} {90}},\ \bibinfo
  {pages} {041302} (\bibinfo {year} {2014})}\BibitemShut {NoStop}%
\bibitem [{\citenamefont {Popov}(1983)}]{Popov1983}%
  \BibitemOpen
  \bibfield  {author} {\bibinfo {author} {\bibfnamefont {V.~N.}\ \bibnamefont
  {Popov}},\ }\href@noop {} {\emph {\bibinfo {title} {Functional integrals in
  quantum field theory and statistical physics}}}\ (\bibinfo  {publisher} {D.
  Reidel Pub. Co.},\ \bibinfo {year} {1983})\BibitemShut {NoStop}%
\bibitem [{\citenamefont {Petrov}\ \emph {et~al.}(2000)\citenamefont {Petrov},
  \citenamefont {Shlyapnikov},\ and\ \citenamefont {Walraven}}]{Petrov2000}%
  \BibitemOpen
  \bibfield  {author} {\bibinfo {author} {\bibfnamefont {D.~S.}\ \bibnamefont
  {Petrov}}, \bibinfo {author} {\bibfnamefont {G.~V.}\ \bibnamefont
  {Shlyapnikov}},\ and\ \bibinfo {author} {\bibfnamefont {J.~T.~M.}\
  \bibnamefont {Walraven}},\ }\bibfield  {title} {\bibinfo {title} {Regimes of
  quantum degeneracy in trapped 1d gases},\ }\href
  {https://doi.org/10.1103/PhysRevLett.85.3745} {\bibfield  {journal} {\bibinfo
   {journal} {Phys. Rev. Lett.}\ }\textbf {\bibinfo {volume} {85}},\ \bibinfo
  {pages} {3745} (\bibinfo {year} {2000})}\BibitemShut {NoStop}%
\bibitem [{\citenamefont {Pethick}\ and\ \citenamefont
  {Smith}(2002)}]{Pethick2002}%
  \BibitemOpen
  \bibfield  {author} {\bibinfo {author} {\bibfnamefont {C.}~\bibnamefont
  {Pethick}}\ and\ \bibinfo {author} {\bibfnamefont {H.}~\bibnamefont
  {Smith}},\ }\href@noop {} {\emph {\bibinfo {title} {{Bose}–{Einstein}
  Condensation in Dilute Gases}}}\ (\bibinfo  {publisher} {Cambridge University
  Press},\ \bibinfo {year} {2002})\BibitemShut {NoStop}%
\bibitem [{\citenamefont {Lieb}\ and\ \citenamefont
  {Liniger}(1963)}]{Lieb1963}%
  \BibitemOpen
  \bibfield  {author} {\bibinfo {author} {\bibfnamefont {E.~H.}\ \bibnamefont
  {Lieb}}\ and\ \bibinfo {author} {\bibfnamefont {W.}~\bibnamefont {Liniger}},\
  }\bibfield  {title} {\bibinfo {title} {Exact analysis of an interacting
  {Bose} gas. i. the general solution and the ground state},\ }\href
  {https://doi.org/10.1103/PhysRev.130.1605} {\bibfield  {journal} {\bibinfo
  {journal} {Phys. Rev.}\ }\textbf {\bibinfo {volume} {130}},\ \bibinfo {pages}
  {1605} (\bibinfo {year} {1963})}\BibitemShut {NoStop}%
\bibitem [{\citenamefont {Lieb}(1963)}]{Lieb1963_2}%
  \BibitemOpen
  \bibfield  {author} {\bibinfo {author} {\bibfnamefont {E.~H.}\ \bibnamefont
  {Lieb}},\ }\bibfield  {title} {\bibinfo {title} {Exact analysis of an
  interacting {Bose} gas. ii. the excitation spectrum},\ }\href
  {https://doi.org/10.1103/PhysRev.130.1616} {\bibfield  {journal} {\bibinfo
  {journal} {Phys. Rev.}\ }\textbf {\bibinfo {volume} {130}},\ \bibinfo {pages}
  {1616} (\bibinfo {year} {1963})}\BibitemShut {NoStop}%
\bibitem [{\citenamefont {Gaudin}(1971)}]{Gaudin1971}%
  \BibitemOpen
  \bibfield  {author} {\bibinfo {author} {\bibfnamefont {M.}~\bibnamefont
  {Gaudin}},\ }\bibfield  {title} {\bibinfo {title} {Boundary energy of a
  {Bose} gas in one dimension},\ }\href
  {https://doi.org/10.1103/PhysRevA.4.386} {\bibfield  {journal} {\bibinfo
  {journal} {Phys. Rev. A}\ }\textbf {\bibinfo {volume} {4}},\ \bibinfo {pages}
  {386} (\bibinfo {year} {1971})}\BibitemShut {NoStop}%
\bibitem [{\citenamefont {Reichert}\ \emph {et~al.}(2019)\citenamefont
  {Reichert}, \citenamefont {Astrakharchik}, \citenamefont
  {Petkovi\ifmmode~\acute{c}\else \'{c}\fi{}},\ and\ \citenamefont
  {Ristivojevic}}]{Reichert2019}%
  \BibitemOpen
  \bibfield  {author} {\bibinfo {author} {\bibfnamefont {B.}~\bibnamefont
  {Reichert}}, \bibinfo {author} {\bibfnamefont {G.~E.}\ \bibnamefont
  {Astrakharchik}}, \bibinfo {author} {\bibfnamefont {A.}~\bibnamefont
  {Petkovi\ifmmode~\acute{c}\else \'{c}\fi{}}},\ and\ \bibinfo {author}
  {\bibfnamefont {Z.}~\bibnamefont {Ristivojevic}},\ }\bibfield  {title}
  {\bibinfo {title} {Exact results for the boundary energy of one-dimensional
  bosons},\ }\href {https://doi.org/10.1103/PhysRevLett.123.250602} {\bibfield
  {journal} {\bibinfo  {journal} {Phys. Rev. Lett.}\ }\textbf {\bibinfo
  {volume} {123}},\ \bibinfo {pages} {250602} (\bibinfo {year}
  {2019})}\BibitemShut {NoStop}%
\bibitem [{\citenamefont {Jager}\ \emph {et~al.}(2020)\citenamefont {Jager},
  \citenamefont {Barnett}, \citenamefont {Will},\ and\ \citenamefont
  {Fleischhauer}}]{Jager2020Deformation}%
  \BibitemOpen
  \bibfield  {author} {\bibinfo {author} {\bibfnamefont {J.}~\bibnamefont
  {Jager}}, \bibinfo {author} {\bibfnamefont {R.}~\bibnamefont {Barnett}},
  \bibinfo {author} {\bibfnamefont {M.}~\bibnamefont {Will}},\ and\ \bibinfo
  {author} {\bibfnamefont {M.}~\bibnamefont {Fleischhauer}},\ }\bibfield
  {title} {\bibinfo {title} {{Strong-coupling {Bose} polarons in one dimension:
  Condensate deformation and modified Bogoliubov phonons}},\ }\href
  {https://doi.org/10.1103/PhysRevResearch.2.033142} {\bibfield  {journal}
  {\bibinfo  {journal} {Phys. Rev. Research}\ }\textbf {\bibinfo {volume}
  {2}},\ \bibinfo {pages} {033142} (\bibinfo {year} {2020})}\BibitemShut
  {NoStop}%
\bibitem [{\citenamefont {Grusdt}\ \emph {et~al.}(2017)\citenamefont {Grusdt},
  \citenamefont {Astrakharchik},\ and\ \citenamefont
  {Demler}}]{Grusdt2017BosePolaron}%
  \BibitemOpen
  \bibfield  {author} {\bibinfo {author} {\bibfnamefont {F.}~\bibnamefont
  {Grusdt}}, \bibinfo {author} {\bibfnamefont {G.~E.}\ \bibnamefont
  {Astrakharchik}},\ and\ \bibinfo {author} {\bibfnamefont {E.}~\bibnamefont
  {Demler}},\ }\bibfield  {title} {\bibinfo {title} {{{Bose} polarons in
  ultracold atoms in one dimension: beyond the Fr\"ohlich paradigm}},\ }\href
  {https://doi.org/10.1088/1367-2630/aa8a2e} {\bibfield  {journal} {\bibinfo
  {journal} {New J. Phys.}\ }\textbf {\bibinfo {volume} {19}},\ \bibinfo
  {pages} {103035} (\bibinfo {year} {2017})}\BibitemShut {NoStop}%
\end{thebibliography}%


\begin{thebibliography}{75}%
\makeatletter
\providecommand \@ifxundefined [1]{%
 \@ifx{#1\undefined}
}%
\providecommand \@ifnum [1]{%
 \ifnum #1\expandafter \@firstoftwo
 \else \expandafter \@secondoftwo
 \fi
}%
\providecommand \@ifx [1]{%
 \ifx #1\expandafter \@firstoftwo
 \else \expandafter \@secondoftwo
 \fi
}%
\providecommand \natexlab [1]{#1}%
\providecommand \enquote  [1]{``#1''}%
\providecommand \bibnamefont  [1]{#1}%
\providecommand \bibfnamefont [1]{#1}%
\providecommand \citenamefont [1]{#1}%
\providecommand \href@noop [0]{\@secondoftwo}%
\providecommand \href [0]{\begingroup \@sanitize@url \@href}%
\providecommand \@href[1]{\@@startlink{#1}\@@href}%
\providecommand \@@href[1]{\endgroup#1\@@endlink}%
\providecommand \@sanitize@url [0]{\catcode `\\12\catcode `\$12\catcode
  `\&12\catcode `\#12\catcode `\^12\catcode `\_12\catcode `\%12\relax}%
\providecommand \@@startlink[1]{}%
\providecommand \@@endlink[0]{}%
\providecommand \url  [0]{\begingroup\@sanitize@url \@url }%
\providecommand \@url [1]{\endgroup\@href {#1}{\urlprefix }}%
\providecommand \urlprefix  [0]{URL }%
\providecommand \Eprint [0]{\href }%
\providecommand \doibase [0]{https://doi.org/}%
\providecommand \selectlanguage [0]{\@gobble}%
\providecommand \bibinfo  [0]{\@secondoftwo}%
\providecommand \bibfield  [0]{\@secondoftwo}%
\providecommand \translation [1]{[#1]}%
\providecommand \BibitemOpen [0]{}%
\providecommand \bibitemStop [0]{}%
\providecommand \bibitemNoStop [0]{.\EOS\space}%
\providecommand \EOS [0]{\spacefactor3000\relax}%
\providecommand \BibitemShut  [1]{\csname bibitem#1\endcsname}%
\let\auto@bib@innerbib\@empty
\bibitem [{\citenamefont {{Aharonov}}\ and\ \citenamefont
  {Bohm}(1959)}]{Aharonov1959}%
  \BibitemOpen
  \bibfield  {author} {\bibinfo {author} {\bibfnamefont {Y.}~\bibnamefont
  {{Aharonov}}}\ and\ \bibinfo {author} {\bibfnamefont {D.}~\bibnamefont
  {Bohm}},\ }\bibfield  {title} {\bibinfo {title} {Significance of
  electromagnetic potentials in the quantum theory},\ }\href
  {https://doi.org/10.1103/PhysRev.115.485} {\bibfield  {journal} {\bibinfo
  {journal} {Phys. Rev.}\ }\textbf {\bibinfo {volume} {115}},\ \bibinfo {pages}
  {485} (\bibinfo {year} {1959})}\BibitemShut {NoStop}%
\bibitem [{\citenamefont {Berry}(1984)}]{Berry1984}%
  \BibitemOpen
  \bibfield  {author} {\bibinfo {author} {\bibfnamefont {M.~V.}\ \bibnamefont
  {Berry}},\ }\bibfield  {title} {\bibinfo {title} {Quantal phase factors
  accompanying adiabatic changes},\ }\href
  {https://doi.org/10.1098/rspa.1984.0023} {\bibfield  {journal} {\bibinfo
  {journal} {Proceedings of the Royal Society of London. A. Mathematical and
  Physical Sciences}\ }\textbf {\bibinfo {volume} {392}},\ \bibinfo {pages}
  {45} (\bibinfo {year} {1984})}\BibitemShut {NoStop}%
\bibitem [{\citenamefont {Bloch}(1968)}]{Bloch1968}%
  \BibitemOpen
  \bibfield  {author} {\bibinfo {author} {\bibfnamefont {F.}~\bibnamefont
  {Bloch}},\ }\bibfield  {title} {\bibinfo {title} {Simple interpretation of
  the {Josephson} effect},\ }\href
  {https://doi.org/10.1103/PhysRevLett.21.1241} {\bibfield  {journal} {\bibinfo
   {journal} {Phys. Rev. Lett.}\ }\textbf {\bibinfo {volume} {21}},\ \bibinfo
  {pages} {1241} (\bibinfo {year} {1968})}\BibitemShut {NoStop}%
\bibitem [{\citenamefont {B{\"u}ttiker}\ \emph {et~al.}(1983)\citenamefont
  {B{\"u}ttiker}, \citenamefont {Imry},\ and\ \citenamefont
  {Landauer}}]{BUTTIKER1983}%
  \BibitemOpen
  \bibfield  {author} {\bibinfo {author} {\bibfnamefont {M.}~\bibnamefont
  {B{\"u}ttiker}}, \bibinfo {author} {\bibfnamefont {Y.}~\bibnamefont {Imry}},\
  and\ \bibinfo {author} {\bibfnamefont {R.}~\bibnamefont {Landauer}},\
  }\bibfield  {title} {\bibinfo {title} {{Josephson} behavior in small normal
  one-dimensional rings},\ }\href
  {https://doi.org/https://doi.org/10.1016/0375-9601(83)90011-7} {\bibfield
  {journal} {\bibinfo  {journal} {Physics Letters A}\ }\textbf {\bibinfo
  {volume} {96}},\ \bibinfo {pages} {365} (\bibinfo {year} {1983})}\BibitemShut
  {NoStop}%
\bibitem [{\citenamefont {Aronov}\ and\ \citenamefont
  {Sharvin}(1987)}]{Aronov1987}%
  \BibitemOpen
  \bibfield  {author} {\bibinfo {author} {\bibfnamefont {A.~G.}\ \bibnamefont
  {Aronov}}\ and\ \bibinfo {author} {\bibfnamefont {Y.~V.}\ \bibnamefont
  {Sharvin}},\ }\bibfield  {title} {\bibinfo {title} {Magnetic flux effects in
  disordered conductors},\ }\href {https://doi.org/10.1103/RevModPhys.59.755}
  {\bibfield  {journal} {\bibinfo  {journal} {Rev. Mod. Phys.}\ }\textbf
  {\bibinfo {volume} {59}},\ \bibinfo {pages} {755} (\bibinfo {year}
  {1987})}\BibitemShut {NoStop}%
\bibitem [{\citenamefont {Viefers}\ \emph {et~al.}(2004)\citenamefont
  {Viefers}, \citenamefont {Koskinen}, \citenamefont {{Singha Deo}},\ and\
  \citenamefont {Manninen}}]{VIEFERS2004}%
  \BibitemOpen
  \bibfield  {author} {\bibinfo {author} {\bibfnamefont {S.}~\bibnamefont
  {Viefers}}, \bibinfo {author} {\bibfnamefont {P.}~\bibnamefont {Koskinen}},
  \bibinfo {author} {\bibfnamefont {P.}~\bibnamefont {{Singha Deo}}},\ and\
  \bibinfo {author} {\bibfnamefont {M.}~\bibnamefont {Manninen}},\ }\bibfield
  {title} {\bibinfo {title} {Quantum rings for beginners: energy spectra and
  persistent currents},\ }\href
  {https://doi.org/https://doi.org/10.1016/j.physe.2003.08.076} {\bibfield
  {journal} {\bibinfo  {journal} {Physica E: Low-dimensional Systems and
  Nanostructures}\ }\textbf {\bibinfo {volume} {21}},\ \bibinfo {pages} {1}
  (\bibinfo {year} {2004})}\BibitemShut {NoStop}%
\bibitem [{\citenamefont {Lorke}\ \emph {et~al.}(2000)\citenamefont {Lorke},
  \citenamefont {Johannes~Luyken}, \citenamefont {Govorov}, \citenamefont
  {Kotthaus}, \citenamefont {Garcia},\ and\ \citenamefont
  {Petroff}}]{Lorke2000}%
  \BibitemOpen
  \bibfield  {author} {\bibinfo {author} {\bibfnamefont {A.}~\bibnamefont
  {Lorke}}, \bibinfo {author} {\bibfnamefont {R.}~\bibnamefont
  {Johannes~Luyken}}, \bibinfo {author} {\bibfnamefont {A.~O.}\ \bibnamefont
  {Govorov}}, \bibinfo {author} {\bibfnamefont {J.~P.}\ \bibnamefont
  {Kotthaus}}, \bibinfo {author} {\bibfnamefont {J.~M.}\ \bibnamefont
  {Garcia}},\ and\ \bibinfo {author} {\bibfnamefont {P.~M.}\ \bibnamefont
  {Petroff}},\ }\bibfield  {title} {\bibinfo {title} {Spectroscopy of
  nanoscopic semiconductor rings},\ }\href
  {https://doi.org/10.1103/PhysRevLett.84.2223} {\bibfield  {journal} {\bibinfo
   {journal} {Phys. Rev. Lett.}\ }\textbf {\bibinfo {volume} {84}},\ \bibinfo
  {pages} {2223} (\bibinfo {year} {2000})}\BibitemShut {NoStop}%
\bibitem [{\citenamefont {M{\"u}ller-Groeling}\ \emph
  {et~al.}(1993)\citenamefont {M{\"u}ller-Groeling}, \citenamefont
  {Weidenm{\"u}ller},\ and\ \citenamefont {Lewenkopf}}]{Groeling1993}%
  \BibitemOpen
  \bibfield  {author} {\bibinfo {author} {\bibfnamefont {A.}~\bibnamefont
  {M{\"u}ller-Groeling}}, \bibinfo {author} {\bibfnamefont {H.~A.}\
  \bibnamefont {Weidenm{\"u}ller}},\ and\ \bibinfo {author} {\bibfnamefont
  {C.~H.}\ \bibnamefont {Lewenkopf}},\ }\bibfield  {title} {\bibinfo {title}
  {Interacting electrons in mesoscopic rings},\ }\href
  {https://doi.org/10.1209/0295-5075/22/3/006} {\bibfield  {journal} {\bibinfo
  {journal} {Europhysics Letters}\ }\textbf {\bibinfo {volume} {22}},\ \bibinfo
  {pages} {193} (\bibinfo {year} {1993})}\BibitemShut {NoStop}%
\bibitem [{\citenamefont {Manninen}\ \emph {et~al.}(2012)\citenamefont
  {Manninen}, \citenamefont {Viefers},\ and\ \citenamefont
  {Reimann}}]{MANNINEN2012}%
  \BibitemOpen
  \bibfield  {author} {\bibinfo {author} {\bibfnamefont {M.}~\bibnamefont
  {Manninen}}, \bibinfo {author} {\bibfnamefont {S.}~\bibnamefont {Viefers}},\
  and\ \bibinfo {author} {\bibfnamefont {S.}~\bibnamefont {Reimann}},\
  }\bibfield  {title} {\bibinfo {title} {Quantum rings for beginners ii: Bosons
  versus fermions},\ }\href
  {https://doi.org/https://doi.org/10.1016/j.physe.2012.09.013} {\bibfield
  {journal} {\bibinfo  {journal} {Physica E: Low-dimensional Systems and
  Nanostructures}\ }\textbf {\bibinfo {volume} {46}},\ \bibinfo {pages} {119}
  (\bibinfo {year} {2012})}\BibitemShut {NoStop}%
\bibitem [{\citenamefont {Naldesi}\ \emph {et~al.}(2022)\citenamefont
  {Naldesi}, \citenamefont {Polo}, \citenamefont {Dunjko}, \citenamefont
  {Perrin}, \citenamefont {Olshanii}, \citenamefont {Amico},\ and\
  \citenamefont {Minguzzi}}]{Naldesi2022}%
  \BibitemOpen
  \bibfield  {author} {\bibinfo {author} {\bibfnamefont {P.}~\bibnamefont
  {Naldesi}}, \bibinfo {author} {\bibfnamefont {J.}~\bibnamefont {Polo}},
  \bibinfo {author} {\bibfnamefont {V.}~\bibnamefont {Dunjko}}, \bibinfo
  {author} {\bibfnamefont {H.}~\bibnamefont {Perrin}}, \bibinfo {author}
  {\bibfnamefont {M.}~\bibnamefont {Olshanii}}, \bibinfo {author}
  {\bibfnamefont {L.}~\bibnamefont {Amico}},\ and\ \bibinfo {author}
  {\bibfnamefont {A.}~\bibnamefont {Minguzzi}},\ }\bibfield  {title} {\bibinfo
  {title} {{Enhancing sensitivity to rotations with quantum solitonic
  currents}},\ }\href {https://doi.org/10.21468/SciPostPhys.12.4.138}
  {\bibfield  {journal} {\bibinfo  {journal} {SciPost Phys.}\ }\textbf
  {\bibinfo {volume} {12}},\ \bibinfo {pages} {138} (\bibinfo {year}
  {2022})}\BibitemShut {NoStop}%
\bibitem [{\citenamefont {Pecci}\ \emph {et~al.}(2022)\citenamefont {Pecci},
  \citenamefont {Aupetit-Diallo}, \citenamefont {Albert}, \citenamefont
  {Vignolo},\ and\ \citenamefont {Minguzzi}}]{Pecci2022}%
  \BibitemOpen
  \bibfield  {author} {\bibinfo {author} {\bibfnamefont {G.}~\bibnamefont
  {Pecci}}, \bibinfo {author} {\bibfnamefont {G.}~\bibnamefont
  {Aupetit-Diallo}}, \bibinfo {author} {\bibfnamefont {M.}~\bibnamefont
  {Albert}}, \bibinfo {author} {\bibfnamefont {P.}~\bibnamefont {Vignolo}},\
  and\ \bibinfo {author} {\bibfnamefont {A.}~\bibnamefont {Minguzzi}},\
  }\bibfield  {title} {\bibinfo {title} {Persistent currents in a strongly
  interacting multicomponent {Bose} gas on a ring},\ }\bibfield  {journal}
  {\bibinfo  {journal} {arXiv:2211.16194}\ }\href
  {https://doi.org/10.48550/ARXIV.2211.16194} {10.48550/ARXIV.2211.16194}
  (\bibinfo {year} {2022})\BibitemShut {NoStop}%
\bibitem [{\citenamefont {R\"omer}\ and\ \citenamefont
  {Raikh}(2000)}]{Romer2000}%
  \BibitemOpen
  \bibfield  {author} {\bibinfo {author} {\bibfnamefont {R.~A.}\ \bibnamefont
  {R\"omer}}\ and\ \bibinfo {author} {\bibfnamefont {M.~E.}\ \bibnamefont
  {Raikh}},\ }\bibfield  {title} {\bibinfo {title} {{Aharonov}-{Bohm} effect
  for an exciton},\ }\href {https://doi.org/10.1103/PhysRevB.62.7045}
  {\bibfield  {journal} {\bibinfo  {journal} {Phys. Rev. B}\ }\textbf {\bibinfo
  {volume} {62}},\ \bibinfo {pages} {7045} (\bibinfo {year}
  {2000})}\BibitemShut {NoStop}%
\bibitem [{\citenamefont {Kyriakou}\ \emph {et~al.}(2010)\citenamefont
  {Kyriakou}, \citenamefont {Moulopoulos}, \citenamefont {Ghazaryan},\ and\
  \citenamefont {Djotyan}}]{Kyriakou2010}%
  \BibitemOpen
  \bibfield  {author} {\bibinfo {author} {\bibfnamefont {K.}~\bibnamefont
  {Kyriakou}}, \bibinfo {author} {\bibfnamefont {K.}~\bibnamefont
  {Moulopoulos}}, \bibinfo {author} {\bibfnamefont {A.~V.}\ \bibnamefont
  {Ghazaryan}},\ and\ \bibinfo {author} {\bibfnamefont {A.~P.}\ \bibnamefont
  {Djotyan}},\ }\bibfield  {title} {\bibinfo {title} {Arbitrary mixture of two
  charged interacting particles in a magnetic {Aharonov}–{Bohm} ring:
  persistent currents and {Berry}'s phases},\ }\href
  {https://doi.org/10.1088/1751-8113/43/35/354018} {\bibfield  {journal}
  {\bibinfo  {journal} {Journal of Physics A: Mathematical and Theoretical}\
  }\textbf {\bibinfo {volume} {43}},\ \bibinfo {pages} {354018} (\bibinfo
  {year} {2010})}\BibitemShut {NoStop}%
\bibitem [{\citenamefont {Moulopoulos}\ and\ \citenamefont
  {Constantinou}(2004)}]{Moulopoulos2004}%
  \BibitemOpen
  \bibfield  {author} {\bibinfo {author} {\bibfnamefont {K.}~\bibnamefont
  {Moulopoulos}}\ and\ \bibinfo {author} {\bibfnamefont {M.}~\bibnamefont
  {Constantinou}},\ }\bibfield  {title} {\bibinfo {title} {Two interacting
  charged particles in an {Aharonov}-{Bohm} ring: Bound state transitions,
  symmetry breaking, persistent currents, and {Berry}'s phase},\ }\href
  {https://doi.org/10.1103/PhysRevB.70.235327} {\bibfield  {journal} {\bibinfo
  {journal} {Phys. Rev. B}\ }\textbf {\bibinfo {volume} {70}},\ \bibinfo
  {pages} {235327} (\bibinfo {year} {2004})}\BibitemShut {NoStop}%
\bibitem [{\citenamefont {Ghazaryan}\ \emph {et~al.}(2011)\citenamefont
  {Ghazaryan}, \citenamefont {Djotyan}, \citenamefont {Moulopoulos},\ and\
  \citenamefont {Kirakosyan}}]{Ghazaryan2011}%
  \BibitemOpen
  \bibfield  {author} {\bibinfo {author} {\bibfnamefont {A.~V.}\ \bibnamefont
  {Ghazaryan}}, \bibinfo {author} {\bibfnamefont {A.~P.}\ \bibnamefont
  {Djotyan}}, \bibinfo {author} {\bibfnamefont {K.}~\bibnamefont
  {Moulopoulos}},\ and\ \bibinfo {author} {\bibfnamefont {A.~A.}\ \bibnamefont
  {Kirakosyan}},\ }\bibfield  {title} {\bibinfo {title} {Linear dynamic
  polarizability and the absorption spectrum of an exciton in a quantum ring in
  a magnetic field},\ }\href {https://doi.org/10.1088/0031-8949/83/03/035703}
  {\bibfield  {journal} {\bibinfo  {journal} {Physica Scripta}\ }\textbf
  {\bibinfo {volume} {83}},\ \bibinfo {pages} {035703} (\bibinfo {year}
  {2011})}\BibitemShut {NoStop}%
\bibitem [{\citenamefont {Spethmann}\ \emph {et~al.}(2012)\citenamefont
  {Spethmann}, \citenamefont {Kindermann}, \citenamefont {John}, \citenamefont
  {Weber}, \citenamefont {Meschede},\ and\ \citenamefont
  {Widera}}]{spethmann2012}%
  \BibitemOpen
  \bibfield  {author} {\bibinfo {author} {\bibfnamefont {N.}~\bibnamefont
  {Spethmann}}, \bibinfo {author} {\bibfnamefont {F.}~\bibnamefont
  {Kindermann}}, \bibinfo {author} {\bibfnamefont {S.}~\bibnamefont {John}},
  \bibinfo {author} {\bibfnamefont {C.}~\bibnamefont {Weber}}, \bibinfo
  {author} {\bibfnamefont {D.}~\bibnamefont {Meschede}},\ and\ \bibinfo
  {author} {\bibfnamefont {A.}~\bibnamefont {Widera}},\ }\bibfield  {title}
  {\bibinfo {title} {Dynamics of {S}ingle {N}eutral {I}mpurity {A}toms
  {I}mmersed in an {U}ltracold {G}as},\ }\href
  {https://doi.org/10.1103/PhysRevLett.109.235301} {\bibfield  {journal}
  {\bibinfo  {journal} {Phys. Rev. Lett.}\ }\textbf {\bibinfo {volume} {109}},\
  \bibinfo {pages} {235301} (\bibinfo {year} {2012})}\BibitemShut {NoStop}%
\bibitem [{\citenamefont {Catani}\ \emph {et~al.}(2012)\citenamefont {Catani},
  \citenamefont {Lamporesi}, \citenamefont {Naik}, \citenamefont {Gring},
  \citenamefont {Inguscio}, \citenamefont {Minardi}, \citenamefont {Kantian},\
  and\ \citenamefont {Giamarchi}}]{Catani2012Experiment}%
  \BibitemOpen
  \bibfield  {author} {\bibinfo {author} {\bibfnamefont {J.}~\bibnamefont
  {Catani}}, \bibinfo {author} {\bibfnamefont {G.}~\bibnamefont {Lamporesi}},
  \bibinfo {author} {\bibfnamefont {D.}~\bibnamefont {Naik}}, \bibinfo {author}
  {\bibfnamefont {M.}~\bibnamefont {Gring}}, \bibinfo {author} {\bibfnamefont
  {M.}~\bibnamefont {Inguscio}}, \bibinfo {author} {\bibfnamefont
  {F.}~\bibnamefont {Minardi}}, \bibinfo {author} {\bibfnamefont
  {A.}~\bibnamefont {Kantian}},\ and\ \bibinfo {author} {\bibfnamefont
  {T.}~\bibnamefont {Giamarchi}},\ }\bibfield  {title} {\bibinfo {title}
  {{Quantum dynamics of impurities in a one-dimensional {Bose} gas}},\ }\href
  {https://doi.org/10.1103/PhysRevA.85.023623} {\bibfield  {journal} {\bibinfo
  {journal} {Phys. Rev. A}\ }\textbf {\bibinfo {volume} {85}},\ \bibinfo
  {pages} {023623} (\bibinfo {year} {2012})}\BibitemShut {NoStop}%
\bibitem [{\citenamefont {Hu}\ \emph {et~al.}(2016)\citenamefont {Hu},
  \citenamefont {Van~de Graaff}, \citenamefont {Kedar}, \citenamefont {Corson},
  \citenamefont {Cornell},\ and\ \citenamefont {Jin}}]{Jin2016}%
  \BibitemOpen
  \bibfield  {author} {\bibinfo {author} {\bibfnamefont {M.-G.}\ \bibnamefont
  {Hu}}, \bibinfo {author} {\bibfnamefont {M.~J.}\ \bibnamefont {Van~de
  Graaff}}, \bibinfo {author} {\bibfnamefont {D.}~\bibnamefont {Kedar}},
  \bibinfo {author} {\bibfnamefont {J.~P.}\ \bibnamefont {Corson}}, \bibinfo
  {author} {\bibfnamefont {E.~A.}\ \bibnamefont {Cornell}},\ and\ \bibinfo
  {author} {\bibfnamefont {D.~S.}\ \bibnamefont {Jin}},\ }\bibfield  {title}
  {\bibinfo {title} {{{Bose} Polarons in the Strongly Interacting Regime}},\
  }\href {https://doi.org/10.1103/PhysRevLett.117.055301} {\bibfield  {journal}
  {\bibinfo  {journal} {Phys. Rev. Lett.}\ }\textbf {\bibinfo {volume} {117}},\
  \bibinfo {pages} {055301} (\bibinfo {year} {2016})}\BibitemShut {NoStop}%
\bibitem [{\citenamefont {J\o{}rgensen}\ \emph {et~al.}(2016)\citenamefont
  {J\o{}rgensen}, \citenamefont {Wacker}, \citenamefont {Skalmstang},
  \citenamefont {Parish}, \citenamefont {Levinsen}, \citenamefont
  {Christensen}, \citenamefont {Bruun},\ and\ \citenamefont {Arlt}}]{Arlt2016}%
  \BibitemOpen
  \bibfield  {author} {\bibinfo {author} {\bibfnamefont {N.~B.}\ \bibnamefont
  {J\o{}rgensen}}, \bibinfo {author} {\bibfnamefont {L.}~\bibnamefont
  {Wacker}}, \bibinfo {author} {\bibfnamefont {K.~T.}\ \bibnamefont
  {Skalmstang}}, \bibinfo {author} {\bibfnamefont {M.~M.}\ \bibnamefont
  {Parish}}, \bibinfo {author} {\bibfnamefont {J.}~\bibnamefont {Levinsen}},
  \bibinfo {author} {\bibfnamefont {R.~S.}\ \bibnamefont {Christensen}},
  \bibinfo {author} {\bibfnamefont {G.~M.}\ \bibnamefont {Bruun}},\ and\
  \bibinfo {author} {\bibfnamefont {J.~J.}\ \bibnamefont {Arlt}},\ }\bibfield
  {title} {\bibinfo {title} {{Observation of Attractive and Repulsive Polarons
  in a {Bose}-{Einstein} Condensate}},\ }\href
  {https://doi.org/10.1103/PhysRevLett.117.055302} {\bibfield  {journal}
  {\bibinfo  {journal} {Phys. Rev. Lett.}\ }\textbf {\bibinfo {volume} {117}},\
  \bibinfo {pages} {055302} (\bibinfo {year} {2016})}\BibitemShut {NoStop}%
\bibitem [{\citenamefont {Pe\~na Ardila}\ \emph {et~al.}(2019)\citenamefont
  {Pe\~na Ardila}, \citenamefont {J\o{}rgensen}, \citenamefont {Pohl},
  \citenamefont {Giorgini}, \citenamefont {Bruun},\ and\ \citenamefont
  {Arlt}}]{Ardilla2019}%
  \BibitemOpen
  \bibfield  {author} {\bibinfo {author} {\bibfnamefont {L.~A.}\ \bibnamefont
  {Pe\~na Ardila}}, \bibinfo {author} {\bibfnamefont {N.~B.}\ \bibnamefont
  {J\o{}rgensen}}, \bibinfo {author} {\bibfnamefont {T.}~\bibnamefont {Pohl}},
  \bibinfo {author} {\bibfnamefont {S.}~\bibnamefont {Giorgini}}, \bibinfo
  {author} {\bibfnamefont {G.~M.}\ \bibnamefont {Bruun}},\ and\ \bibinfo
  {author} {\bibfnamefont {J.~J.}\ \bibnamefont {Arlt}},\ }\bibfield  {title}
  {\bibinfo {title} {Analyzing a {Bose} polaron across resonant interactions},\
  }\href {https://doi.org/10.1103/PhysRevA.99.063607} {\bibfield  {journal}
  {\bibinfo  {journal} {Phys. Rev. A}\ }\textbf {\bibinfo {volume} {99}},\
  \bibinfo {pages} {063607} (\bibinfo {year} {2019})}\BibitemShut {NoStop}%
\bibitem [{\citenamefont {Yan}\ \emph {et~al.}(2020)\citenamefont {Yan},
  \citenamefont {Ni}, \citenamefont {Robens},\ and\ \citenamefont
  {Zwierlein}}]{Yan2020}%
  \BibitemOpen
  \bibfield  {author} {\bibinfo {author} {\bibfnamefont {Z.~Z.}\ \bibnamefont
  {Yan}}, \bibinfo {author} {\bibfnamefont {Y.}~\bibnamefont {Ni}}, \bibinfo
  {author} {\bibfnamefont {C.}~\bibnamefont {Robens}},\ and\ \bibinfo {author}
  {\bibfnamefont {M.~W.}\ \bibnamefont {Zwierlein}},\ }\bibfield  {title}
  {\bibinfo {title} {{Bose} polarons near quantum criticality},\ }\href
  {https://doi.org/10.1126/science.aax5850} {\bibfield  {journal} {\bibinfo
  {journal} {Science}\ }\textbf {\bibinfo {volume} {368}},\ \bibinfo {pages}
  {190} (\bibinfo {year} {2020})}\BibitemShut {NoStop}%
\bibitem [{\citenamefont {Skou}\ \emph {et~al.}(2021)\citenamefont {Skou},
  \citenamefont {Skov}, \citenamefont {J{\o}rgensen}, \citenamefont {Nielsen},
  \citenamefont {Camacho-Guardian}, \citenamefont {Pohl}, \citenamefont
  {Bruun},\ and\ \citenamefont {Arlt}}]{Skou2021Polaron}%
  \BibitemOpen
  \bibfield  {author} {\bibinfo {author} {\bibfnamefont {M.~G.}\ \bibnamefont
  {Skou}}, \bibinfo {author} {\bibfnamefont {T.~G.}\ \bibnamefont {Skov}},
  \bibinfo {author} {\bibfnamefont {N.~B.}\ \bibnamefont {J{\o}rgensen}},
  \bibinfo {author} {\bibfnamefont {K.~K.}\ \bibnamefont {Nielsen}}, \bibinfo
  {author} {\bibfnamefont {A.}~\bibnamefont {Camacho-Guardian}}, \bibinfo
  {author} {\bibfnamefont {T.}~\bibnamefont {Pohl}}, \bibinfo {author}
  {\bibfnamefont {G.~M.}\ \bibnamefont {Bruun}},\ and\ \bibinfo {author}
  {\bibfnamefont {J.~J.}\ \bibnamefont {Arlt}},\ }\bibfield  {title} {\bibinfo
  {title} {Non-equilibrium quantum dynamics and formation of the {Bose}
  polaron},\ }\href {https://doi.org/10.1038%2Fs41567-021-01184-5} {\bibfield
  {journal} {\bibinfo  {journal} {Nature Physics}\ }\textbf {\bibinfo {volume}
  {17}} (\bibinfo {year} {2021})}\BibitemShut {NoStop}%
\bibitem [{\citenamefont {Dalibard}\ \emph {et~al.}(2011)\citenamefont
  {Dalibard}, \citenamefont {Gerbier}, \citenamefont
  {Juzeli\ifmmode~\bar{u}\else \={u}\fi{}nas},\ and\ \citenamefont
  {\"Ohberg}}]{Dalibard2011}%
  \BibitemOpen
  \bibfield  {author} {\bibinfo {author} {\bibfnamefont {J.}~\bibnamefont
  {Dalibard}}, \bibinfo {author} {\bibfnamefont {F.}~\bibnamefont {Gerbier}},
  \bibinfo {author} {\bibfnamefont {G.}~\bibnamefont
  {Juzeli\ifmmode~\bar{u}\else \={u}\fi{}nas}},\ and\ \bibinfo {author}
  {\bibfnamefont {P.}~\bibnamefont {\"Ohberg}},\ }\bibfield  {title} {\bibinfo
  {title} {Colloquium: Artificial gauge potentials for neutral atoms},\ }\href
  {https://doi.org/10.1103/RevModPhys.83.1523} {\bibfield  {journal} {\bibinfo
  {journal} {Rev. Mod. Phys.}\ }\textbf {\bibinfo {volume} {83}},\ \bibinfo
  {pages} {1523} (\bibinfo {year} {2011})}\BibitemShut {NoStop}%
\bibitem [{\citenamefont {Amico}\ \emph {et~al.}(2021)\citenamefont {Amico},
  \citenamefont {Boshier}, \citenamefont {Birkl}, \citenamefont {Minguzzi},
  \citenamefont {Miniatura}, \citenamefont {Kwek}, \citenamefont {Aghamalyan},
  \citenamefont {Ahufinger}, \citenamefont {Anderson}, \citenamefont {Andrei},
  \citenamefont {Arnold}, \citenamefont {Baker}, \citenamefont {Bell},
  \citenamefont {Bland}, \citenamefont {Brantut}, \citenamefont {Cassettari},
  \citenamefont {Chetcuti}, \citenamefont {Chevy}, \citenamefont {Citro},
  \citenamefont {De~Palo}, \citenamefont {Dumke}, \citenamefont {Edwards},
  \citenamefont {Folman}, \citenamefont {Fortagh}, \citenamefont {Gardiner},
  \citenamefont {Garraway}, \citenamefont {Gauthier}, \citenamefont {Günther},
  \citenamefont {Haug}, \citenamefont {Hufnagel}, \citenamefont {Keil},
  \citenamefont {Ireland}, \citenamefont {Lebrat}, \citenamefont {Li},
  \citenamefont {Longchambon}, \citenamefont {Mompart}, \citenamefont {Morsch},
  \citenamefont {Naldesi}, \citenamefont {Neely}, \citenamefont {Olshanii},
  \citenamefont {Orignac}, \citenamefont {Pandey}, \citenamefont
  {Pérez-Obiol}, \citenamefont {Perrin}, \citenamefont {Piroli}, \citenamefont
  {Polo}, \citenamefont {Pritchard}, \citenamefont {Proukakis}, \citenamefont
  {Rylands}, \citenamefont {Rubinsztein-Dunlop}, \citenamefont {Scazza},
  \citenamefont {Stringari}, \citenamefont {Tosto}, \citenamefont
  {Trombettoni}, \citenamefont {Victorin}, \citenamefont {Klitzing},
  \citenamefont {Wilkowski}, \citenamefont {Xhani},\ and\ \citenamefont
  {Yakimenko}}]{Amico2021}%
  \BibitemOpen
  \bibfield  {author} {\bibinfo {author} {\bibfnamefont {L.}~\bibnamefont
  {Amico}}, \bibinfo {author} {\bibfnamefont {M.}~\bibnamefont {Boshier}},
  \bibinfo {author} {\bibfnamefont {G.}~\bibnamefont {Birkl}}, \bibinfo
  {author} {\bibfnamefont {A.}~\bibnamefont {Minguzzi}}, \bibinfo {author}
  {\bibfnamefont {C.}~\bibnamefont {Miniatura}}, \bibinfo {author}
  {\bibfnamefont {L.-C.}\ \bibnamefont {Kwek}}, \bibinfo {author}
  {\bibfnamefont {D.}~\bibnamefont {Aghamalyan}}, \bibinfo {author}
  {\bibfnamefont {V.}~\bibnamefont {Ahufinger}}, \bibinfo {author}
  {\bibfnamefont {D.}~\bibnamefont {Anderson}}, \bibinfo {author}
  {\bibfnamefont {N.}~\bibnamefont {Andrei}}, \bibinfo {author} {\bibfnamefont
  {A.~S.}\ \bibnamefont {Arnold}}, \bibinfo {author} {\bibfnamefont
  {M.}~\bibnamefont {Baker}}, \bibinfo {author} {\bibfnamefont {T.~A.}\
  \bibnamefont {Bell}}, \bibinfo {author} {\bibfnamefont {T.}~\bibnamefont
  {Bland}}, \bibinfo {author} {\bibfnamefont {J.~P.}\ \bibnamefont {Brantut}},
  \bibinfo {author} {\bibfnamefont {D.}~\bibnamefont {Cassettari}}, \bibinfo
  {author} {\bibfnamefont {W.~J.}\ \bibnamefont {Chetcuti}}, \bibinfo {author}
  {\bibfnamefont {F.}~\bibnamefont {Chevy}}, \bibinfo {author} {\bibfnamefont
  {R.}~\bibnamefont {Citro}}, \bibinfo {author} {\bibfnamefont
  {S.}~\bibnamefont {De~Palo}}, \bibinfo {author} {\bibfnamefont
  {R.}~\bibnamefont {Dumke}}, \bibinfo {author} {\bibfnamefont
  {M.}~\bibnamefont {Edwards}}, \bibinfo {author} {\bibfnamefont
  {R.}~\bibnamefont {Folman}}, \bibinfo {author} {\bibfnamefont
  {J.}~\bibnamefont {Fortagh}}, \bibinfo {author} {\bibfnamefont {S.~A.}\
  \bibnamefont {Gardiner}}, \bibinfo {author} {\bibfnamefont {B.~M.}\
  \bibnamefont {Garraway}}, \bibinfo {author} {\bibfnamefont {G.}~\bibnamefont
  {Gauthier}}, \bibinfo {author} {\bibfnamefont {A.}~\bibnamefont {Günther}},
  \bibinfo {author} {\bibfnamefont {T.}~\bibnamefont {Haug}}, \bibinfo {author}
  {\bibfnamefont {C.}~\bibnamefont {Hufnagel}}, \bibinfo {author}
  {\bibfnamefont {M.}~\bibnamefont {Keil}}, \bibinfo {author} {\bibfnamefont
  {P.}~\bibnamefont {Ireland}}, \bibinfo {author} {\bibfnamefont
  {M.}~\bibnamefont {Lebrat}}, \bibinfo {author} {\bibfnamefont
  {W.}~\bibnamefont {Li}}, \bibinfo {author} {\bibfnamefont {L.}~\bibnamefont
  {Longchambon}}, \bibinfo {author} {\bibfnamefont {J.}~\bibnamefont
  {Mompart}}, \bibinfo {author} {\bibfnamefont {O.}~\bibnamefont {Morsch}},
  \bibinfo {author} {\bibfnamefont {P.}~\bibnamefont {Naldesi}}, \bibinfo
  {author} {\bibfnamefont {T.~W.}\ \bibnamefont {Neely}}, \bibinfo {author}
  {\bibfnamefont {M.}~\bibnamefont {Olshanii}}, \bibinfo {author}
  {\bibfnamefont {E.}~\bibnamefont {Orignac}}, \bibinfo {author} {\bibfnamefont
  {S.}~\bibnamefont {Pandey}}, \bibinfo {author} {\bibfnamefont
  {A.}~\bibnamefont {Pérez-Obiol}}, \bibinfo {author} {\bibfnamefont
  {H.}~\bibnamefont {Perrin}}, \bibinfo {author} {\bibfnamefont
  {L.}~\bibnamefont {Piroli}}, \bibinfo {author} {\bibfnamefont
  {J.}~\bibnamefont {Polo}}, \bibinfo {author} {\bibfnamefont {A.~L.}\
  \bibnamefont {Pritchard}}, \bibinfo {author} {\bibfnamefont {N.~P.}\
  \bibnamefont {Proukakis}}, \bibinfo {author} {\bibfnamefont {C.}~\bibnamefont
  {Rylands}}, \bibinfo {author} {\bibfnamefont {H.}~\bibnamefont
  {Rubinsztein-Dunlop}}, \bibinfo {author} {\bibfnamefont {F.}~\bibnamefont
  {Scazza}}, \bibinfo {author} {\bibfnamefont {S.}~\bibnamefont {Stringari}},
  \bibinfo {author} {\bibfnamefont {F.}~\bibnamefont {Tosto}}, \bibinfo
  {author} {\bibfnamefont {A.}~\bibnamefont {Trombettoni}}, \bibinfo {author}
  {\bibfnamefont {N.}~\bibnamefont {Victorin}}, \bibinfo {author}
  {\bibfnamefont {W.~v.}\ \bibnamefont {Klitzing}}, \bibinfo {author}
  {\bibfnamefont {D.}~\bibnamefont {Wilkowski}}, \bibinfo {author}
  {\bibfnamefont {K.}~\bibnamefont {Xhani}},\ and\ \bibinfo {author}
  {\bibfnamefont {A.}~\bibnamefont {Yakimenko}},\ }\bibfield  {title} {\bibinfo
  {title} {Roadmap on atomtronics: State of the art and perspective},\ }\href
  {https://doi.org/10.1116/5.0026178} {\bibfield  {journal} {\bibinfo
  {journal} {AVS Quantum Science}\ }\textbf {\bibinfo {volume} {3}},\ \bibinfo
  {pages} {039201} (\bibinfo {year} {2021})}\BibitemShut {NoStop}%
\bibitem [{\citenamefont {Amico}\ \emph {et~al.}(2022)\citenamefont {Amico},
  \citenamefont {Anderson}, \citenamefont {Boshier}, \citenamefont {Brantut},
  \citenamefont {Kwek}, \citenamefont {Minguzzi},\ and\ \citenamefont {von
  Klitzing}}]{Amico_2022}%
  \BibitemOpen
  \bibfield  {author} {\bibinfo {author} {\bibfnamefont {L.}~\bibnamefont
  {Amico}}, \bibinfo {author} {\bibfnamefont {D.}~\bibnamefont {Anderson}},
  \bibinfo {author} {\bibfnamefont {M.}~\bibnamefont {Boshier}}, \bibinfo
  {author} {\bibfnamefont {J.-P.}\ \bibnamefont {Brantut}}, \bibinfo {author}
  {\bibfnamefont {L.-C.}\ \bibnamefont {Kwek}}, \bibinfo {author}
  {\bibfnamefont {A.}~\bibnamefont {Minguzzi}},\ and\ \bibinfo {author}
  {\bibfnamefont {W.}~\bibnamefont {von Klitzing}},\ }\bibfield  {title}
  {\bibinfo {title} {Colloquium: Atomtronic circuits: From many-body physics to
  quantum technologies},\ }\bibfield  {journal} {\bibinfo  {journal} {Reviews
  of Modern Physics}\ }\textbf {\bibinfo {volume} {94}},\ \href
  {https://doi.org/10.1103/revmodphys.94.041001} {10.1103/revmodphys.94.041001}
  (\bibinfo {year} {2022})\BibitemShut {NoStop}%
\bibitem [{\citenamefont {Aladinskaia}\ \emph {et~al.}(2023)\citenamefont
  {Aladinskaia}, \citenamefont {Cherbunin}, \citenamefont {Sedov},
  \citenamefont {Liubomirov}, \citenamefont {Kavokin}, \citenamefont
  {Khramtsov}, \citenamefont {Petrov}, \citenamefont {Savvidis},\ and\
  \citenamefont {Kavokin}}]{RingShapedPolaritons1}%
  \BibitemOpen
  \bibfield  {author} {\bibinfo {author} {\bibfnamefont {E.}~\bibnamefont
  {Aladinskaia}}, \bibinfo {author} {\bibfnamefont {R.}~\bibnamefont
  {Cherbunin}}, \bibinfo {author} {\bibfnamefont {E.}~\bibnamefont {Sedov}},
  \bibinfo {author} {\bibfnamefont {A.}~\bibnamefont {Liubomirov}}, \bibinfo
  {author} {\bibfnamefont {K.}~\bibnamefont {Kavokin}}, \bibinfo {author}
  {\bibfnamefont {E.}~\bibnamefont {Khramtsov}}, \bibinfo {author}
  {\bibfnamefont {M.}~\bibnamefont {Petrov}}, \bibinfo {author} {\bibfnamefont
  {P.~G.}\ \bibnamefont {Savvidis}},\ and\ \bibinfo {author} {\bibfnamefont
  {A.}~\bibnamefont {Kavokin}},\ }\bibfield  {title} {\bibinfo {title} {Spatial
  quantization of exciton-polariton condensates in optically induced traps},\
  }\href {https://doi.org/10.1103/PhysRevB.107.045302} {\bibfield  {journal}
  {\bibinfo  {journal} {Phys. Rev. B}\ }\textbf {\bibinfo {volume} {107}},\
  \bibinfo {pages} {045302} (\bibinfo {year} {2023})}\BibitemShut {NoStop}%
\bibitem [{\citenamefont {Lukoshkin}\ \emph {et~al.}(2018)\citenamefont
  {Lukoshkin}, \citenamefont {Kalevich}, \citenamefont {Afanasiev},
  \citenamefont {Kavokin}, \citenamefont {Hatzopoulos}, \citenamefont
  {Savvidis}, \citenamefont {Sedov},\ and\ \citenamefont
  {Kavokin}}]{RingShapedPolaritons2}%
  \BibitemOpen
  \bibfield  {author} {\bibinfo {author} {\bibfnamefont {V.~A.}\ \bibnamefont
  {Lukoshkin}}, \bibinfo {author} {\bibfnamefont {V.~K.}\ \bibnamefont
  {Kalevich}}, \bibinfo {author} {\bibfnamefont {M.~M.}\ \bibnamefont
  {Afanasiev}}, \bibinfo {author} {\bibfnamefont {K.~V.}\ \bibnamefont
  {Kavokin}}, \bibinfo {author} {\bibfnamefont {Z.}~\bibnamefont
  {Hatzopoulos}}, \bibinfo {author} {\bibfnamefont {P.~G.}\ \bibnamefont
  {Savvidis}}, \bibinfo {author} {\bibfnamefont {E.~S.}\ \bibnamefont
  {Sedov}},\ and\ \bibinfo {author} {\bibfnamefont {A.~V.}\ \bibnamefont
  {Kavokin}},\ }\bibfield  {title} {\bibinfo {title} {Persistent circular
  currents of exciton-polaritons in cylindrical pillar microcavities},\ }\href
  {https://doi.org/10.1103/PhysRevB.97.195149} {\bibfield  {journal} {\bibinfo
  {journal} {Phys. Rev. B}\ }\textbf {\bibinfo {volume} {97}},\ \bibinfo
  {pages} {195149} (\bibinfo {year} {2018})}\BibitemShut {NoStop}%
\bibitem [{\citenamefont {Kalevich}\ \emph {et~al.}(2014)\citenamefont
  {Kalevich}, \citenamefont {Afanasiev}, \citenamefont {Lukoshkin},
  \citenamefont {Kavokin}, \citenamefont {Tsintzos}, \citenamefont {Savvidis},\
  and\ \citenamefont {Kavokin}}]{RingShapedPolaritons3}%
  \BibitemOpen
  \bibfield  {author} {\bibinfo {author} {\bibfnamefont {V.}~\bibnamefont
  {Kalevich}}, \bibinfo {author} {\bibfnamefont {M.}~\bibnamefont {Afanasiev}},
  \bibinfo {author} {\bibfnamefont {V.}~\bibnamefont {Lukoshkin}}, \bibinfo
  {author} {\bibfnamefont {K.}~\bibnamefont {Kavokin}}, \bibinfo {author}
  {\bibfnamefont {S.}~\bibnamefont {Tsintzos}}, \bibinfo {author}
  {\bibfnamefont {P.}~\bibnamefont {Savvidis}},\ and\ \bibinfo {author}
  {\bibfnamefont {A.}~\bibnamefont {Kavokin}},\ }\bibfield  {title} {\bibinfo
  {title} {Ring-shaped polariton lasing in pillar microcavities},\ }\href@noop
  {} {\bibfield  {journal} {\bibinfo  {journal} {Journal of Applied Physics}\
  }\textbf {\bibinfo {volume} {115}},\ \bibinfo {pages} {094304} (\bibinfo
  {year} {2014})}\BibitemShut {NoStop}%
\bibitem [{\citenamefont {Lim}\ \emph {et~al.}(2017)\citenamefont {Lim},
  \citenamefont {Togan}, \citenamefont {Kroner}, \citenamefont
  {Miguel-Sanchez},\ and\ \citenamefont {Imamo{\u{g}}lu}}]{PolaritonsGauge}%
  \BibitemOpen
  \bibfield  {author} {\bibinfo {author} {\bibfnamefont {H.-T.}\ \bibnamefont
  {Lim}}, \bibinfo {author} {\bibfnamefont {E.}~\bibnamefont {Togan}}, \bibinfo
  {author} {\bibfnamefont {M.}~\bibnamefont {Kroner}}, \bibinfo {author}
  {\bibfnamefont {J.}~\bibnamefont {Miguel-Sanchez}},\ and\ \bibinfo {author}
  {\bibfnamefont {A.}~\bibnamefont {Imamo{\u{g}}lu}},\ }\bibfield  {title}
  {\bibinfo {title} {Electrically tunable artificial gauge potential for
  polaritons},\ }\href {https://doi.org/10.1038/ncomms14540} {\bibfield
  {journal} {\bibinfo  {journal} {Nature Communications}\ }\textbf {\bibinfo
  {volume} {8}},\ \bibinfo {pages} {14540} (\bibinfo {year}
  {2017})}\BibitemShut {NoStop}%
\bibitem [{\citenamefont {Casteels}\ \emph {et~al.}(2012)\citenamefont
  {Casteels}, \citenamefont {Tempere},\ and\ \citenamefont
  {Devreese}}]{Casteels2012Impurity}%
  \BibitemOpen
  \bibfield  {author} {\bibinfo {author} {\bibfnamefont {W.}~\bibnamefont
  {Casteels}}, \bibinfo {author} {\bibfnamefont {J.}~\bibnamefont {Tempere}},\
  and\ \bibinfo {author} {\bibfnamefont {J.~T.}\ \bibnamefont {Devreese}},\
  }\bibfield  {title} {\bibinfo {title} {{Polaronic properties of an impurity
  in a {Bose}-{Einstein} condensate in reduced dimensions}},\ }\href
  {https://doi.org/10.1103/PhysRevA.86.043614} {\bibfield  {journal} {\bibinfo
  {journal} {Phys. Rev. A}\ }\textbf {\bibinfo {volume} {86}},\ \bibinfo
  {pages} {043614} (\bibinfo {year} {2012})}\BibitemShut {NoStop}%
\bibitem [{\citenamefont {Petkovi\ifmmode~\acute{c}\else \'{c}\fi{}}\ and\
  \citenamefont {Ristivojevic}(2016)}]{Petkovic2016}%
  \BibitemOpen
  \bibfield  {author} {\bibinfo {author} {\bibfnamefont {A.}~\bibnamefont
  {Petkovi\ifmmode~\acute{c}\else \'{c}\fi{}}}\ and\ \bibinfo {author}
  {\bibfnamefont {Z.}~\bibnamefont {Ristivojevic}},\ }\bibfield  {title}
  {\bibinfo {title} {Dynamics of a mobile impurity in a one-dimensional {Bose}
  liquid},\ }\href {https://doi.org/10.1103/PhysRevLett.117.105301} {\bibfield
  {journal} {\bibinfo  {journal} {Phys. Rev. Lett.}\ }\textbf {\bibinfo
  {volume} {117}},\ \bibinfo {pages} {105301} (\bibinfo {year}
  {2016})}\BibitemShut {NoStop}%
\bibitem [{\citenamefont {Schecter}\ \emph {et~al.}(2016)\citenamefont
  {Schecter}, \citenamefont {Gangardt},\ and\ \citenamefont
  {Kamenev}}]{Schecter2016}%
  \BibitemOpen
  \bibfield  {author} {\bibinfo {author} {\bibfnamefont {M.}~\bibnamefont
  {Schecter}}, \bibinfo {author} {\bibfnamefont {D.~M.}\ \bibnamefont
  {Gangardt}},\ and\ \bibinfo {author} {\bibfnamefont {A.}~\bibnamefont
  {Kamenev}},\ }\bibfield  {title} {\bibinfo {title} {Quantum impurities: from
  mobile {Josephson} junctions to depletons},\ }\href
  {https://doi.org/10.1088/1367-2630/18/6/065002} {\bibfield  {journal}
  {\bibinfo  {journal} {New Journal of Physics}\ }\textbf {\bibinfo {volume}
  {18}},\ \bibinfo {pages} {065002} (\bibinfo {year} {2016})}\BibitemShut
  {NoStop}%
\bibitem [{\citenamefont {Parisi}\ and\ \citenamefont
  {Giorgini}(2017)}]{Parisi2017BosePolaron}%
  \BibitemOpen
  \bibfield  {author} {\bibinfo {author} {\bibfnamefont {L.}~\bibnamefont
  {Parisi}}\ and\ \bibinfo {author} {\bibfnamefont {S.}~\bibnamefont
  {Giorgini}},\ }\bibfield  {title} {\bibinfo {title} {{Quantum Monte Carlo
  study of the {Bose}-polaron problem in a one-dimensional gas with contact
  interactions}},\ }\href {https://doi.org/10.1103/PhysRevA.95.023619}
  {\bibfield  {journal} {\bibinfo  {journal} {Phys. Rev. A}\ }\textbf {\bibinfo
  {volume} {95}},\ \bibinfo {pages} {023619} (\bibinfo {year}
  {2017})}\BibitemShut {NoStop}%
\bibitem [{\citenamefont {Grusdt}\ \emph {et~al.}(2017)\citenamefont {Grusdt},
  \citenamefont {Astrakharchik},\ and\ \citenamefont
  {Demler}}]{Grusdt2017BosePolaron}%
  \BibitemOpen
  \bibfield  {author} {\bibinfo {author} {\bibfnamefont {F.}~\bibnamefont
  {Grusdt}}, \bibinfo {author} {\bibfnamefont {G.~E.}\ \bibnamefont
  {Astrakharchik}},\ and\ \bibinfo {author} {\bibfnamefont {E.}~\bibnamefont
  {Demler}},\ }\bibfield  {title} {\bibinfo {title} {{{Bose} polarons in
  ultracold atoms in one dimension: beyond the Fr\"ohlich paradigm}},\ }\href
  {https://doi.org/10.1088/1367-2630/aa8a2e} {\bibfield  {journal} {\bibinfo
  {journal} {New J. Phys.}\ }\textbf {\bibinfo {volume} {19}},\ \bibinfo
  {pages} {103035} (\bibinfo {year} {2017})}\BibitemShut {NoStop}%
\bibitem [{\citenamefont {Volosniev}\ and\ \citenamefont
  {Hammer}(2017{\natexlab{a}})}]{Volosniev2017BosePolaron}%
  \BibitemOpen
  \bibfield  {author} {\bibinfo {author} {\bibfnamefont {A.~G.}\ \bibnamefont
  {Volosniev}}\ and\ \bibinfo {author} {\bibfnamefont {H.-W.}\ \bibnamefont
  {Hammer}},\ }\bibfield  {title} {\bibinfo {title} {{Analytical approach to
  the {Bose}-polaron problem in one dimension}},\ }\href
  {https://doi.org/10.1103/PhysRevA.96.031601} {\bibfield  {journal} {\bibinfo
  {journal} {Phys. Rev. A}\ }\textbf {\bibinfo {volume} {96}},\ \bibinfo
  {pages} {031601} (\bibinfo {year} {2017}{\natexlab{a}})}\BibitemShut
  {NoStop}%
\bibitem [{\citenamefont {Pastukhov}(2017)}]{Pastukhov2017Impurity}%
  \BibitemOpen
  \bibfield  {author} {\bibinfo {author} {\bibfnamefont {V.}~\bibnamefont
  {Pastukhov}},\ }\bibfield  {title} {\bibinfo {title} {{Impurity states in the
  one-dimensional {Bose} gas}},\ }\href
  {https://doi.org/10.1103/PhysRevA.96.043625} {\bibfield  {journal} {\bibinfo
  {journal} {Phys. Rev. A}\ }\textbf {\bibinfo {volume} {96}},\ \bibinfo
  {pages} {043625} (\bibinfo {year} {2017})}\BibitemShut {NoStop}%
\bibitem [{\citenamefont {Kain}\ and\ \citenamefont
  {Ling}(2018)}]{Kain2018Static}%
  \BibitemOpen
  \bibfield  {author} {\bibinfo {author} {\bibfnamefont {B.}~\bibnamefont
  {Kain}}\ and\ \bibinfo {author} {\bibfnamefont {H.~Y.}\ \bibnamefont
  {Ling}},\ }\bibfield  {title} {\bibinfo {title} {{Analytical study of static
  beyond-Fr\"ohlich {Bose} polarons in one dimension}},\ }\href
  {https://doi.org/10.1103/PhysRevA.98.033610} {\bibfield  {journal} {\bibinfo
  {journal} {Phys. Rev. A}\ }\textbf {\bibinfo {volume} {98}},\ \bibinfo
  {pages} {033610} (\bibinfo {year} {2018})}\BibitemShut {NoStop}%
\bibitem [{\citenamefont {Mistakidis}\ \emph
  {et~al.}(2019{\natexlab{a}})\citenamefont {Mistakidis}, \citenamefont
  {Katsimiga}, \citenamefont {Koutentakis}, \citenamefont {Busch},\ and\
  \citenamefont {Schmelcher}}]{Mistakidis2019QuenchBosePolarons}%
  \BibitemOpen
  \bibfield  {author} {\bibinfo {author} {\bibfnamefont {S.~I.}\ \bibnamefont
  {Mistakidis}}, \bibinfo {author} {\bibfnamefont {G.~C.}\ \bibnamefont
  {Katsimiga}}, \bibinfo {author} {\bibfnamefont {G.~M.}\ \bibnamefont
  {Koutentakis}}, \bibinfo {author} {\bibfnamefont {T.}~\bibnamefont {Busch}},\
  and\ \bibinfo {author} {\bibfnamefont {P.}~\bibnamefont {Schmelcher}},\
  }\bibfield  {title} {\bibinfo {title} {{Quench Dynamics and Orthogonality
  Catastrophe of {Bose} Polarons}},\ }\href
  {https://doi.org/10.1103/PhysRevLett.122.183001} {\bibfield  {journal}
  {\bibinfo  {journal} {Phys. Rev. Lett.}\ }\textbf {\bibinfo {volume} {122}},\
  \bibinfo {pages} {183001} (\bibinfo {year} {2019}{\natexlab{a}})}\BibitemShut
  {NoStop}%
\bibitem [{\citenamefont {Jager}\ \emph {et~al.}(2020)\citenamefont {Jager},
  \citenamefont {Barnett}, \citenamefont {Will},\ and\ \citenamefont
  {Fleischhauer}}]{Jager2020Deformation}%
  \BibitemOpen
  \bibfield  {author} {\bibinfo {author} {\bibfnamefont {J.}~\bibnamefont
  {Jager}}, \bibinfo {author} {\bibfnamefont {R.}~\bibnamefont {Barnett}},
  \bibinfo {author} {\bibfnamefont {M.}~\bibnamefont {Will}},\ and\ \bibinfo
  {author} {\bibfnamefont {M.}~\bibnamefont {Fleischhauer}},\ }\bibfield
  {title} {\bibinfo {title} {{Strong-coupling {Bose} polarons in one dimension:
  Condensate deformation and modified Bogoliubov phonons}},\ }\href
  {https://doi.org/10.1103/PhysRevResearch.2.033142} {\bibfield  {journal}
  {\bibinfo  {journal} {Phys. Rev. Research}\ }\textbf {\bibinfo {volume}
  {2}},\ \bibinfo {pages} {033142} (\bibinfo {year} {2020})}\BibitemShut
  {NoStop}%
\bibitem [{\citenamefont {Monisha}\ \emph {et~al.}(2016)\citenamefont
  {Monisha}, \citenamefont {Sankar}, \citenamefont {Sil},\ and\ \citenamefont
  {Chatterjee}}]{Monisha2016}%
  \BibitemOpen
  \bibfield  {author} {\bibinfo {author} {\bibfnamefont {P.}~\bibnamefont
  {Monisha}}, \bibinfo {author} {\bibfnamefont {I.}~\bibnamefont {Sankar}},
  \bibinfo {author} {\bibfnamefont {S.}~\bibnamefont {Sil}},\ and\ \bibinfo
  {author} {\bibfnamefont {A.}~\bibnamefont {Chatterjee}},\ }\bibfield  {title}
  {\bibinfo {title} {Persistent current in a correlated quantum ring with
  electron-phonon interaction in the presence of {Rashba} interaction and
  {Aharonov}-{Bohm} flux},\ }\href {https://doi.org/10.1038/srep20056}
  {\bibfield  {journal} {\bibinfo  {journal} {Sci Rep}\ }\textbf {\bibinfo
  {volume} {6}},\ \bibinfo {pages} {20056} (\bibinfo {year}
  {2016})}\BibitemShut {NoStop}%
\bibitem [{\citenamefont {Wenz}\ \emph {et~al.}(2013)\citenamefont {Wenz},
  \citenamefont {Z{\"u}rn}, \citenamefont {Murmann}, \citenamefont {Brouzos},
  \citenamefont {Lompe},\ and\ \citenamefont {Jochim}}]{wenz2013}%
  \BibitemOpen
  \bibfield  {author} {\bibinfo {author} {\bibfnamefont {A.}~\bibnamefont
  {Wenz}}, \bibinfo {author} {\bibfnamefont {G.}~\bibnamefont {Z{\"u}rn}},
  \bibinfo {author} {\bibfnamefont {S.}~\bibnamefont {Murmann}}, \bibinfo
  {author} {\bibfnamefont {I.}~\bibnamefont {Brouzos}}, \bibinfo {author}
  {\bibfnamefont {T.}~\bibnamefont {Lompe}},\ and\ \bibinfo {author}
  {\bibfnamefont {S.}~\bibnamefont {Jochim}},\ }\bibfield  {title} {\bibinfo
  {title} {From few to many: Observing the formation of a {Fermi} sea one atom
  at a time},\ }\href@noop {} {\bibfield  {journal} {\bibinfo  {journal}
  {Science}\ }\textbf {\bibinfo {volume} {342}},\ \bibinfo {pages} {457}
  (\bibinfo {year} {2013})}\BibitemShut {NoStop}%
\bibitem [{\citenamefont {Goldman}\ \emph {et~al.}(2014)\citenamefont
  {Goldman}, \citenamefont {Juzeliūnas}, \citenamefont {Öhberg},\ and\
  \citenamefont {Spielman}}]{Goldman2014}%
  \BibitemOpen
  \bibfield  {author} {\bibinfo {author} {\bibfnamefont {N.}~\bibnamefont
  {Goldman}}, \bibinfo {author} {\bibfnamefont {G.}~\bibnamefont
  {Juzeliūnas}}, \bibinfo {author} {\bibfnamefont {P.}~\bibnamefont
  {Öhberg}},\ and\ \bibinfo {author} {\bibfnamefont {I.~B.}\ \bibnamefont
  {Spielman}},\ }\bibfield  {title} {\bibinfo {title} {Light-induced gauge
  fields for ultracold atoms},\ }\href
  {https://doi.org/10.1088/0034-4885/77/12/126401} {\bibfield  {journal}
  {\bibinfo  {journal} {Reports on Progress in Physics}\ }\textbf {\bibinfo
  {volume} {77}},\ \bibinfo {pages} {126401} (\bibinfo {year}
  {2014})}\BibitemShut {NoStop}%
\bibitem [{\citenamefont {Gunn}\ and\ \citenamefont {Gunn}(1988)}]{Gunn1988}%
  \BibitemOpen
  \bibfield  {author} {\bibinfo {author} {\bibfnamefont {J.~C.}\ \bibnamefont
  {Gunn}}\ and\ \bibinfo {author} {\bibfnamefont {J.~M.~F.}\ \bibnamefont
  {Gunn}},\ }\bibfield  {title} {\bibinfo {title} {An exactly soluble hartree
  problem in an external potential},\ }\href
  {https://doi.org/10.1088/0143-0807/9/1/009} {\bibfield  {journal} {\bibinfo
  {journal} {European Journal of Physics}\ }\textbf {\bibinfo {volume} {9}},\
  \bibinfo {pages} {51} (\bibinfo {year} {1988})}\BibitemShut {NoStop}%
\bibitem [{\citenamefont {Kolomeisky}\ \emph {et~al.}(2004)\citenamefont
  {Kolomeisky}, \citenamefont {Straley},\ and\ \citenamefont
  {Kalas}}]{Kolomeisky2004}%
  \BibitemOpen
  \bibfield  {author} {\bibinfo {author} {\bibfnamefont {E.~B.}\ \bibnamefont
  {Kolomeisky}}, \bibinfo {author} {\bibfnamefont {J.~P.}\ \bibnamefont
  {Straley}},\ and\ \bibinfo {author} {\bibfnamefont {R.~M.}\ \bibnamefont
  {Kalas}},\ }\bibfield  {title} {\bibinfo {title} {Ground-state properties of
  artificial bosonic atoms, {Bose} interaction blockade, and the single-atom
  pipette},\ }\href {https://doi.org/10.1103/PhysRevA.69.063401} {\bibfield
  {journal} {\bibinfo  {journal} {Phys. Rev. A}\ }\textbf {\bibinfo {volume}
  {69}},\ \bibinfo {pages} {063401} (\bibinfo {year} {2004})}\BibitemShut
  {NoStop}%
\bibitem [{\citenamefont {Brauneis}\ \emph {et~al.}(2022)\citenamefont
  {Brauneis}, \citenamefont {Backert}, \citenamefont {Mistakidis},
  \citenamefont {Lemeshko}, \citenamefont {Hammer},\ and\ \citenamefont
  {Volosniev}}]{Brauneis2022}%
  \BibitemOpen
  \bibfield  {author} {\bibinfo {author} {\bibfnamefont {F.}~\bibnamefont
  {Brauneis}}, \bibinfo {author} {\bibfnamefont {T.~G.}\ \bibnamefont
  {Backert}}, \bibinfo {author} {\bibfnamefont {S.~I.}\ \bibnamefont
  {Mistakidis}}, \bibinfo {author} {\bibfnamefont {M.}~\bibnamefont
  {Lemeshko}}, \bibinfo {author} {\bibfnamefont {H.-W.}\ \bibnamefont
  {Hammer}},\ and\ \bibinfo {author} {\bibfnamefont {A.~G.}\ \bibnamefont
  {Volosniev}},\ }\bibfield  {title} {\bibinfo {title} {Artificial atoms from
  cold bosons in one dimension},\ }\href
  {https://doi.org/10.1088/1367-2630/ac78d8} {\bibfield  {journal} {\bibinfo
  {journal} {New Journal of Physics}\ }\textbf {\bibinfo {volume} {24}},\
  \bibinfo {pages} {063036} (\bibinfo {year} {2022})}\BibitemShut {NoStop}%
\bibitem [{\citenamefont {Yang}\ \emph {et~al.}(2022)\citenamefont {Yang},
  \citenamefont {Čufar}, \citenamefont {Pahl},\ and\ \citenamefont
  {Brand}}]{Yang2022}%
  \BibitemOpen
  \bibfield  {author} {\bibinfo {author} {\bibfnamefont {M.}~\bibnamefont
  {Yang}}, \bibinfo {author} {\bibfnamefont {M.}~\bibnamefont {Čufar}},
  \bibinfo {author} {\bibfnamefont {E.}~\bibnamefont {Pahl}},\ and\ \bibinfo
  {author} {\bibfnamefont {J.}~\bibnamefont {Brand}},\ }\bibfield  {title}
  {\bibinfo {title} {Polaron-depleton transition in the yrast excitations of a
  one-dimensional {Bose} gas with a mobile impurity},\ }\bibfield  {journal}
  {\bibinfo  {journal} {Condensed Matter}\ }\textbf {\bibinfo {volume} {7}},\
  \href {https://doi.org/10.3390/condmat7010015} {10.3390/condmat7010015}
  (\bibinfo {year} {2022})\BibitemShut {NoStop}%
\bibitem [{\citenamefont {Byers}\ and\ \citenamefont {Yang}(1961)}]{Byers1961}%
  \BibitemOpen
  \bibfield  {author} {\bibinfo {author} {\bibfnamefont {N.}~\bibnamefont
  {Byers}}\ and\ \bibinfo {author} {\bibfnamefont {C.~N.}\ \bibnamefont
  {Yang}},\ }\bibfield  {title} {\bibinfo {title} {Theoretical considerations
  concerning quantized magnetic flux in superconducting cylinders},\ }\href
  {https://doi.org/10.1103/PhysRevLett.7.46} {\bibfield  {journal} {\bibinfo
  {journal} {Phys. Rev. Lett.}\ }\textbf {\bibinfo {volume} {7}},\ \bibinfo
  {pages} {46} (\bibinfo {year} {1961})}\BibitemShut {NoStop}%
\bibitem [{\citenamefont {Imry}(1986)}]{Imry1986}%
  \BibitemOpen
  \bibfield  {author} {\bibinfo {author} {\bibfnamefont {Y.}~\bibnamefont
  {Imry}},\ }\bibinfo {title} {Physics of mesoscopic systems},\ in\ \href
  {https://doi.org/10.1142/9789814415309_0004} {\emph {\bibinfo {booktitle}
  {Directions in Condensed Matter Physics}}}\ (\bibinfo {year} {1986})\ pp.\
  \bibinfo {pages} {101--163}\BibitemShut {NoStop}%
\bibitem [{\citenamefont {Cazalilla}\ \emph {et~al.}(2011)\citenamefont
  {Cazalilla}, \citenamefont {Citro}, \citenamefont {Giamarchi}, \citenamefont
  {Orignac},\ and\ \citenamefont {Rigol}}]{cazalilla2011}%
  \BibitemOpen
  \bibfield  {author} {\bibinfo {author} {\bibfnamefont {M.~A.}\ \bibnamefont
  {Cazalilla}}, \bibinfo {author} {\bibfnamefont {R.}~\bibnamefont {Citro}},
  \bibinfo {author} {\bibfnamefont {T.}~\bibnamefont {Giamarchi}}, \bibinfo
  {author} {\bibfnamefont {E.}~\bibnamefont {Orignac}},\ and\ \bibinfo {author}
  {\bibfnamefont {M.}~\bibnamefont {Rigol}},\ }\bibfield  {title} {\bibinfo
  {title} {One dimensional bosons: From condensed matter systems to ultracold
  gases},\ }\href {https://doi.org/10.1103/RevModPhys.83.1405} {\bibfield
  {journal} {\bibinfo  {journal} {Rev. Mod. Phys.}\ }\textbf {\bibinfo {volume}
  {83}},\ \bibinfo {pages} {1405} (\bibinfo {year} {2011})}\BibitemShut
  {NoStop}%
\bibitem [{\citenamefont {Guan}\ \emph {et~al.}(2013)\citenamefont {Guan},
  \citenamefont {Batchelor},\ and\ \citenamefont {Lee}}]{Guan2013Review}%
  \BibitemOpen
  \bibfield  {author} {\bibinfo {author} {\bibfnamefont {X.-W.}\ \bibnamefont
  {Guan}}, \bibinfo {author} {\bibfnamefont {M.~T.}\ \bibnamefont
  {Batchelor}},\ and\ \bibinfo {author} {\bibfnamefont {C.}~\bibnamefont
  {Lee}},\ }\bibfield  {title} {\bibinfo {title} {{Fermi} gases in one
  dimension: From {B}ethe ansatz to experiments},\ }\href
  {https://doi.org/10.1103/RevModPhys.85.1633} {\bibfield  {journal} {\bibinfo
  {journal} {Rev. Mod. Phys.}\ }\textbf {\bibinfo {volume} {85}},\ \bibinfo
  {pages} {1633} (\bibinfo {year} {2013})}\BibitemShut {NoStop}%
\bibitem [{\citenamefont {Sowi{\'{n}}ski}\ and\ \citenamefont
  {Garc{\'{\i}}a-March}(2019)}]{Sowinski2019Review}%
  \BibitemOpen
  \bibfield  {author} {\bibinfo {author} {\bibfnamefont {T.}~\bibnamefont
  {Sowi{\'{n}}ski}}\ and\ \bibinfo {author} {\bibfnamefont {M.~{\'{A}}.}\
  \bibnamefont {Garc{\'{\i}}a-March}},\ }\bibfield  {title} {\bibinfo {title}
  {One-dimensional mixtures of several ultracold atoms: a review},\ }\href
  {https://doi.org/10.1088/1361-6633/ab3a80} {\bibfield  {journal} {\bibinfo
  {journal} {Reports on Progress in Physics}\ }\textbf {\bibinfo {volume}
  {82}},\ \bibinfo {pages} {104401} (\bibinfo {year} {2019})}\BibitemShut
  {NoStop}%
\bibitem [{\citenamefont {Mistakidis}\ \emph {et~al.}(2022)\citenamefont
  {Mistakidis}, \citenamefont {Volosniev}, \citenamefont {Barfknecht},
  \citenamefont {Fogarty}, \citenamefont {Busch}, \citenamefont {Foerster},
  \citenamefont {Schmelcher},\ and\ \citenamefont {Zinner}}]{Mistakidis2022}%
  \BibitemOpen
  \bibfield  {author} {\bibinfo {author} {\bibfnamefont {S.~I.}\ \bibnamefont
  {Mistakidis}}, \bibinfo {author} {\bibfnamefont {A.~G.}\ \bibnamefont
  {Volosniev}}, \bibinfo {author} {\bibfnamefont {R.~E.}\ \bibnamefont
  {Barfknecht}}, \bibinfo {author} {\bibfnamefont {T.}~\bibnamefont {Fogarty}},
  \bibinfo {author} {\bibfnamefont {T.}~\bibnamefont {Busch}}, \bibinfo
  {author} {\bibfnamefont {A.}~\bibnamefont {Foerster}}, \bibinfo {author}
  {\bibfnamefont {P.}~\bibnamefont {Schmelcher}},\ and\ \bibinfo {author}
  {\bibfnamefont {N.~T.}\ \bibnamefont {Zinner}},\ }\href
  {https://doi.org/10.48550/ARXIV.2202.11071} {\bibinfo {title} {Cold atoms in
  low dimensions -- a laboratory for quantum dynamics}} (\bibinfo {year}
  {2022})\BibitemShut {NoStop}%
\bibitem [{\citenamefont {Chen}\ and\ \citenamefont
  {Chen}(1998)}]{CHEN1998537}%
  \BibitemOpen
  \bibfield  {author} {\bibinfo {author} {\bibfnamefont {H.}~\bibnamefont
  {Chen}}\ and\ \bibinfo {author} {\bibfnamefont {Y.}~\bibnamefont {Chen}},\
  }\bibfield  {title} {\bibinfo {title} {Influence of the {Aharonov}-{Bohm}
  flux on the optical polarons in the molecular-crystal model with the
  dispersion term in a ring},\ }\href
  {https://doi.org/https://doi.org/10.1016/S0038-1098(97)10188-0} {\bibfield
  {journal} {\bibinfo  {journal} {Solid State Communications}\ }\textbf
  {\bibinfo {volume} {105}},\ \bibinfo {pages} {537} (\bibinfo {year}
  {1998})}\BibitemShut {NoStop}%
\bibitem [{\citenamefont {Zhou}\ \emph {et~al.}(1996)\citenamefont {Zhou},
  \citenamefont {Chen},\ and\ \citenamefont {Yu}}]{ZHOU1996167}%
  \BibitemOpen
  \bibfield  {author} {\bibinfo {author} {\bibfnamefont {Y.-C.}\ \bibnamefont
  {Zhou}}, \bibinfo {author} {\bibfnamefont {H.}~\bibnamefont {Chen}},\ and\
  \bibinfo {author} {\bibfnamefont {C.-F.}\ \bibnamefont {Yu}},\ }\bibfield
  {title} {\bibinfo {title} {Effect of the {Aharonov}-{Bohm} potential on the
  acoustical polaron in one-dimensional rings},\ }\href
  {https://doi.org/https://doi.org/10.1016/0375-9601(96)00035-7} {\bibfield
  {journal} {\bibinfo  {journal} {Physics Letters A}\ }\textbf {\bibinfo
  {volume} {212}},\ \bibinfo {pages} {167} (\bibinfo {year}
  {1996})}\BibitemShut {NoStop}%
\bibitem [{\citenamefont {Gross}(1962)}]{Gross1962}%
  \BibitemOpen
  \bibfield  {author} {\bibinfo {author} {\bibfnamefont {E.}~\bibnamefont
  {Gross}},\ }\bibfield  {title} {\bibinfo {title} {Motion of foreign bodies in
  boson systems},\ }\href
  {https://doi.org/https://doi.org/10.1016/0003-4916(62)90217-8} {\bibfield
  {journal} {\bibinfo  {journal} {Annals of Physics}\ }\textbf {\bibinfo
  {volume} {19}},\ \bibinfo {pages} {234} (\bibinfo {year} {1962})}\BibitemShut
  {NoStop}%
\bibitem [{\citenamefont {Lee}\ \emph {et~al.}(1953)\citenamefont {Lee},
  \citenamefont {Low},\ and\ \citenamefont {Pines}}]{Lee1953}%
  \BibitemOpen
  \bibfield  {author} {\bibinfo {author} {\bibfnamefont {T.~D.}\ \bibnamefont
  {Lee}}, \bibinfo {author} {\bibfnamefont {F.~E.}\ \bibnamefont {Low}},\ and\
  \bibinfo {author} {\bibfnamefont {D.}~\bibnamefont {Pines}},\ }\bibfield
  {title} {\bibinfo {title} {The motion of slow electrons in a polar crystal},\
  }\href {https://doi.org/10.1103/PhysRev.90.297} {\bibfield  {journal}
  {\bibinfo  {journal} {Phys. Rev.}\ }\textbf {\bibinfo {volume} {90}},\
  \bibinfo {pages} {297} (\bibinfo {year} {1953})}\BibitemShut {NoStop}%
\bibitem [{\citenamefont {Mistakidis}\ \emph
  {et~al.}(2019{\natexlab{b}})\citenamefont {Mistakidis}, \citenamefont
  {Volosniev}, \citenamefont {Zinner},\ and\ \citenamefont
  {Schmelcher}}]{Mistakidis2019Effective}%
  \BibitemOpen
  \bibfield  {author} {\bibinfo {author} {\bibfnamefont {S.~I.}\ \bibnamefont
  {Mistakidis}}, \bibinfo {author} {\bibfnamefont {A.~G.}\ \bibnamefont
  {Volosniev}}, \bibinfo {author} {\bibfnamefont {N.~T.}\ \bibnamefont
  {Zinner}},\ and\ \bibinfo {author} {\bibfnamefont {P.}~\bibnamefont
  {Schmelcher}},\ }\bibfield  {title} {\bibinfo {title} {Effective approach to
  impurity dynamics in one-dimensional trapped {Bose} gases},\ }\href
  {https://doi.org/10.1103/PhysRevA.100.013619} {\bibfield  {journal} {\bibinfo
   {journal} {Phys. Rev. A}\ }\textbf {\bibinfo {volume} {100}},\ \bibinfo
  {pages} {013619} (\bibinfo {year} {2019}{\natexlab{b}})}\BibitemShut
  {NoStop}%
\bibitem [{\citenamefont {Astrakharchik}\ and\ \citenamefont
  {Brouzos}(2013)}]{Astrakharchik2013Impurity}%
  \BibitemOpen
  \bibfield  {author} {\bibinfo {author} {\bibfnamefont {G.~E.}\ \bibnamefont
  {Astrakharchik}}\ and\ \bibinfo {author} {\bibfnamefont {I.}~\bibnamefont
  {Brouzos}},\ }\bibfield  {title} {\bibinfo {title} {Trapped one-dimensional
  ideal {Fermi} gas with a single impurity},\ }\href
  {https://doi.org/10.1103/PhysRevA.88.021602} {\bibfield  {journal} {\bibinfo
  {journal} {Phys. Rev. A}\ }\textbf {\bibinfo {volume} {88}},\ \bibinfo
  {pages} {021602} (\bibinfo {year} {2013})}\BibitemShut {NoStop}%
\bibitem [{\citenamefont {Levinsen}\ \emph {et~al.}(2015)\citenamefont
  {Levinsen}, \citenamefont {Massignan}, \citenamefont {Bruun},\ and\
  \citenamefont {Parish}}]{Levinsen_2015}%
  \BibitemOpen
  \bibfield  {author} {\bibinfo {author} {\bibfnamefont {J.}~\bibnamefont
  {Levinsen}}, \bibinfo {author} {\bibfnamefont {P.}~\bibnamefont {Massignan}},
  \bibinfo {author} {\bibfnamefont {G.~M.}\ \bibnamefont {Bruun}},\ and\
  \bibinfo {author} {\bibfnamefont {M.~M.}\ \bibnamefont {Parish}},\ }\bibfield
   {title} {\bibinfo {title} {Strong-coupling ansatz for the one-dimensional
  {Fermi} gas in a harmonic potential},\ }\bibfield  {journal} {\bibinfo
  {journal} {Science Advances}\ }\textbf {\bibinfo {volume} {1}},\ \href
  {https://doi.org/10.1126/sciadv.1500197} {10.1126/sciadv.1500197} (\bibinfo
  {year} {2015})\BibitemShut {NoStop}%
\bibitem [{\citenamefont {Scazza}\ \emph {et~al.}(2017)\citenamefont {Scazza},
  \citenamefont {Valtolina}, \citenamefont {Massignan}, \citenamefont {Recati},
  \citenamefont {Amico}, \citenamefont {Burchianti}, \citenamefont {Fort},
  \citenamefont {Inguscio}, \citenamefont {Zaccanti},\ and\ \citenamefont
  {Roati}}]{Scazza2017}%
  \BibitemOpen
  \bibfield  {author} {\bibinfo {author} {\bibfnamefont {F.}~\bibnamefont
  {Scazza}}, \bibinfo {author} {\bibfnamefont {G.}~\bibnamefont {Valtolina}},
  \bibinfo {author} {\bibfnamefont {P.}~\bibnamefont {Massignan}}, \bibinfo
  {author} {\bibfnamefont {A.}~\bibnamefont {Recati}}, \bibinfo {author}
  {\bibfnamefont {A.}~\bibnamefont {Amico}}, \bibinfo {author} {\bibfnamefont
  {A.}~\bibnamefont {Burchianti}}, \bibinfo {author} {\bibfnamefont
  {C.}~\bibnamefont {Fort}}, \bibinfo {author} {\bibfnamefont {M.}~\bibnamefont
  {Inguscio}}, \bibinfo {author} {\bibfnamefont {M.}~\bibnamefont {Zaccanti}},\
  and\ \bibinfo {author} {\bibfnamefont {G.}~\bibnamefont {Roati}},\ }\bibfield
   {title} {\bibinfo {title} {Repulsive {Fermi} polarons in a resonant mixture
  of ultracold $^{6}\mathrm{Li}$ atoms},\ }\href
  {https://doi.org/10.1103/PhysRevLett.118.083602} {\bibfield  {journal}
  {\bibinfo  {journal} {Phys. Rev. Lett.}\ }\textbf {\bibinfo {volume} {118}},\
  \bibinfo {pages} {083602} (\bibinfo {year} {2017})}\BibitemShut {NoStop}%
\bibitem [{\citenamefont {Hakim}(1997)}]{Hakim1997}%
  \BibitemOpen
  \bibfield  {author} {\bibinfo {author} {\bibfnamefont {V.}~\bibnamefont
  {Hakim}},\ }\bibfield  {title} {\bibinfo {title} {Nonlinear schr\"odinger
  flow past an obstacle in one dimension},\ }\href
  {https://doi.org/10.1103/PhysRevE.55.2835} {\bibfield  {journal} {\bibinfo
  {journal} {Phys. Rev. E}\ }\textbf {\bibinfo {volume} {55}},\ \bibinfo
  {pages} {2835} (\bibinfo {year} {1997})}\BibitemShut {NoStop}%
\bibitem [{\citenamefont {Smith}\ and\ \citenamefont
  {Volosniev}(2019)}]{Smith2019}%
  \BibitemOpen
  \bibfield  {author} {\bibinfo {author} {\bibfnamefont {D.~H.}\ \bibnamefont
  {Smith}}\ and\ \bibinfo {author} {\bibfnamefont {A.~G.}\ \bibnamefont
  {Volosniev}},\ }\bibfield  {title} {\bibinfo {title} {Engineering momentum
  profiles of cold-atom beams},\ }\href
  {https://doi.org/10.1103/PhysRevA.100.033604} {\bibfield  {journal} {\bibinfo
   {journal} {Phys. Rev. A}\ }\textbf {\bibinfo {volume} {100}},\ \bibinfo
  {pages} {033604} (\bibinfo {year} {2019})}\BibitemShut {NoStop}%
\bibitem [{\citenamefont {Lieb}(1963)}]{Lieb1963_2}%
  \BibitemOpen
  \bibfield  {author} {\bibinfo {author} {\bibfnamefont {E.~H.}\ \bibnamefont
  {Lieb}},\ }\bibfield  {title} {\bibinfo {title} {Exact analysis of an
  interacting {Bose} gas. ii. the excitation spectrum},\ }\href
  {https://doi.org/10.1103/PhysRev.130.1616} {\bibfield  {journal} {\bibinfo
  {journal} {Phys. Rev.}\ }\textbf {\bibinfo {volume} {130}},\ \bibinfo {pages}
  {1616} (\bibinfo {year} {1963})}\BibitemShut {NoStop}%
\bibitem [{\citenamefont {Syrwid}(2021)}]{Syrwid2021}%
  \BibitemOpen
  \bibfield  {author} {\bibinfo {author} {\bibfnamefont {A.}~\bibnamefont
  {Syrwid}},\ }\bibfield  {title} {\bibinfo {title} {Quantum dark solitons in
  ultracold one-dimensional {Bose} and {Fermi} gases},\ }\href
  {https://doi.org/10.1088/1361-6455/abd37f} {\bibfield  {journal} {\bibinfo
  {journal} {Journal of Physics B: Atomic, Molecular and Optical Physics}\
  }\textbf {\bibinfo {volume} {54}},\ \bibinfo {pages} {103001} (\bibinfo
  {year} {2021})}\BibitemShut {NoStop}%
\bibitem [{\citenamefont {Wright}\ \emph {et~al.}(2013)\citenamefont {Wright},
  \citenamefont {Blakestad}, \citenamefont {Lobb}, \citenamefont {Phillips},\
  and\ \citenamefont {Campbell}}]{Wright2013}%
  \BibitemOpen
  \bibfield  {author} {\bibinfo {author} {\bibfnamefont {K.~C.}\ \bibnamefont
  {Wright}}, \bibinfo {author} {\bibfnamefont {R.~B.}\ \bibnamefont
  {Blakestad}}, \bibinfo {author} {\bibfnamefont {C.~J.}\ \bibnamefont {Lobb}},
  \bibinfo {author} {\bibfnamefont {W.~D.}\ \bibnamefont {Phillips}},\ and\
  \bibinfo {author} {\bibfnamefont {G.~K.}\ \bibnamefont {Campbell}},\
  }\bibfield  {title} {\bibinfo {title} {Driving phase slips in a superfluid
  atom circuit with a rotating weak link},\ }\href
  {https://doi.org/10.1103/PhysRevLett.110.025302} {\bibfield  {journal}
  {\bibinfo  {journal} {Phys. Rev. Lett.}\ }\textbf {\bibinfo {volume} {110}},\
  \bibinfo {pages} {025302} (\bibinfo {year} {2013})}\BibitemShut {NoStop}%
\bibitem [{\citenamefont {Volosniev}\ and\ \citenamefont
  {Hammer}(2017{\natexlab{b}})}]{Volosniev2017}%
  \BibitemOpen
  \bibfield  {author} {\bibinfo {author} {\bibfnamefont {A.~G.}\ \bibnamefont
  {Volosniev}}\ and\ \bibinfo {author} {\bibfnamefont {H.-W.}\ \bibnamefont
  {Hammer}},\ }\bibfield  {title} {\bibinfo {title} {Flow equations for cold
  {Bose} gases},\ }\href {https://doi.org/10.1088/1367-2630/aa9011} {\bibfield
  {journal} {\bibinfo  {journal} {New Journal of Physics}\ }\textbf {\bibinfo
  {volume} {19}},\ \bibinfo {pages} {113051} (\bibinfo {year}
  {2017}{\natexlab{b}})}\BibitemShut {NoStop}%
\bibitem [{\citenamefont {Thouless}(1974)}]{Thouless1974}%
  \BibitemOpen
  \bibfield  {author} {\bibinfo {author} {\bibfnamefont {D.}~\bibnamefont
  {Thouless}},\ }\bibfield  {title} {\bibinfo {title} {Electrons in disordered
  systems and the theory of localization},\ }\href
  {https://doi.org/https://doi.org/10.1016/0370-1573(74)90029-5} {\bibfield
  {journal} {\bibinfo  {journal} {Physics Reports}\ }\textbf {\bibinfo {volume}
  {13}},\ \bibinfo {pages} {93} (\bibinfo {year} {1974})}\BibitemShut {NoStop}%
\bibitem [{\citenamefont {Cominotti}\ \emph {et~al.}(2014)\citenamefont
  {Cominotti}, \citenamefont {Rossini}, \citenamefont {Rizzi}, \citenamefont
  {Hekking},\ and\ \citenamefont {Minguzzi}}]{Cominotti2014}%
  \BibitemOpen
  \bibfield  {author} {\bibinfo {author} {\bibfnamefont {M.}~\bibnamefont
  {Cominotti}}, \bibinfo {author} {\bibfnamefont {D.}~\bibnamefont {Rossini}},
  \bibinfo {author} {\bibfnamefont {M.}~\bibnamefont {Rizzi}}, \bibinfo
  {author} {\bibfnamefont {F.}~\bibnamefont {Hekking}},\ and\ \bibinfo {author}
  {\bibfnamefont {A.}~\bibnamefont {Minguzzi}},\ }\bibfield  {title} {\bibinfo
  {title} {Optimal persistent currents for interacting bosons on a ring with a
  gauge field},\ }\href {https://doi.org/10.1103/PhysRevLett.113.025301}
  {\bibfield  {journal} {\bibinfo  {journal} {Phys. Rev. Lett.}\ }\textbf
  {\bibinfo {volume} {113}},\ \bibinfo {pages} {025301} (\bibinfo {year}
  {2014})}\BibitemShut {NoStop}%
\bibitem [{\citenamefont {Panochko}\ and\ \citenamefont
  {Pastukhov}(2019)}]{PANOCHKO2019}%
  \BibitemOpen
  \bibfield  {author} {\bibinfo {author} {\bibfnamefont {G.}~\bibnamefont
  {Panochko}}\ and\ \bibinfo {author} {\bibfnamefont {V.}~\bibnamefont
  {Pastukhov}},\ }\bibfield  {title} {\bibinfo {title} {Mean-field construction
  for spectrum of one-dimensional {Bose} polaron},\ }\href
  {https://doi.org/https://doi.org/10.1016/j.aop.2019.167933} {\bibfield
  {journal} {\bibinfo  {journal} {Annals of Physics}\ }\textbf {\bibinfo
  {volume} {409}},\ \bibinfo {pages} {167933} (\bibinfo {year}
  {2019})}\BibitemShut {NoStop}%
\bibitem [{\citenamefont {Brauneis}\ \emph {et~al.}(2021)\citenamefont
  {Brauneis}, \citenamefont {Hammer}, \citenamefont {Lemeshko},\ and\
  \citenamefont {Volosniev}}]{Brauneis2021}%
  \BibitemOpen
  \bibfield  {author} {\bibinfo {author} {\bibfnamefont {F.}~\bibnamefont
  {Brauneis}}, \bibinfo {author} {\bibfnamefont {H.-W.}\ \bibnamefont
  {Hammer}}, \bibinfo {author} {\bibfnamefont {M.}~\bibnamefont {Lemeshko}},\
  and\ \bibinfo {author} {\bibfnamefont {A.~G.}\ \bibnamefont {Volosniev}},\
  }\bibfield  {title} {\bibinfo {title} {{Impurities in a one-dimensional
  {Bose} gas: the flow equation approach}},\ }\href
  {https://doi.org/10.21468/SciPostPhys.11.1.008} {\bibfield  {journal}
  {\bibinfo  {journal} {SciPost Phys.}\ }\textbf {\bibinfo {volume} {11}},\
  \bibinfo {pages} {008} (\bibinfo {year} {2021})}\BibitemShut {NoStop}%
\bibitem [{\citenamefont {Jager}\ and\ \citenamefont
  {Barnett}(2021)}]{Jager2021}%
  \BibitemOpen
  \bibfield  {author} {\bibinfo {author} {\bibfnamefont {J.}~\bibnamefont
  {Jager}}\ and\ \bibinfo {author} {\bibfnamefont {R.}~\bibnamefont
  {Barnett}},\ }\bibfield  {title} {\bibinfo {title} {Stochastic-field approach
  to the quench dynamics of the one-dimensional {Bose} polaron},\ }\href
  {https://doi.org/10.1103/PhysRevResearch.3.033212} {\bibfield  {journal}
  {\bibinfo  {journal} {Phys. Rev. Research}\ }\textbf {\bibinfo {volume}
  {3}},\ \bibinfo {pages} {033212} (\bibinfo {year} {2021})}\BibitemShut
  {NoStop}%
\bibitem [{\citenamefont {Koutentakis}\ \emph {et~al.}(2022)\citenamefont
  {Koutentakis}, \citenamefont {Mistakidis},\ and\ \citenamefont
  {Schmelcher}}]{Koutentakis2022}%
  \BibitemOpen
  \bibfield  {author} {\bibinfo {author} {\bibfnamefont {G.~M.}\ \bibnamefont
  {Koutentakis}}, \bibinfo {author} {\bibfnamefont {S.~I.}\ \bibnamefont
  {Mistakidis}},\ and\ \bibinfo {author} {\bibfnamefont {P.}~\bibnamefont
  {Schmelcher}},\ }\bibfield  {title} {\bibinfo {title} {Pattern formation in
  one-dimensional polaron systems and temporal orthogonality catastrophe},\
  }\bibfield  {journal} {\bibinfo  {journal} {Atoms}\ }\textbf {\bibinfo
  {volume} {10}},\ \href {https://doi.org/10.3390/atoms10010003}
  {10.3390/atoms10010003} (\bibinfo {year} {2022})\BibitemShut {NoStop}%
\bibitem [{\citenamefont {Kehrein}(2006)}]{Kehrein2006}%
  \BibitemOpen
  \bibfield  {author} {\bibinfo {author} {\bibfnamefont {S.}~\bibnamefont
  {Kehrein}},\ }\href@noop {} {\emph {\bibinfo {title} {The Flow Equation
  Approach to Many-Particle Systems}}}\ (\bibinfo  {publisher}
  {Springer(Berlin)},\ \bibinfo {year} {2006})\BibitemShut {NoStop}%
\bibitem [{\citenamefont {Tsukiyama}\ \emph {et~al.}(2011)\citenamefont
  {Tsukiyama}, \citenamefont {Bogner},\ and\ \citenamefont
  {Schwenk}}]{Tsukiyama2011}%
  \BibitemOpen
  \bibfield  {author} {\bibinfo {author} {\bibfnamefont {K.}~\bibnamefont
  {Tsukiyama}}, \bibinfo {author} {\bibfnamefont {S.~K.}\ \bibnamefont
  {Bogner}},\ and\ \bibinfo {author} {\bibfnamefont {A.}~\bibnamefont
  {Schwenk}},\ }\bibfield  {title} {\bibinfo {title} {In-medium similarity
  renormalization group for nuclei},\ }\href
  {https://doi.org/10.1103/PhysRevLett.106.222502} {\bibfield  {journal}
  {\bibinfo  {journal} {Phys. Rev. Lett.}\ }\textbf {\bibinfo {volume} {106}},\
  \bibinfo {pages} {222502} (\bibinfo {year} {2011})}\BibitemShut {NoStop}%
\bibitem [{\citenamefont {Hergert}\ \emph {et~al.}(2016)\citenamefont
  {Hergert}, \citenamefont {Bogner}, \citenamefont {Morris}, \citenamefont
  {Schwenk},\ and\ \citenamefont {Tsukiyama}}]{HERGERT2016}%
  \BibitemOpen
  \bibfield  {author} {\bibinfo {author} {\bibfnamefont {H.}~\bibnamefont
  {Hergert}}, \bibinfo {author} {\bibfnamefont {S.}~\bibnamefont {Bogner}},
  \bibinfo {author} {\bibfnamefont {T.}~\bibnamefont {Morris}}, \bibinfo
  {author} {\bibfnamefont {A.}~\bibnamefont {Schwenk}},\ and\ \bibinfo {author}
  {\bibfnamefont {K.}~\bibnamefont {Tsukiyama}},\ }\bibfield  {title} {\bibinfo
  {title} {The in-medium similarity renormalization group: A novel ab initio
  method for nuclei},\ }\href
  {https://doi.org/https://doi.org/10.1016/j.physrep.2015.12.007} {\bibfield
  {journal} {\bibinfo  {journal} {Physics Reports}\ }\textbf {\bibinfo {volume}
  {621}},\ \bibinfo {pages} {165} (\bibinfo {year} {2016})},\ \bibinfo {note}
  {memorial Volume in Honor of Gerald E. Brown}\BibitemShut {NoStop}%
\end{thebibliography}%
\end{document}


\title{Supplementary Information for: Emergence of a Bose polaron in a small ring threaded by the Aharonov-Bohm flux}

	\author{Fabian Brauneis}
	\affiliation{Technische Universit\"{a}t Darmstadt$,$ Department of Physics$,$ 64289 Darmstadt$,$ Germany}
	
	\author{Areg Ghazaryan}
	\affiliation{Institute of Science and Technology Austria (ISTA)$,$ Am Campus 1$,$ 3400 Klosterneuburg$,$ Austria} 
	
	\author{Hans-Werner Hammer}
	\affiliation{Technische Universit\"{a}t Darmstadt$,$ Department of Physics$,$ 64289 Darmstadt$,$ Germany}
	\affiliation{ExtreMe Matter Institute EMMI and Helmholtz Forschungsakademie
  Hessen f\"ur FAIR (HFHF)$,$ GSI Helmholtzzentrum f\"ur Schwerionenforschung GmbH$,$ 64291 Darmstadt$,$ Germany}
	
	\author{Artem G. Volosniev}
	\affiliation{Institute of Science and Technology Austria (ISTA)$,$ Am Campus 1$,$ 3400 Klosterneuburg$,$ Austria}

\maketitle

\appendix
\renewcommand{\appendixname}{}
\renewcommand{\thesection}{Supplementary Note \arabic{section}}
\renewcommand{\theequation}{S\arabic{equation}}
\renewcommand{\thefigure}{S\arabic{figure}}
\renewcommand{\thetable}{S\arabic{table}}

\section{The AB flux coupled to bosons}
\label{App:CouplingToBosons}

In the main part of the paper, we consider the AB flux coupled to the impurity. Here, we discuss a more general scenario in which the AB flux is coupled also to the bosons. The corresponding Hamiltonian reads:
\begin{equation}
\mathcal{H}=h+H+V_{ib}+V_{bb},
\end{equation}
where $h = \frac{1}{2L^2}\left(-i \partial/\partial y+ \Phi_I\right)^2$ and $H=\frac{1}{2 L^2}\sum_i\left(-i\partial/\partial x_i+\Phi_B\right)^2$. As in the main part, we analyze the Schr{\"o}dinger equation in the frame co-moving with the impurity 
\begin{equation}
    \begin{split}
        &\bigg[-\frac{1}{2}\left(\sum_i\frac{\partial}{\partial z_i}\right)^2+\frac{1}{2}\sum_i\frac{\partial^2}{\partial z_i^2}+ \frac{(P+\Phi_I)^2}{2}+\frac{N}{2}\Phi_B^2\\
        &+L^2V_{ib} + L^2 V_{bb} + i(P+\Phi_I-\Phi_B)\sum_i\frac{\partial}{\partial z_i}\bigg]\Tilde{\Psi} =  E\Tilde{\Psi},
        \end{split}
        \label{eq:ABflux_bosons}
\end{equation}
Note the differences between this equation and the one where the AB flux coupled only to the impurity. In particular, one cannot introduce a single effective variable (cf.~$\mathcal{P}$) that parameterizes the energy of the system. Still, one can cast Eq.~(\ref{eq:ABflux_bosons}) into the form discussed in the main part
\begin{equation}
    \begin{split}
        &\bigg[-\frac{1}{2}\left(\sum_i\frac{\partial}{\partial z_i}\right)^2+\frac{1}{2}\sum_i\frac{\partial^2}{\partial z_i^2}+ \frac{k^2}{2}\\
        &+L^2V_{ib} + L^2 V_{bb} + i k\sum_i\frac{\partial}{\partial z_i}\bigg]\Tilde{\Psi} =  E_k\Tilde{\Psi},
        \end{split}
\end{equation}
where $k=P+\Phi_I-\Phi_B$ and $E_k=E-(P+\Phi_I)\Phi_B-(N-1)\Phi_B^2/2$. 
Therefore, one can use the methods and results of the main part to study a system with the AB flux coupled to bosons. 

\begin{figure}
    \centering
    \includegraphics[width=1\linewidth]{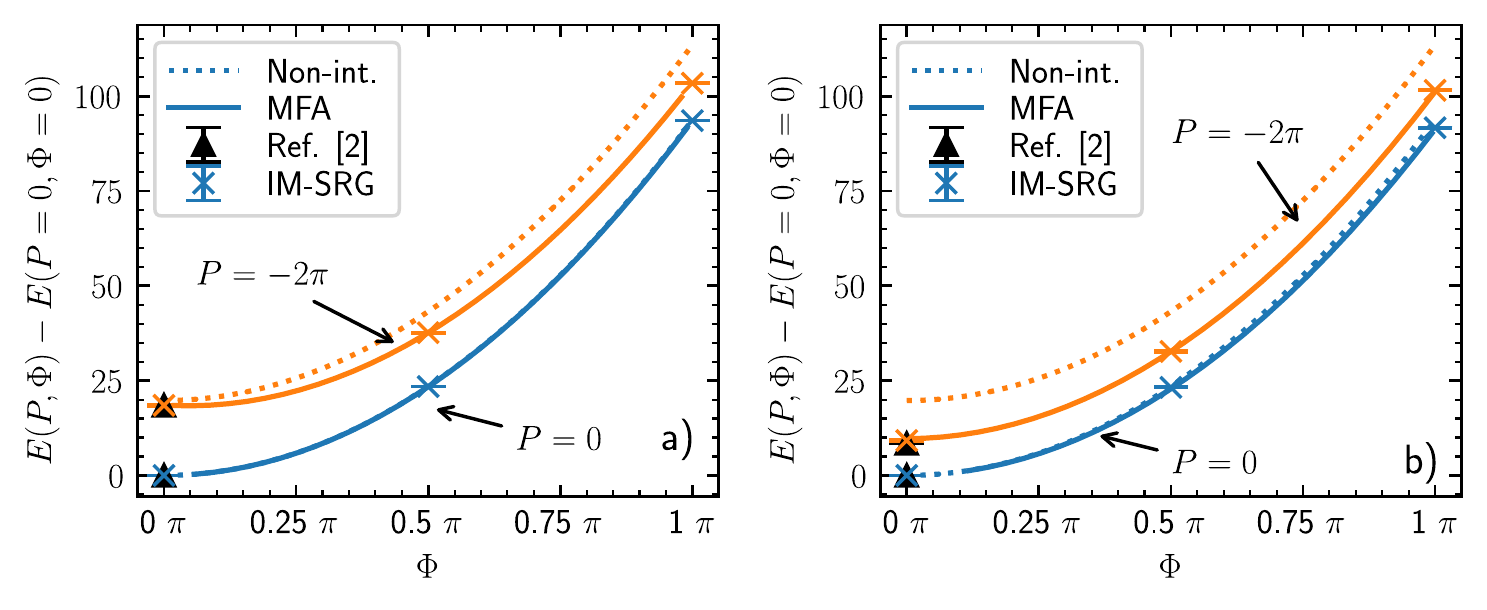}
    \caption{{\bf The Yrast energy spectrum as a function of the AB flux coupled to the bosons: $\Phi_I=0$, $\Phi_B=\Phi$.}\\
    Blue color shows $P=0$; orange color is for $P=-2\pi$. The parameters of the system are $N=19$ and $\gamma=0.2$.  Panel a) is for $c/g=1$; panel b) is for $c/g=5$. The data are obtained using the mean-field ansatz (solid curves) and the IM-SRG (crosses with errorbars), black triangles are the results of Ref.~\cite{Yang2022} for $\Phi=0$. The dotted curves show the energy of the non-interacting system.}
    \label{fig:Energy_coupled_bosons}
\end{figure}

We illustrate this in Fig.~\ref{fig:Energy_coupled_bosons}, which presents the energy for the parameters of Fig.~2 assuming that $\Phi_I=0$. The energies at $\Phi_B=0$ are not modified, but there is dramatic change for $\Phi_B>0$. First, the term $N\Phi_B^2/2$ leads to a faster change of the energy with the flux. Second, we see that  there is no energy level crossing for the Yrast states with $P=0$ and $|P|=2\pi$. While the energy spectrum must still be periodic in $\Phi_B$ with the period $\Phi_B/2\pi$, the energy is no longer parameterized by an effective total momentum $\mathcal{P}$, which means that the states with $0<\Phi_B<\pi$ are not connected to the states with $P=-2\pi$ and $\Phi_B>\pi$. 

Finally, we mention the special case of equal fluxes $\Phi_I=\Phi_B$.  This scenario corresponds to a `rotation' of all particles around the ring with the same `velocity'.  In this case, the energy depends on the AB flux only via a constant energy shift. The flux couples exclusively to the center-of-mass motion of the system, in agreement with previous studies, see, e.g.,~\cite{VIEFERS2004}.

\section{Mass imbalance}
\label{App:MassImbalance}

In the main part of the paper, we assume that the masses of the impurity and the bosons are equal. Here, we briefly discuss a mass-imbalanced system. In this case, the Schr{\"o}dinger equation in the co-moving frame reads:
\begin{equation}
    \begin{split}
        &\bigg[-\frac{1}{2m}\left(\sum_i\frac{\partial}{\partial z_i}\right)^2+\frac{1}{2}\sum_i\frac{\partial^2}{\partial z_i^2}+ \frac{(P+\Phi)^2}{2m}\\
        &+L^2V_{ib} + L^2 V_{bb} + i\frac{(P+\Phi)}{m}\sum_i\frac{\partial}{\partial z_i}\bigg]\Tilde{\Psi} =  E\Tilde{\Psi}.
        \end{split}
\end{equation}
The corresponding Gross-Pitaevskii equation is:
\begin{equation}
	-\frac{1}{2\kappa}\frac{\partial^2 f}{\partial z^2}+i\frac{\mathcal{P}}{m}\frac{\partial f}{\partial z}-i(N-1)\frac{P_{\mathrm{bos}}}{m}\frac{\partial f}{\partial z}+g(N-1)|f|^2f=\mu f
\end{equation}
with $\kappa=m/(m+1)$ the reduced mass. (Recall that we use a system of units in which $\hbar=M=1$.) Note that $\kappa$ is the only parameter that contains information about mass imbalance.  One can write the solution to the mean-field equation using the same approach (see \ref{App:Methods}), allowing us to transfer the conclusions of the main part of the paper to the mass-imbalanced system.

\begin{figure}
    \centering
    \includegraphics[width=1\linewidth]{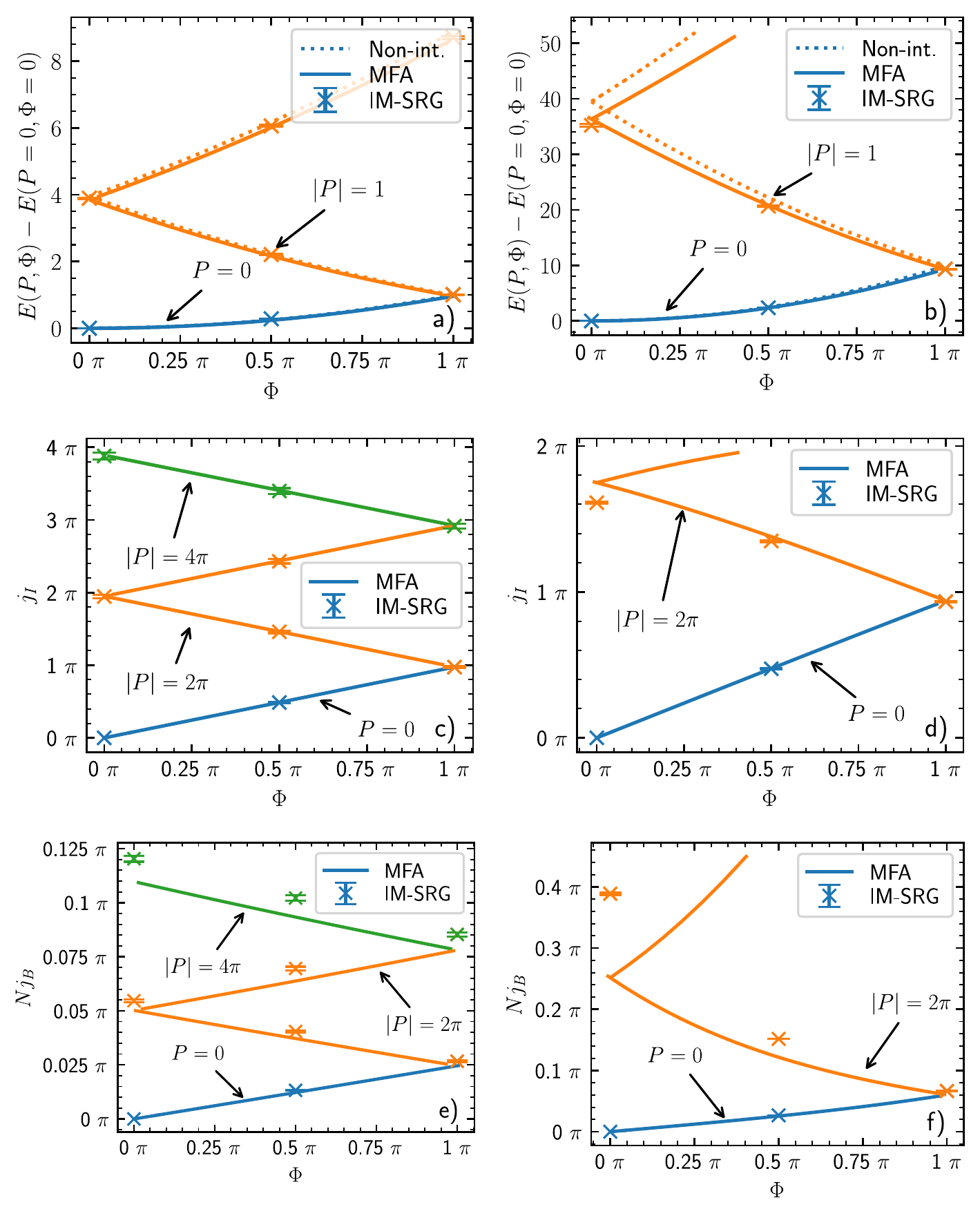}
\caption{{\bf Energy, impurity and bosonic current for mass imbalance.}\\
Energy [a), b)], impurity current [c), d)] and bosonic current [e), f)]. Panels a), c), e) are for a heavy impurity ($m=5$); b), d), f) are for a light impurity ($m=0.5$). The other parameters are $N=19$, $\gamma=0.2$ and $c/g=1$. Solid curves show the MFA results; symbols are the IM-SRG calculations. Dotted curves in the upper panels show the energies for the non-interacting case.}
\label{fig:MassIMbalanced}
\end{figure}

In Fig.~\ref{fig:MassIMbalanced}, we present the energies as well as the currents for a heavy ($m=5$) and a light ($m=0.5$) impurities. [Note that the definition of the impurity current should contain an additional factor $1/m$ in Eq.~8 for the mass-imbalanced case. In practice, to calculate $j_I$, we use conservation of momentum, i.e., we calculate the bosonic current first, and then use that $j_I=\mathcal{P}-Nj_B$.] First of all, note that the qualitative behavior of the observables is similar to that in the equal-mass case. The main difference is that a light (heavy) impurity generates stronger (weaker) bosonic currents: the kinetic energy of the impurity is large (small) and therefore it is energetically more (less) favorable to excite the Bose gas.  

Finally, we remark that for a heavy impurity, the mean-field results are in agreement with IM-SRG over a large range of $\Phi$ and $P$. This might be connected to the two observations: First, like discussed above, a heavy impurity leads to a weak bosonic current. Second, the mean-field approximation neglects the mixed derivative terms which scale as $1/m$, therefore, a higher impurity mass leads to a weaker effect of this simplification.  For a light impurity, the mean-field approximation is accurate only for low values of $\mathcal{P}$.

\section{Methods}
\label{App:Methods}

Here, we provide some technical details about the methods used to investigate the system. In particular, we discuss the semi-analytical mean-field ansatz, its validity, and the numerical IM-SRG method.

\subsection{Mean-field ansatz}
\label{App:Methods:MFA}

For the mean-field approximation, we use a product state in the co-moving frame 
\begin{equation}
\Tilde{\Psi}(z_1, z_2, ..., z_N)=\prod\limits_{i=1}^Nf(z_i),
\end{equation}
 where $f$ minimizes the expectation value of the Hamiltonian. Note that this ansatz is conceptually different from the strong-coupling approach in which the product state is written in a laboratory frame, and which predicts self-localization of the impurity~\cite{Sacha2006}. $f$ is computed using the Gross-Pitaevskii equation:
\begin{equation}
	-\frac{1}{2\kappa}\frac{\partial^2 f}{\partial z^2}+i\frac{\mathcal{P}}{m}\frac{\partial f}{\partial z}-i(N-1)\frac{P_{\mathrm{bos}}}{m}\frac{\partial f}{\partial z}+g(N-1)|f|^2f=\mu f,
\end{equation}
where $\mu$ is the chemical potential, $\mathcal{P}=P+\Phi$ is the `total' momentum, $\kappa=m/(m+1)$ is the reduced mass, and 
\begin{equation}
	P_{\mathrm{bos}}=-i\int f^*\frac{\partial f}{\partial z}dz
	\label{eq:PbosFromPhase}
\end{equation}
is the momentum of one boson. The impurity-boson interaction leads to the boundary condition:
\begin{equation}
	f'(0^+)-f'(0^-)= 2 c\kappa f(0).
\end{equation}
The derived Gross-Pitaevskii equation has been discussed before, see, e.g., Ref.~\cite{Cominotti2014}. For convenience of the reader, we summarize here the steps to find its solution semi-analytically.
To simplify the notation, we introduce the momentum
\begin{equation}
	P_{\mathrm{imp}}=\mathcal{P}-(N-1)P_{\mathrm{bos}}.
\end{equation}

We assume that 
\begin{equation}
f(z)=h(z)e^{i\theta(z)},
\end{equation} 
which leads to the two coupled differential equations:
\begin{align}
    0&=-h''+(\theta')^2h-2 P_{\mathrm{imp}}\theta' h \kappa/m + 2\kappa g(N-1)h^3-2 h\kappa \mu,
    \label{eq:GPEreal}\\
    0&=-\theta''h-2\theta'h'+2P_{\mathrm{imp}}h'\kappa/m,
    \label{eq:GPEimg}
\end{align}
whose solutions read:
\begin{widetext}
\begin{align}
	h(z)&=\sqrt{(s_2-s_1) \,\text{sn}\left(\left.\alpha +2 z
   (K(p)-\alpha )\right|p\right)^2+s_1}, \\
   \theta(z)&=\frac{\text{dn}\left(\left.\alpha +2 z (K(p)-\alpha
   )\right|p\right) \sqrt{\kappa g  (N-1) s_1 s_2 s_3}
   \,\Pi \left(1-\frac{s_2}{s_1};\left.\text{am}\left(\left.\alpha
   +2 z  (K(p)-\alpha )\right|p\right)\right|p\right)}{2
   s_1 (K(p)-\alpha ) \sqrt{1-p \,\text{sn}\left(\left.\alpha +2
   z (K(p)-\alpha )\right|p\right)^2}}+z P_{\mathrm{imp}}\kappa/m.
\end{align}
\end{widetext}
with $p=\frac{s_2-s_1}{s_3-s_1}$, $\alpha=F\left(\left.\sin
   ^{-1}\left(\sqrt{\frac{s_{min}-s_{1}}{s_{2}-s_{1}}}\right)\right|p\right)$ and $\beta=E\left(\left.\sin
   ^{-1}\left(\sqrt{\frac{s_{min}-s_{1}}{s_{2}-s_{1}}}\right)\right|p\right)$. $\text{sn}$ and $\text{dn}$ are Jacobi elliptic functions, $F$ ($K$) is the (complete) elliptic integral of first kind, $E$ the elliptic integral of second kind, $\Pi$ the elliptic integral of third kind and $\text{am}$ is the Jacobi amplitude~\cite{abramowitz1972handbook}.

The parameters $s_1, s_2, s_3$ determine the solution depending on the particle number and interactions. The boundary condition due to the impurity interaction gives an expression for $s_{min}$ as a function of $s_1, s_2, s_3$. Note that there exist three solutions. We take the physical solution that reproduces the behavior of the Bose polaron in the limit $\mathcal{P}\to0$ discussed in Ref.~\cite{Volosniev2017BosePolaron}.
   
To find the parameters $s_1, s_2, s_3$, we solve a set of equations that demand that the state solves the Gross-Pitaevskii equation, it is normalized, and its phase is periodic:
\begin{align}
1&=-\frac{(s_{1}-s_{3}) (\alpha -\beta -K(p)+E(p))}{\alpha -K(p)}+s_{1},\\
\theta(0)&=\theta(1),\\
0&=4 p (K(p)-\alpha )^2-\kappa g (N-1) (s_{2}-s_{1}).
\end{align}
These equations are solved numerically with $s_1\leq s_2\leq s_3$. (Other orderings produce solutions that do not reduce to the mean-field solution of Ref.~\cite{Volosniev2017BosePolaron} in the limit $\mathcal{P}\to0$.) 
The chemical potential is given by:
   \begin{equation}
   \mu=\frac{1}{2} g (s_1+s_2+s_3).
   \end{equation}
Having obtained the solution, we can calculate properties of the system, e.g. the energy of the system:
\begin{equation}
    E=\frac{\mathcal{P}^2}{2m}+\mu N-\frac{P_{\mathrm{bos}}^2N(N-1)}{2m}-\frac{gN(N-1)}{2}\int |f|^4 dz. \nonumber
\end{equation}

\subsection{Flow Equation Approach (IM-SRG)}
\label{App:IMSRG}

{\it Flow equations.} The flow equation approach (also called in-medium similarity renormalization group or IM-SRG) is a beyond-mean-field  method that (block)-diagonalizes the Hamiltonian in second quantization,
\begin{equation}
    H=\sum\limits_{i,j}A_{ij}a_i^\dagger a_j + \sum\limits_{i,j,k,l}B_{ijkl}a_i^\dagger a_j^\dagger a_k a_l,
\end{equation}
via the so-called flow equation
\begin{equation}
    \frac{dH}{ds}=[\eta, H].
    \label{eq:FlowEquation}
\end{equation}
Here, $s$ is the flow parameter, which formally plays a role of (imaginary) time. The generator of the flow $\eta$ should be chosen such that the off-diagonal matrix elements vanish in the limit $s\to\infty$~\cite{Kehrein2006}. 

As we are only interested in the lowest energy state of the system for a given value of $\mathcal{P}$, we normal order the Hamiltonian using a  condensate as a reference state, see Ref.~\cite{Volosniev2017}. This leads to the normal-ordered Hamiltonian
\begin{equation}
    H=E\mathbb{I}+\sum\limits_{i,j}f_{i,j}:a_i^\dagger a_j: + \sum\limits_{i,j,k,l}\Gamma_{ijkl}:a_i^\dagger a_j^\dagger a_k a_l:,\nonumber
\end{equation}
where we denote normal ordered operators with $:O:$. The matrix elements $f_{ij}$ and $\Gamma_{ijkl}$ describe one- and two-particle excitations from the reference state. For the generator, we use
\begin{equation}
    \eta(s)=\sum\limits_if_{i0}(s):a_i^\dagger a_0:+\sum\limits_{i,j}\Gamma_{ij00}(s):a_i^\dagger a_j^\dagger a_0 a_0: -\mathrm{h.c.},\nonumber
\end{equation}
these are the matrix elements that need to vanish in order to decouple the lowest energy state from the excitations. Once the flow equation reaches a steady state, the lowest energy state state is decoupled.

To calculate observables other than the energy, we evolve the operator $O$ together with the Hamiltonian, i.e., we solve the flow equation
\begin{align}
    \frac{dO}{ds}&=[\eta, O],
\end{align}
in addition to Eq.~(\ref{eq:FlowEquation}). We can calculate different observables, e.g., the current of a single boson,
\begin{equation}
    j_{\mathrm{bos}}=-i\sum\limits_{k,l}\int\phi_k(x)^*\frac{d\phi_l(x)}{dx}dx~ a^\dagger_ka_l,
\end{equation}
where $\phi_k$ is the one-body basis function in which we expand the Hamiltonian and $a_k$ ($a_k^\dagger$) are annihilation (creation) operators.

{\it Reference state.} The transformation governed by the flow equations can be understood as a mapping between the reference state and an eigenstate of the system. Since we are interested in a system of bosons, it is reasonable to use condensate as reference state. Our reference state is constructed iteratively: Starting from the ground state solution of the non-interacting Hamiltonian for $\mathcal{P}=0$, the density and phase are calculated with IM-SRG and used as a new reference state. This procedure is repeated until both, the phase and the density, do not change upon iteration. Then, $\mathcal{P}$ is increased by a small amount and the reference state is calculated again. We repeat this until the desired value of $\mathcal{P}$ is reached. 

Note that other choices for the reference state are possible in principle, e.g., the mean-field solution discussed in \ref{App:Methods}, see Ref.~\cite{Brauneis2021}. However, we observed that for the system under consideration the procedure outlined above allowed us to calculate properties of the system for a larger range of $\mathcal{P}$. 

{\it Accuracy of IM-SRG.} Higher order terms induced in evolution of Eq.~\eqref{eq:FlowEquation} make it impossible to find a numerically exact solution for the considered values of $N$, and should be truncated. In our truncation scheme, we truncate at the two-body level. Three-body operators that contain at least one $a_0^\dagger a_0$ operator are also kept, however, $a_0^\dagger a_0$ is treated as a $c$-number. This means that we work only with zero-, one- and two-body operators in Eq.~\eqref{eq:FlowEquation} which leads to a system of coupled, closed, non-linear differential equations, which we solve numerically~\cite{Volosniev2017, Brauneis2021, Brauneis2022}.

We estimate the error due to the neglected pieces (called $W$) using second order perturbation theory
\begin{equation}
    \delta E\simeq\sum\limits_p\frac{\left(\braket{\Phi_p|\int_0^\infty W(s)ds|\Phi_{\mathrm{ref}}}\right)}{\braket{\Phi_p|H|\Phi_p}-\braket{\Phi_{\mathrm{ref}}|H|\Phi_{\mathrm{ref}}}}\,,
\end{equation}
where $\Phi_p$ is a state that contains three-body excitations and $\Phi_{\mathrm{ref}}$ is our reference state. 
 
 We construct the Hamiltonian in second quantization using the eigenfunctions of the one-body Hamiltonian without momentum or flux ($\mathcal{P}=0$). Since we can only work with a finite Hilbert space, we solve the flow equations for different numbers of basis states (in our case $n\in[11, 13, 15, 17, 19, 21]$). For the energy, we fit these values with
\begin{equation}
    E(n)=E(n\to\infty)+\frac{b_1}{n^{b_2}},
\end{equation}
to estimate the result in infinite Hilbert space, $E(n\to\infty)$; $b_1$ and $b_2$ are fitting parameters. For other observables, such a fit is not always possible. In such cases, we  use instead the result for the largest Hilbert space ($n=21$) and estimate its error from the largest deviation to smaller Hilbert spaces ($n=11, 13, 15, 17, 19$). 

Therefore, there are in total two contributions to our error bars: The truncation error from neglecting higher order terms in the flow equation and the truncation error due to a finite Hilbert space. We assume that both errors are not correlated, and add their absolute values for our final error estimation

For a more detailed description of the method we refer to Ref.~\cite{Volosniev2017}, where the flow equations and our estimate of the truncation error are introduced, see also Ref.~\cite{Brauneis2021} for information about calculation of observables and a detailed explanation of our estimate of error bars.

To further validate our numerical results, we benchmark them against Ref.~\cite{Yang2022}. Overall we observe a good agreement between IM-SRG and Ref.~\cite{Yang2022}, see the figures in the main part of the paper. Only for $c/g=1$ and $\mathcal{P}=4\pi$ there are some small deviations. We interpret this deviation as follows: IM-SRG uses a condensate as a reference state, and fails to describe excited states of the Bose gas,
which become important when the impurity current reaches its critical value.
This argument is supported by the fact that the bosonic currents calculated with IM-SRG are always smaller than those in Ref.~\cite{Yang2022}. 
To improve the IM-SRG method, a reference state that includes boson-boson interactions could be used, like it is done in nuclear physics for fermionic systems, see e.g. Ref.~\cite{Hergert2013, Hergert2014}. Such references states are outside the scope of the present work.

\subsection{Validity of mean-field ansatz}
\label{App:ValidityMFA}

In the main part of the paper, we showed results for the energy (Fig.~2) and the currents (Fig.~4). For  $\mathcal{P}\leq 3\pi$, we observed a good agreement between the MFA and IM-SRG, justifying the use of the MFA. For larger values of $\mathcal{P}$, the MFA was less accurate, see, in particular the currents for $P=4\pi$.
Here, we use IM-SRG results to further test the validity of the MFA. 
 
\begin{figure}
    \centering
    \includegraphics[width=1\linewidth]{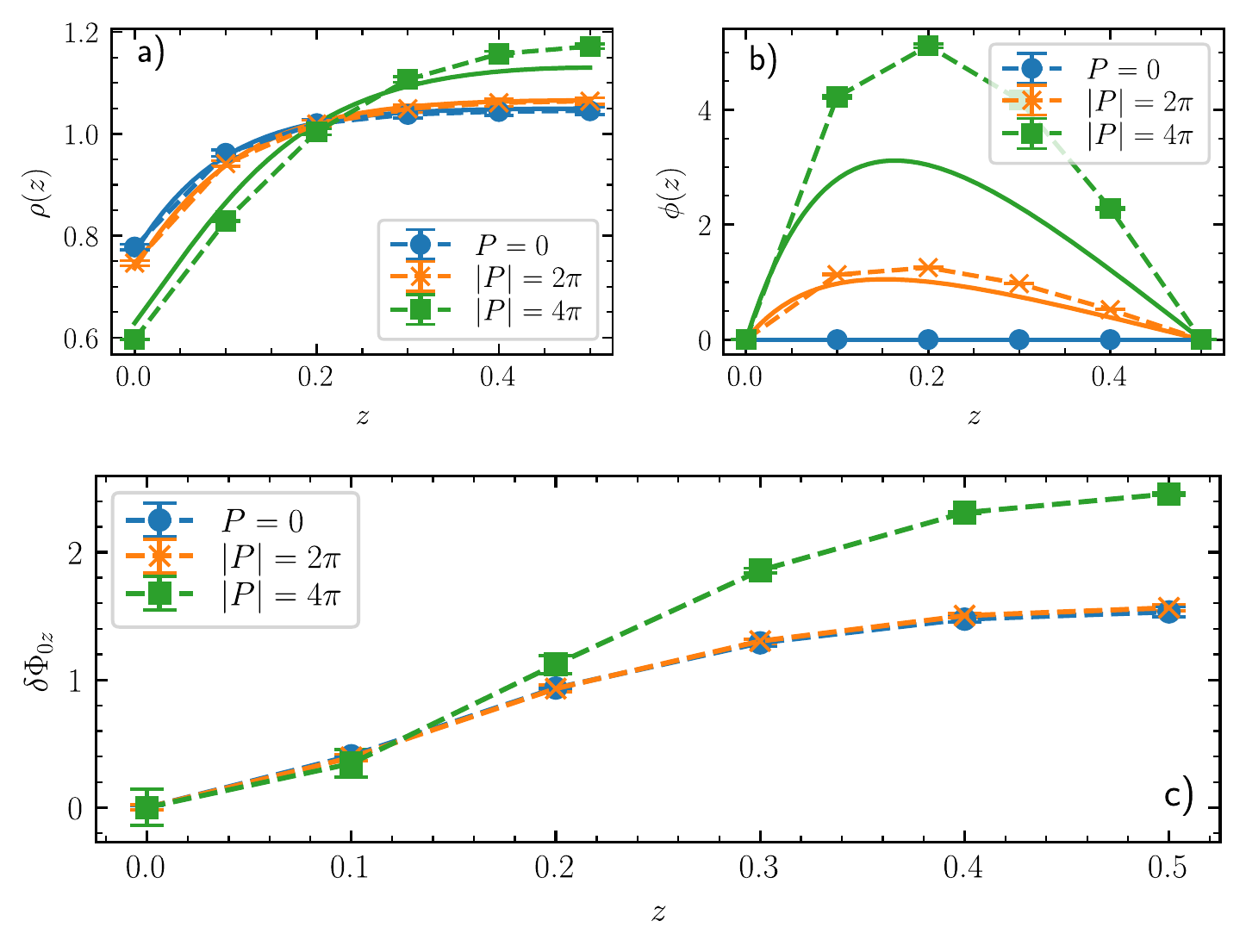}
\caption{{\bf Mean-field validation of density. phase and phase fluctuations with IM-SRG.}\\
Density (a)), phase (b)) and phase fluctuations (c)) for $N=19$, $M=m=1$, $\gamma=0.2$, $c/g=1$ and different total momenta $P$ (given in the legend) and $\Phi=0$. Solid lines show the results from the MFA; the symbols demonstrate the IM-SRG results. Dashed lines between these symbols are added to guide the eye.}
\label{fig:DensityPhasePhasefluc}
\end{figure}

To this end, we calculate the densities, $\rho(z)=|f(z)|^2$, and phases, $\theta(z)$, of the Bose gas  for $\Phi=0$ and $|P|=0, 2\pi, 4\pi$, see Fig.~\ref{fig:DensityPhasePhasefluc}. In our IM-SRG calculations these quantities can be expressed as follows:
\begin{align*}
    \rho(z)&=\sum_{i,j}\phi_i(z)\phi_j(z)a^\dagger_ia_j\,,\\
    \theta(z)&= \mathrm{Im}\left[\log\left(\sum_{i,j}\phi_i^*(0)\phi_j(z)a_i^\dagger a_j\right)\right].
\end{align*}
For $P=0$ and $|P|=2\pi$, there is a good agreement between the two methods. For $|P|=4\pi$, we start to see a difference, which is particularly noticeable in the phase. This is in agreement with our observation (see the main text) that  the IM-SRG and mean-field results for the currents disagree when the value of $\mathcal{P}$ is large. Recall that the bosonic and impurity currents can be related to the gradient of the phase using Eq.~\eqref{eq:PbosFromPhase}.

We can go one step further with the beyond-mean-field IM-SRG method and estimate directly the off-diagonal coherence of the bosons by calculating a property called phase fluctuations given by the one-body density matrix (see, e.g.,~\cite{Popov1983,Petrov2000,Pethick2002}):
\begin{equation}
    \rho(z,z')\equiv \braket{\Phi_{gr}|\rho(z,z')|\Phi_{gr}}=\sqrt{\rho(z)\rho(z')}\exp\left\{-\frac{\delta\Phi_{zz'}}{2}\right\}.
    \label{eq:phase_def}
\end{equation}
The quantity $\delta\Phi_{zz'}$ vanishes for a condensate. It is therefore a direct measure of the validity of the mean-field approximation. A plot for the same parameters as before is also shown in Fig.~\ref{fig:DensityPhasePhasefluc}. We see that for $P=0$ and $|P|=2\pi$ phase fluctuations are nearly identical and that for $|P|=4\pi$ there is a considerable increase. Together with the behavior of the energies, currents, phases and densities, we come to the conclusion that the mean-field approximation works well only for small values of $P$ and $\Phi$ such that $\mathcal{P}<3\pi$. 

Let us provide some physical intuition into why the MFA results become less accurate when we increase $P+\Phi$. To this end, we consider  the case of vanishing impurity-boson interaction strength, $c\to0$. In this limit, the bosons are described by the well-known Lieb-Liniger  gas~\cite{Lieb1963, Lieb1963_2}. There exist two different types of excitations in the Lieb-Liniger model: type-I and type-II excitations. 

\begin{figure}[t]
    \centering
    \includegraphics[width=1\linewidth]{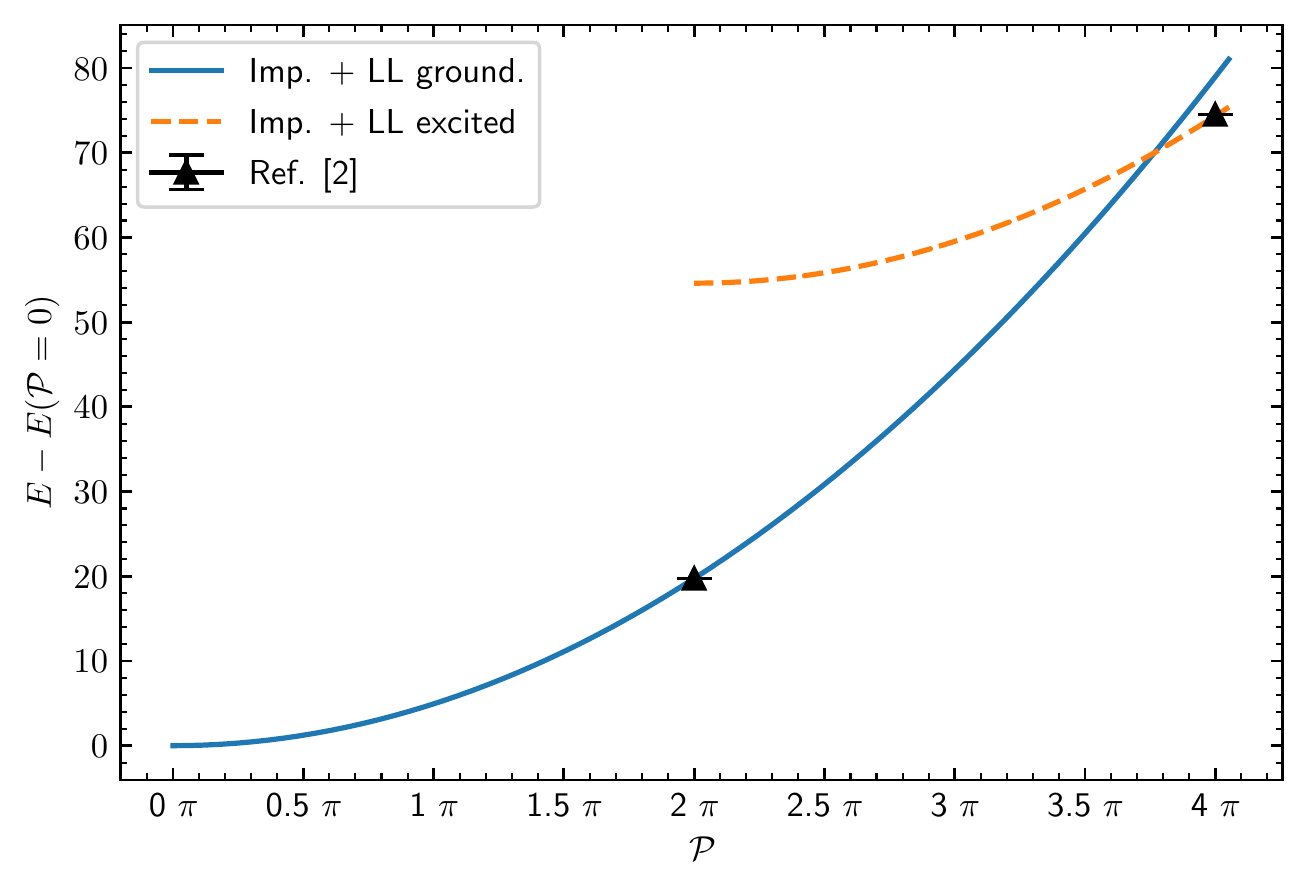}
\caption{{\bf Energy for a system of a non-interacting impurity ($c\to0$) in the Lieb-Liniger gas.}\\
The Lieb-Liniger gas is either in the ground (blue solid curve) or in the excited (orange dashed curve) states. The parameters of the system are $N=19$, $\gamma=0.2$, $m=M=1$. The triangles show 
the results from Ref.~\cite{Yang2022} for these parameters.}
\label{fig:LLLimit}
\end{figure}

To illustrate this discussion, we plot the energy of the system with $c=0$ (non-interacting impurity) as a function of $\mathcal{P}$ in Fig.~\ref{fig:LLLimit}. The blue (solid) curve represents the case when the bosons are in the ground state and all of the momentum is distributed to the impurity such that
\begin{equation}
    E(\mathcal{P})-E(\mathcal{P}=0)=\frac{\mathcal{P}^2}{2}.
\end{equation}
The orange (dashed) curve is obtained using the lowest energy state of the Bose gas with a quantized value of $|P|=2\pi$. The rest of the total momentum is given to the impurity
\begin{equation}
    E(\mathcal{P})-E(\mathcal{P}=0)=E_{\mathrm{bosons}}+\frac{(\mathcal{P}-2\pi/L)^2}{2}.
\end{equation}
We use $E_{\mathrm{bosons}}$ from the Yrast curve presented in Ref.~\cite{Yang2022}. For comparison in Fig.~\ref{fig:LLLimit}, we also show the results from Ref.~\cite{Yang2022} for $c=0$. The point where the curves cross imply a strong momentum exchange if $c\neq0$, which might be beyond IM-SRG and MFA. Note that this crossing can be observed directly only if $\Phi\neq0$.

We see that at small values of $\mathcal{P}$, it is energetically favorable to have all `vorticity' in the impurity. For some parameters $\mathcal{P}<4\pi$, however, it is more favorable to excite the Bose gas. This analysis is in agreement with the calculations of ~\cite{Yang2022} for quantized values of $P$. It becomes clear now why the mean-field results presented in the main part of the paper are accurate only for  $P=0$ and $|P|=2\pi$: The MFA  does not describe the excitations of the bosons for $c=0$ well.

The IM-SRG method faces a similar problem since we use a condensate as a reference state. For example, our flow-equation calculations for $c/g\ll 1$ diverge close to the point where it is energetically favorable to excite the bosons (e.g., $\mathcal{P}\simeq 3.8\pi$ in Fig.~\ref{fig:LLLimit}). In this way IM-SRG signals that a simple condensate description of the Bose gas in the co-rotating frame is no longer valid. It is worth noting that IM-SRG converges over a larger range of $\mathcal{P}$ if the impurity-boson interactions are stronger. In particular, IM-SRG is in agreement with Ref.~\cite{Yang2022} for $|P|=4\pi$ for strong impurity-boson interactions. This leads us to the question about the effect of the impurity-boson interactions on the excitations of the system, which determines the overlap with the reference state of the IM-SRG method, a condensate. Note that this question might be connected to the critical velocity of an impurity in a one-dimensional Bose gas. We leave this question to future studies.

\section{Convergence towards thermodynamic limit}
\label{App:ConvEnergy}

As the main part of the paper argues, an AB ring allows us to study the convergence of the effective mass to the thermodynamic limit ($N, L \to \infty$, $N/L=\rho=\text{const}$). Here, we discuss this few- to many-body crossover for the effective mass and the energy in more detail.
Note that the energy has been discussed already in Ref.~\cite{Volosniev2017BosePolaron}. In particular, it was found there that the energy decays as $E-E(c=0)=\rho^2\pi^2/(2N\kappa)$ for a non-interacting Bose gas with $P=\Phi=0$. For an interacting Bose-gas, the energy converges to a finite value that is determined by the polaron energy.

Below, we first use the results of Ref.~\cite{Volosniev2017BosePolaron} to provide some further analytical insight. After that, we use a fitting procedure to find the convergence of the energy and the effective mass in our numerical simulations.

\subsection{Energy}

Reference~\cite{Volosniev2017BosePolaron} shows that the energy of a system with an impenetrable impurity ($c\to\infty$) is

\begin{widetext}
\begin{equation}
\frac{E-E(c=0)}{\rho^2} = \frac{8K^4(p)p+2 K^2(p) \kappa \gamma N(N-1)(p+1)}{3\kappa^2\gamma N^2(N-1)}-\frac{\gamma(N-1)}{2},
\label{eq:energy_finiteN}
\end{equation}
\end{widetext}
where $K$ is the complete elliptic integral of first kind and $p$ is a parameter determined from the equation:
\begin{equation}
\frac{4K(p)(K(p)-E(p))}{\kappa\gamma N (N-1)}=1,
\end{equation}
with the complete elliptic integral of second kind, $E$, \cite{abramowitz1972handbook}. In the limit of $N\to\infty$, $p\to1$ and therefore $E(p)\to1$. This allows us to derive an analytic expression for $K(p)$ in terms of $ \gamma$ and $N$, which can be inserted into Eq.~(\ref{eq:energy_finiteN}) producing 
\begin{equation}
\frac{E-E(c=0)}{\rho^2}=\frac{2}{\kappa N}+\frac{4}{3\kappa}\sqrt{\kappa\gamma+\frac{1}{N^2}} + \mathcal{O}\left(\frac{1}{N^{3/2}}\right).
\end{equation}
In the limit of large $N$, the energy decays as  $2/\kappa N$ to the value $\sqrt{16\gamma/9\kappa}$, which corresponds to the boundary energy of the Bose gas (cf.~Refs.~\cite{Gaudin1971,Reichert2019} for an infinitely heavy impurity with $\kappa=1$).

\begin{figure}
    \centering
    \includegraphics[width=\linewidth]{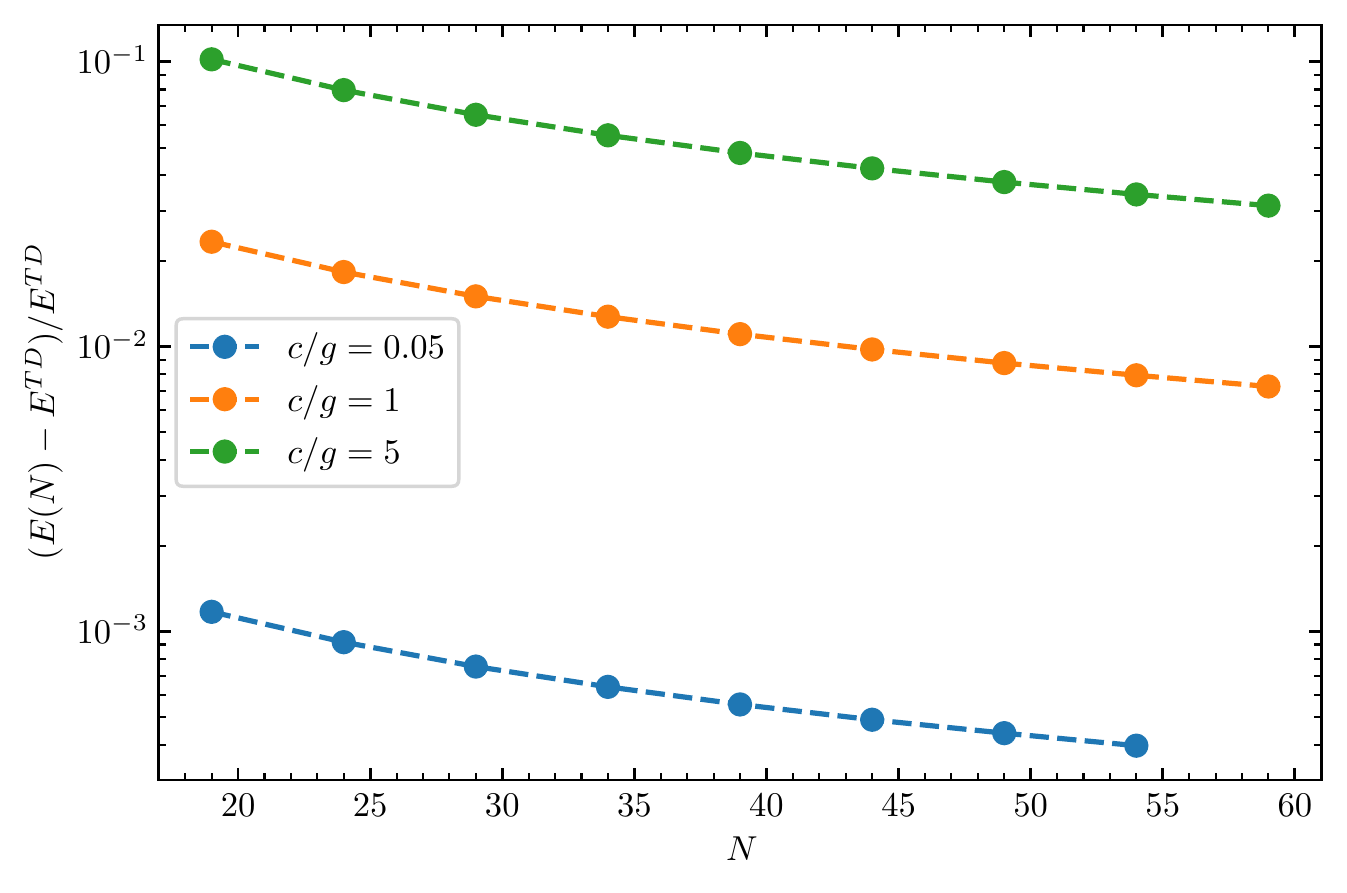}
    \caption{{\bf Convergence of energy to thermodynamic limit.}\\
    Energy as a function of $N$ with respect to its thermodynamic value for $m=M=1$, $\gamma=0.2$. The values of $c/g$ are quoted in the inset.}
    \label{fig:ConvergenceEnergy}
\end{figure}

Let us now consider finite impurity-boson interactions. To this end, we fit the IM-SRG results with
\begin{equation}
	f(N)=f_\infty+\frac{A}{N^\sigma}.
	\label{eq:app:fN}
\end{equation}
In Fig.~\ref{fig:ConvergenceEnergy}, we show the energy for $\gamma=0.2$ and $c/g=0.05, 1, 5$. Fitting reveals that the parameter $\sigma$ is in the range of $\sigma=1.065-1.08$ suggesting that also for finite impurity-boson interactions the energy decays like $1/N$. Small deviations from the $1/N$ behavior are probably due to relatively small values of $N$. We checked this statement by adjusting the number of particles.

\subsection{Effective mass}
It appears complicated to gain analytic insight into the few- to many-body crossover of the effective mass.  Therefore, we fit the results from Fig.~3 with Eq.~(\ref{eq:app:fN}). We find that contrary to the self-energy of the impurity the convergence seems to strongly depend on the strength of the impurity-boson interactions. For weak interactions ($c/g=0.05$), we find the fastest convergence with $\sigma=1.07\pm0.15$. For $c/g=1$, the effective mass convergences with $\sigma=0.99\pm0.01$. For strong interactions ($c/g=5$), the convergence is significantly slower $\sigma=0.67\pm0.03$.
To understand this, note that for an impenetrable impurity ($c\to\infty$), the impurity acts as a wall moving through the Bose gas. This means that the effective mass should be the mass of the whole system, i.e., the effective mass should actually increase with $N$ in this case. 

Our results demonstrate that the effective mass is highly sensitive to the value of $c/g$, unlike the energy.
In particular for strong impurity-boson interactions, one requires more bosons to observe convergence of the effective mass.

\section{Definition of effective mass}
\label{App:EffMass}

For convenience of the reader, here we review and explain equations that define the effective mass in the thermodynamic limit with $\Phi=0$, cf.~Refs.~\cite{Jager2020Deformation, Grusdt2017BosePolaron}. These equations also appear in the main part of the paper, although in a somewhat modified form.

The problem of a quantum impurity moving through a medium is simplified by mapping this many-body system to an effective one-body set-up -- polaron.
This quasi-particle is intuitively understood as the impurity dressed by the excitations of the medium, which lead to an effective mass higher than the bare mass.

In the polaron picture, the low-lying states of the system have (approximately) the energy
\begin{equation}
    E=E_0+\frac{P^2}{2m_{\mathrm{eff}}},
    \label{eq:app:E_P_meff}
\end{equation}
where $m_{\mathrm{eff}}$ is the effective mass and $P$ is total momentum of the system. Similar to Eq.~10, we can write the velocity (probability current) of the impurity as  \begin{equation}
    v_{\mathrm{imp}}=P/m_{\mathrm{eff}}. 
    \label{eq:app:v_imp}
\end{equation}
Sometimes, one employs this expression to define the probability current of the polaron as $v_{\mathrm{pol}}=v_{\mathrm{imp}}$. 

Using Eq.~(\ref{eq:app:v_imp}), we write the energy of the system as
\begin{equation}
    E=E_0+\frac{m_{\mathrm{eff}}v_{\mathrm{imp}}^2}{2},
    \label{eq:app:E_v_imp_meff}
\end{equation}
where $m_{\mathrm{eff}}$ is the effective mass and $v_{\mathrm{imp}}$ is the velocity of the impurity. 

Finally, we connect the effective mass to the momentum carried by the medium (in our case, the Bose gas). To this end, we use the fact that the total momentum is distributed between the impurity and the bosons $P=P_{\mathrm{imp}}+P_{\mathrm{bos}}$, and that (by definition) the impurity momentum is given by $P_{\mathrm{imp}}=m v_{\mathrm{imp}}$. With this, we derive 
\begin{equation}
    m/m_{\mathrm{eff}}=1-P_{\mathrm{bos}}/P.
    \label{eq:app:m_eff_P_bos}
\end{equation}

Eqs.~(\ref{eq:app:E_P_meff})-(\ref{eq:app:m_eff_P_bos}) are equivalent and can be used to compute the effective mass in numerical simulations. We noticed in our IM-SRG calculations that the last equation is numerically the most stable and therefore it was used in our study.

\newpage

\section*{Supplementary references}
\bibliography{refs}